\pdfoutput=1
\documentclass[10pt,a4paper]{article}
\usepackage{jheppub,bbold,graphicx,hyperref,multirow,subfig}
\usepackage[dvipsnames]{xcolor}
\usepackage[utf8]{inputenc}
\usepackage{listings}
\usepackage{booktabs}
\usepackage{float,ulem}
\usepackage{mathrsfs}
\usepackage{indentfirst}
\usepackage{verbatim}
\usepackage{color}
\usepackage{ulem}
\usepackage{amsmath, graphicx}
\usepackage{stackengine}
\newcommand{\beq}{\begin{equation}}
\newcommand{\eeq}{\end{equation}}
\newcommand{\bea}{\begin{eqnarray}}
\newcommand{\eea}{\end{eqnarray}}

\DeclareMathOperator*{\SumInt}{%
	\mathchoice%
	{\ooalign{$\displaystyle\sum$\cr\hidewidth$\displaystyle\int$\hidewidth\cr}}%
	{\ooalign{\raisebox{.14\height}{\scalebox{.7}{$\textstyle\sum$}}\cr\hidewidth$\textstyle\int$\hidewidth\cr}}%
	{\ooalign{\raisebox{.2\height}{\scalebox{.6}{$\scriptstyle\sum$}}\cr$\scriptstyle\int$\cr}}%
	{\ooalign{\raisebox{.2\height}{\scalebox{.6}{$\scriptstyle\sum$}}\cr$\scriptstyle\int$\cr}}%
}

\def\eps{\epsilon}
\def\lamhs{\lambda_{HS}}
\def\lams{\lambda_{S}}
\def\ztwo{\mathbb{Z}_2}

\newcommand{\vTs}{{\langle s \rangle}_{T_{\rm high}}}
\newcommand{\vTh}{{\langle h \rangle}_{T_{\rm high}}}
\newcommand{\vT}{{\langle h \rangle}_T}
\newcommand{\vs}{\langle s \rangle}

\newcommand{\eq}[1]{Eq.~(\ref{#1})}

\def\mh{m_h}
\def\ms{m_S}
\def\lamhs{\lambda_{HS}}
\def\lams{\lambda_{S}}
\def\lamh{\lambda_{H}}
\def\gev{~\text{GeV}}

\author[a]{Haibin~Chen}
\author[a,b,1]{\!\!, Yun~Jiang\note{Corresponding author.}}
\emailAdd{chenhb66@mail2.sysu.edu.cn}
\emailAdd{jiangyun5@sysu.edu.cn} 

\affiliation[a]{\textit{Department of Physics, School of Physics and Astronomy, \\Sun Yat-sen University (Zhuhai Campus), Zhuhai 519082, China}}

\affiliation[b]{\textit{Guangdong Provincial Key Laboratory of Quantum Metrology and Sensing, \\Sun Yat-sen University, Zhuhai, 519082, China.}}
\title{A Comprehensive Framework for Electroweak Phase Transitions: Thermal History and Dynamics from Bubble Nucleation to Percolation}

\date{\today}

\abstract{ 
The electroweak phase transition (EWPT) is crucial for cosmology and particle physics, with a profound impact on electroweak baryogenesis, symmetry breaking, and gravitational wave (GW) signals. However, many studies overlook key aspects of EWPT dynamics, leading to misidentified patterns and overestimated GW signals. 
To address these gaps, we present a comprehensive framework for analyzing EWPTs, focusing on the vacuum's thermal history and dynamics from bubble nucleation to percolation. 
Using the $\mathbb{Z}_2$-odd real scalar singlet model, we demonstrate the occurrence of spontaneous $\mathbb{Z}_2$ symmetry breaking in the high-temperature vacuum, leading to diverse EWPT processes, including multi-step transitions and inverse symmetry breaking. 
We identify four distinct EWPT patterns, each characterized by unique symmetry-breaking mechanisms and associated with bubbles exhibiting distinct field configurations, which can be analyzed using a formalism based on energy density distributions developed here. 
A key finding is that bubble nucleation fails in extremely strong phase transitions (PTs) with low nucleation rates, or in ultra-fast PTs involving inverse $s$-bubbles that collapse instantly upon formation, both of which lead to false vacuum trapping and the absence of observable GW signals. 
In first-order PTs where nucleation succeeds, stronger transitions occur later in the universe's evolution, while weaker transitions proceed more rapidly. 
Multi-step transitions involving (inverse) $\mathbb{Z}_2$ symmetry breaking give rise to complex transition sequences and exotic bubble dynamics, such as sequential nucleation or the coexistence of bubbles from different vacua---phenomena with significant implications for GW spectra, dark matter and baryogenesis. 
This work advances our understanding of EWPT dynamics and lays the groundwork for future studies of EWPTs in beyond-the-Standard-Model physics.
}

\begin{document}
\maketitle
\flushbottom

\newpage

%%%%%%%%%%%%%%%%%%%%%%%%%%%%%%%%%
\section{Introduction}
%%%%%%%%%%%%%%%%%%%%%%%%%%%%%%%%%

The Standard Model (SM) of particle physics, despite its remarkable success in explaining a wide range of phenomena, fails to account for the observed baryon asymmetry of the universe (BAU)~\cite{WMAP:2006bqn,Planck:2018vyg}. 
motivating the search for new physics beyond the SM (BSM). 
One of the compelling solutions is electroweak baryogenesis~\cite{Kuzmin:1985mm,Cohen:1993nk,Rubakov:1996vz,Morrissey:2012db}, which dynamically generates this asymmetry during the electroweak phase transition (EWPT) in the early universe. This mechanism is particularly appealing due to its intrinsic connection to electroweak symmetry breaking (EWSB) and its potential for experimental verification at particle colliders. For electroweak baryogenesis to succeed, the EWPT must be a first-order phase transition (PT)~\cite{Sakharov:1967dj}. Such a transition could also produce significant abundance of primordial black holes (PBHs), initially  proposed in~\cite{Hawking:1982ga,Kodama:1982sf} and developed through various generation mechanisms in recent works~\cite{Jung:2021mku,Liu:2021svg,Kawana:2021tde,Lewicki:2023ioy,Gouttenoire:2023naa,Flores:2024lng,Lewicki:2024ghw}, as well as detectable gravitational wave (GW) signals~\cite{Kosowsky:1992rz,Caprini:2015zlo,Ruan:2018tsw,Caprini:2019egz,Liang:2021bde}. These phenomena provide a unique opportunity to probe EWPTs and test the underlying mechanism of EWSB.

However, lattice simulations indicate that a first-order EWPT within the SM is only possible if the Higgs boson mass is below approximately 70 GeV~\cite{Kajantie:1996mn,Fodor:1998ji}. The observed Higgs boson mass of 125.09 GeV, as measured at the Large Hadron Collider (LHC)~\cite{particle2024review}, confirms that the EWPT in the SM proceeds as a smooth crossover at temperature of $159.5 \pm 1.5$ GeV~\cite{DOnofrio:2014rug,DOnofrio:2015gop}. Consequently, achieving a first-order PT necessitates theoretical extensions to the SM. A broad range of BSM models, including singlet-extended models~\cite{Espinosa:1993bs,Profumo:2007wc,Ahriche:2007jp,Espinosa:2011ax,Curtin:2014jma,Profumo:2014opa,Kotwal:2016tex,Vaskonen:2016yiu,Beniwal:2017eik,Kurup:2017dzf,Alves:2018jsw,Carena:2019une,Kozaczuk:2019pet,Ellis:2022lft,Niemi:2021qvp,Harigaya:2022ptp}, two-Higgs doublet models (2HDMs)~\cite{Cline:1996mga,Fromme:2006cm,Ginzburg:2009dp,Cline:2011mm,Tranberg:2012jp,Dorsch:2013wja,Dorsch:2014qja,Basler:2016obg,Dorsch:2016nrg,Dorsch:2017nza,Bernon:2017jgv,Su:2020pjw,Fabian:2020hny,Zhou:2020irf,Aoki:2021oez,Goncalves:2021egx,Biekotter:2021ysx,Basler:2021kgq,Wang:2022yhm,Biekotter:2022kgf}, triplet models~\cite{Patel:2012pi,Inoue:2015pza,Chala:2018opy,Niemi:2020hto,Zhou:2022mlz}, 
supersymmetric models~\cite{Huet:1995sh,Cline:1996cr,Worah:1997ni,Cline:1998hy,Cline:2000nw,Ham:2004nv,Huber:2006wf,Funakubo:2009eg,Carena:2011jy,Huang:2014ifa,Bi:2015qva,Baum:2020vfl}, 
and models with effective operators~\cite{Ham:2004zs,Bodeker:2004ws,Grojean:2004xa,Delaunay:2007wb,Cai:2017tmh,Chala:2018ari,Ellis:2018mja,Phong:2020ybr,Camargo-Molina:2021zgz,Cai:2022bcf,Kanemura:2022txx,Qin:2024dfp,Chala:2024xll,Oikonomou:2024jms},  have been shown to support first-order transitions within specific parameter spaces. 

The study of cosmological phase transitions has evolved significantly over time. 
Early research primarily focused on identifying first-order EWPTs in various BSM models, as well as evaluating the critical temperature $T_c$ and the associated parameter $\xi_c\equiv v_c/T_c$, which quantifies the strength of the PT. It is well established that a strong first-order PT (characterized by $\xi_c>1$) is essential for generating sufficient baryon number asymmetry through sphaleron processes~\cite{Cohen:1993nk,Patel:2011th,Morrissey:2012db}.
In recent years, however, the focus has shifted towards studying the dynamics of PTs, beginning with the verification of bubble nucleation~\cite{Anderson:1991zb}. The nucleation temperature, $T_n$, at which bubble formation becomes statistically probable, is determined by the nucleation rate. This rate can be estimated semi-analytically using the thin-wall approximation~\cite{Coleman:1977py,Linde:1981zj}, which assumes that the size of nucleated bubbles is much larger than the thickness of their walls. While useful, the applicability of the thin-wall approximation and the theoretical uncertainties in calculating $T_n$ remains inadequately explored. 
These uncertainties must be carefully controlled, as $T_n$ plays a critical role in predicting the GW power spectrum~\cite{Caprini:2015zlo}. 
Furthermore, it has become clear that even when a strong first-order PT ($\xi_c>1$) is theoretically expected, a low nucleation rate can prevent the transition from proceeding as anticipated~\cite{Kurup:2017dzf}. In such cases, the universe could remain trapped in a false vacuum state~\cite{Kurup:2017dzf,Baum:2020vfl,Biekotter:2021ysx,Biekotter:2022kgf}, 
potentially leading to catastrophic inflation~\cite{Guth:1982pn,Guth:2007ng} and an improper electroweak (EW) vacuum that fails to describe the  present universe.  
Recent numerical studies within the context of the 2HDM~\cite{Biekotter:2022kgf} has demonstrated that the condition for effective bubble nucleation imposes an upper limit on $\xi_c$, with a maximum value of approximately 1.8. 
This constraint sharply narrows the parameter space in which a successful first-order PT could occur and be experimentally testable at particle colliders~\cite{Kurup:2017dzf,Biekotter:2022kgf}. 

Another important insight emerging from recent studies is the recognition of high-temperature vacuum phases where symmetry breaking occurs~\cite{Espinosa:2004pn,Baldes:2018nel,Meade:2018saz,Carena:2019une,Matsedonskyi:2020mlz,Biekotter:2021ysx,Angelescu:2021pcd}, challenging the long-held assumption that no symmetry was broken in the early universe. The existence of such vacuum phases prior to EWPTs introduces fascinating phenomena, including inverse symmetry breaking (ISB)~\cite{Meade:2018saz,Carena:2019une,Matsedonskyi:2020mlz,Biekotter:2021ysx} and absolute symmetry non-restoration (SNR)~\cite{Meade:2018saz,Carena:2019une,Matsedonskyi:2020mlz,Biekotter:2021ysx,Biekotter:2022kgf}. 
These scenarios suggest a more intricate picture of PTs than the traditional single-step EWPT, especially in the context of multi-step PTs~\cite{Chung:2010cd,Patel:2012pi,Croon:2018new,Angelescu:2018dkk,Morais:2019fnm,Fabian:2020hny,Aoki:2021oez,Benincasa:2022jka,Cao:2022ocg,Liu:2023sey}, where distinct vacuum phases emerge sequentially. 
Typically, multi-step transitions introduce significantly more complex dynamics, and these additional complexities demand a detailed analysis of each step, from bubble nucleation to percolation, to fully understand the underlying processes.
In extreme cases, the interplay between successive transitions may lead to exotic bubble configurations, such as the coexistence of bubbles corresponding to different true vacuum states~\cite{Aguirre:2007an,Croon:2018new} or the formation of nested bubbles~\cite{Aguirre:2007an,Croon:2018new,Morais:2019fnm}. Moreover, multi-step PTs are likely to give rise to topological defects, such as domain walls, which can have a profound impact on the EWPT dynamics. 
For instance,  domain walls may catalyze PTs by enhancing nucleation rates~\cite{Blasi:2022woz,Agrawal:2023cgp}, or even introduce novel mechanisms that enable FOPTs by bypassing traditional bubble formation~\cite{Wei:2024qpy}. These complex phenomena greatly expand the theoretical landscape of PTs, offering profound implication for our understanding of the EWPT dynamics.

In this work, we consider an extension of the SM with a $\mathbb{Z}_2$-odd real scalar singlet as a case study, aiming to establish a comprehensive framework for analyzing PTs. The simplicity and broad applicability of this model make it an ideal prototype for systematically exploring a wide range of theoretical contexts. Our primary objective is to investigate high-temperature vacuum phases in which either the EW symmetry or the $\mathbb{Z}_2$ symmetry is broken, and to analyze the diverse EWPT processes that emerge from these scenarios, with a particular focus on multi-step PTs. 
Such transitions may involve the formation of bubbles whose exterior resides in a symmetry-broken phase, or bubbles consisting of two distinct fields. To accurately characterize the properties of these bubbles, we attempt to develop a new formalism tailored to multi-field bubble configurations. Additionally, we assess the validity of the thin-wall approximation and classify the various mechanisms by which bubbles are nucleated. 
Through this detailed analysis, we seek to deepen our understanding of PT dynamics throughout the thermal history of the universe and identify potential observational signatures that could reveal the EWPT processes the universe may have experienced.

This paper is organized as follows. After a brief description of the model under investigation in Section~\ref{sec:theory}, Section~\ref{sec:ana} analyzes symmetry breaking in the vacuum at high temperatures (above the EW scale) and summarizes the diverse EWPT processes that originate from this vacuum and occur throughout the thermal history of the universe. 
Section~\ref{sec:dyn} focuses on the dynamics of the EWPT and bubble evolution from nucleation to percolation. 
Special emphasis is placed on constraints that prevent vacuum trapping and on exotic bubble configurations resulting from successive FOPTs, which are further discussed in Section~\ref{ssec:twostep}. 
In Section~\ref{sec:bubbles}, we examine the properties of nucleated bubbles, including their radius and thickness. The applicability of the thin-wall approximation is also assessed, and the nucleation mode is discussed.  Section~\ref{sec:gwparas} explores the potential of GWs as signals to distinguish between different patterns of PTs, where EWSB occurs differently. Finally, in Section~\ref{sec:concl} we summarize the key findings of this work and offer perspectives for future research.

%%%%%%%%%%%%%%%%%%%%%%%%%%%%%%%%%
\section{Effective potential description of the model} \label{sec:theory}
%%%%%%%%%%%%%%%%%%%%%%%%%%%%%%%%%
\subsection{Zero-temperature model description}\label{sec:model}
%%%%%%%%%%%%%%%%%%%%%%%%%%%%%%%%%

We consider the simple extension of the SM with a real scalar singlet $S$ and impose a $\mathbb{Z}_2$ symmetry, under which $S \to -S$. Provided that $S$ does not acquire a vacuum expectation value (vev), it is absolutely stable and thereby provides a possible candidate for dark matter~\cite{Cline:2013gha,GAMBIT:2017gge}. 
The renormalizable tree-level scalar potential is
\beq
	V_0(H,S)=-\mu_H^2 H^{\dagger}H+\lambda_H(H^{\dagger}H)^2+\lambda_{HS}H^{\dagger}HS^2-\frac{1}{2}\mu_S^2S^2+\frac{1}{4}\lambda_SS^4
	\label{Vtree}
\eeq

This potential is bounded from below if and only if the following conditions are satisfied:
\beq
	\begin{aligned}
		\lambda_H&>0, \quad 
		\lambda_S&>0, \quad
		-\sqrt{\lambda_H\lambda_S}&<\lambda_{HS}
	\end{aligned}
\eeq

In addition, perturbative unitarity imposes an upper bound on the quartic couplings, requiring~\cite{Hashino:2016rvx}
\beq
	6\lamh+3\lams+\sqrt{(6\lamh-3\lams)^2+16\lamhs^2}<16\pi
\eeq

After spontaneous breaking of the electroweak symmetry, the Higgs field $H$ acquires a vev. We parametrize the $H$ and $S$ fields as follows:
\beq
	H=\frac{1}{\sqrt{2}}	
	\left(\begin{array}{c}  
		G^+  \\ 
		v+\hat{h}+iG^0  \\  
	\end{array}
	\right), \quad S= \hat{s}
	\label{eq:Hfield}
\eeq
where $v=246.22~{\rm GeV}$ corresponds to the Higgs vev at zero temperature. This parametrization gives rise to two scalar mass eigenstates $(\hat{h},\hat{s})$ with squared-masses given by $\mh^2=2\lamh v^2=2 \mu_H^2$ and $\ms^2=\lamhs v^2-\mu_S^2$, along with three massless Goldstone bosons, $G^0, G^\pm$, which are absorbed by the corresponding gauge bosons $Z, W^\pm$. 
The $\hat{h}$ state is identified with the SM-like Higgs observed at the LHC, and we fix its mass at $\mh=125\gev$~\cite{ATLAS:2022vkf}. 
To facilitate the following analysis, we trade the bare mass-squared parameter $\mu_S^2$ for the physical mass of the singlet scalar $\ms$.
Consequently, this model is left with three independent parameters $(\ms,\lamhs,\lams)$. 
To escape the experimental bounds from the Higgs invisible decay, in this work we consider the mass of the singlet scalar in the range $\ms > \mh/2$. 

%%%%%%%%%%%%%%%%%%%%%%%%%%%%%%%%%
\subsection{Finite-temperature effective potential}
%%%%%%%%%%%%%%%%%%%%%%%

To investigate the EWPT of the model, we utilize the effective potential expressed in terms of the field condensates $h \equiv \langle h \rangle_T$ and $s \equiv \langle s \rangle_T$ at finite temperature. Schematically,  the one-loop effective potential takes the form:
\begin{align}
\label{eq:Veff}
	V_\text{eff,1-loop}(h,s,T)=V_0(h,s)+V_\text{CW}(h,s)+V_\text{CT}(h,s)+V_\text{1T}(h,s,T)+V_\text{daisy}(h,s,T)
\end{align}
where the first three terms are independent of $T$ and are given in Appendix~\ref{sec:vac}. In this section, we present the last two terms which are dependent on temperature $T$. 

The term $V_\text{1T}(h,s,T)$ accounts for the thermal effect at one-loop level and is given by~\cite{Dolan:1973qd}, 
\begin{align*}
	V_\text{1T}(h,s,T)=\frac{T^4}{2\pi^2}\sum_i n_i J_{B,F}\Big(\frac{m_i^2(h,s)}{T^2}\Big)
\end{align*}
Here the summation index $i$ runs over the contributions from the top quark, $W^\pm$, $Z$ gauge bosons, as well as all Higgs bosons and Goldstone bosons. For each particle species, $n_i$ denotes the number of degrees of freedom, and $m_i$ is the field-dependent mass, whose explicit form is given in Appendix~\ref{sec:fielddepmass}. The thermal integrals $J_{B,F}$ are defined as
\begin{align}
	J_{B,F}(y)=\pm \int^\infty_0 dx \ x^2\ln\Big[1\mp\exp\Big(-\sqrt{x^2+y}\Big)\Big].
	\label{eq:JBF}
\end{align}
where the upper (lower) sign corresponds to bosonic (fermionic) contributions. 

The term $V_\text{daisy}(h,s,T)$ deals with the leading-order resummation of the ring diagrams that are introduced to fix the infra-red divergence. There are two common approaches that are used to evaluate the daisy diagrams proposed by Parwani~\cite{Parwani:1991gq} and Arnold-Espinosa~\cite{Arnold:1992rz}. For practical reason, we use the latter approach in which $V_\text{daisy}(h,s,T)$ is given by,  
\begin{align}
	V_{\text{daisy}}(h,s,T)=-\frac{T}{12\pi}\sum_i n_i\left[\left( M_{i}^2(h,s,T)\right)^{\frac{3}{2}}-\left(m^2_{i}(h,s)\right)^{\frac{3}{2}} \right]
\end{align}
where $M_{i}^2$ are the thermal Deybe masses of the bosons 
\beq
	M_{i}^2(h,s,T)=\text{eigenvalues}(\widehat{M}^2_{i}(h,s)+\Pi_{i}(T))
\eeq
Here $\widehat{M}^2_{i}$ denotes the field-dependent mass matrix, as defined in Appendix~\ref{sec:fielddepmass}, while $\Pi_{i}(T)$ encodes the thermal corrections. The explicit form of $\Pi_{i}(T)$ for each degree of freedom is provided in Appendix~\ref{app:thermal}.

%%%%%%%%%%%%%%%%%%%%%%%%%%%%%%%%%
\section{Vacuum phases in the thermal history}\label{sec:ana}
%%%%%%%%%%%%%%%%%%%%%%%
\subsection{Symmetry breaking and high-temperature vacuum structures}
\label{ssec:SNR}
%%%%%%%%%%%%%%%%%%%%%%%

While it is commonly assumed that all the symmetries (including the EW symmetry) are preserved in the hot, early universe, it remains unclear how the EW symmetry evolves as the universe cools, or how it eventually breaks to reach a proper EW vacuum that is consistent with the measured Higgs mass at the LHC. Moreover, this assumption itself may be incorrect. 
By using \eq{eq:Veff}, one can numerically trace the trajectory of the true vacuum (the global minimum of \eq{eq:Veff} in field space) and reconstruct the thermal history of the EW symmetry. 

%%%%%%%%%%%%%%%%%%%%%%%%%%%%%%%%
\begin{figure}[t]
	\centering
	\includegraphics[width=0.65\textwidth]{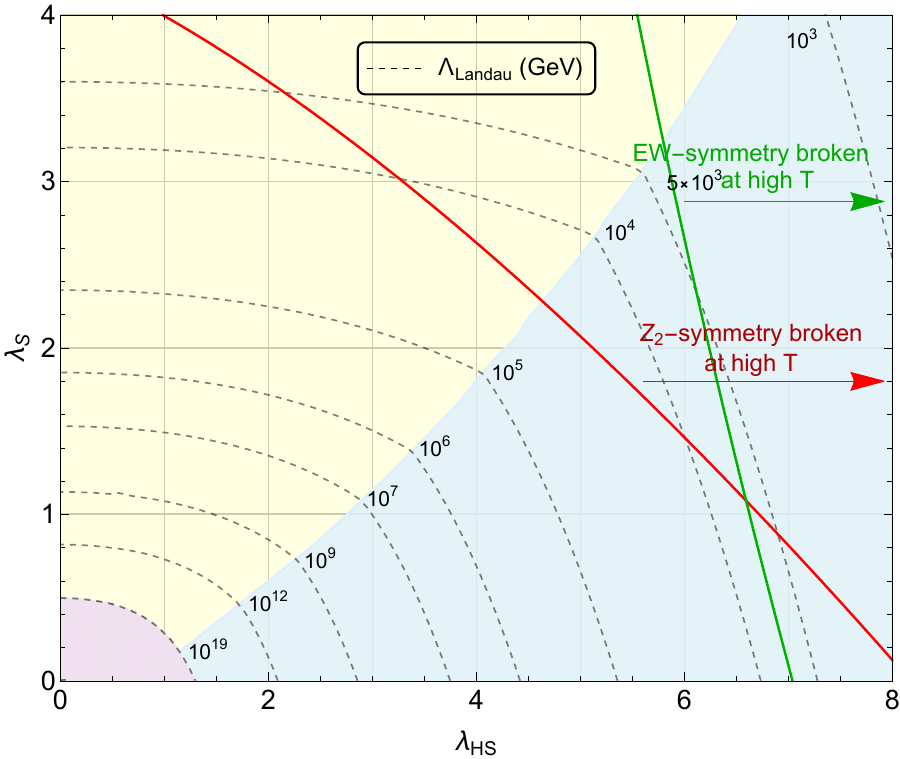}
	\caption{In the region bounded by the red and green curves, the high-temperature vacuum preserves both the EW and $\ztwo$ symmetries. Outside this region, a vacuum that breaks either EW- or $\ztwo$ symmetry emerges at sufficiently high temperatures. 
The dash lines indicate the mass scales above which the Landau pole arises due to the divergence or effectively exceeding the perturbative limit of $\lams$ (shaded in yellow) or $\lamhs$ (shaded in blue). In contrast, in the purple-shaded region the Landau pole remains absent up to the Planck scale, $m_{\rm Pl}=10^{19}~{\rm GeV}$. The renormalization group equations (RGEs) for the relevant couplings are provided in Appendix~\ref{app:RGEs}. 
	}
\label{fig:ldpole}
\end{figure}
%%%%%%%%%%%%%%%%%%%%%%%%%%%%%%%%

In fact, the assumption that the EW vacuum, when heated, resides at the origin holds only in the region bounded by the red and green curves in Fig.~\ref{fig:ldpole}.  
This results from the decoupling of thermal effects, and the bound associated with the relevant couplings, $\lamhs$ and  $\lams$, can be easily derived from the existence of a (local) minimum at the origin at very high temperature, $T_{\rm high}$. Mathematically, this condition requires that the Hessian matrix at the origin 
\beq
\label{eq:hen}
\mathbb{H}_{(0,0)}=\begin{pmatrix}
		\frac{\partial^2 V_\text{eff}}{\partial h^2} & \frac{\partial^2 V_\text{eff}}{\partial h \partial s}\\
		\frac{\partial^2 V_\text{eff}}{\partial s \partial h} & \frac{\partial^2 V_\text{eff}}{\partial s^2}
	\end{pmatrix}\Bigg|_{(0,0)}
\eeq
be positive definite at $T_{\rm high}$.\footnote{In general, this is a necessary but not sufficient condition, as the effective potential may have a global minimum away from the origin. However, we did not find such a scenario in this model because the contribution from the temperature-dependent terms significantly raises the effective potential in all directions around the origin.}
Since the off-diagonal terms are vanishing at the origin, a sufficient and necessary condition for \eq{eq:hen} to be positive definite is 
\beq
\label{eq:bound}
\frac{\partial^2 V_\text{eff}}{\partial h^2}\Bigg|_{(0,0)} >0,\quad \frac{\partial^2 V_\text{eff}}{\partial s^2}\Bigg|_{(0,0)} >0.
\eeq 
Keeping the dominant $T^2$ term in $V_{\rm 1T}$ and $V_{\rm daisy}$ components at $T_{\rm high}$, we find
\begin{align}
	\frac{\partial^2 V_\text{eff}}{\partial s^2}\Bigg|_{(0,0)}&\stackrel{\text{high T}}{\longrightarrow}c_s-\frac{1}{4\pi}\Big(4\lambda_{HS}\sqrt{c_h}+3\lambda_{S}\sqrt{c_s}\Big)>0 \label{eq:lamSsnrbound_s}\\
	\frac{\partial^2 V_\text{eff}}{\partial h^2}\Bigg|_{(0,0)}&\stackrel{\text{high T}}{\longrightarrow}c_h -\frac{1}{96\pi}\Big(3\sqrt{66}g^3+\sqrt{66}gg'^2+24(6\sqrt{c_h}\lambda_H+\sqrt{c_s}\lambda_{HS})\Big)>0	\label{eq:lamSsnrbound_h}
\end{align}
where $J'_{B(F)}$ denote the derivate of the bosonic (fermionic) thermal integral $J_{B(F)}(y)$ (defined in Eq.~\eqref{eq:JBF}) with respect to $y$ and  
\begin{align}
	c_h=\frac{1}{48}\Big(9g^2+3g'^2+12y_t^2+24\lambda_{H}+4\lambda_{HS}\Big),\quad
	c_s=\frac{1}{12}\Big(4\lambda_{HS}+3\lambda_{S}\Big).
\end{align}

In contrast, outside this bound, at $T \gtrsim T_{\rm high}$, either the $h$-field or the $s$-field can develop a non-zero condensate, leading to an EW-broken or $\ztwo$-broken vacuum.
This typically happens when large couplings induce substantial thermal effect. 
To simplify the analysis we fix one of the couplings, i.e. $\lams=3$ as an example. 
As the other coupling, $\lamhs$, increases, the $\ztwo$-broken vacuum does not appears until $\lamhs>3.2$. However, the situation becomes complicated once $\lamhs\gtrsim 5.8$, where a new local minimum of the effective potential appears in both the $h$-field and $s$-field directions. 
In most cases, it is not straightforward to directly determine the vacuum phase in which the universe resided. This determination depends on which minimum has the lowest value of the effective potential, which must be numerically evaluated at $T_{\rm high}$ to ascertain whether the EW or $\ztwo$ symmetry was broken. 
The opposite situation occurs for $\lams \lesssim 1$, but the $\ztwo$-symmetry was never broken at $T \gtrsim T_{\rm high}$. We will examine this further in Fig.~\ref{fig:thigheq2}.

%%%%%%%%%%%%%%%%%%%%%%%
\subsection{Diverse EWPTs to the EW vacuum}
\label{ssec:PTbreak}
%%%%%%%%%%%%%%%%%%%%%%%

%%%%%%%%%%%%%%%%%%%%%%%%%%%%%%%%
\begin{figure}[t]
	\centering
	\includegraphics[width=\textwidth, height=5cm]{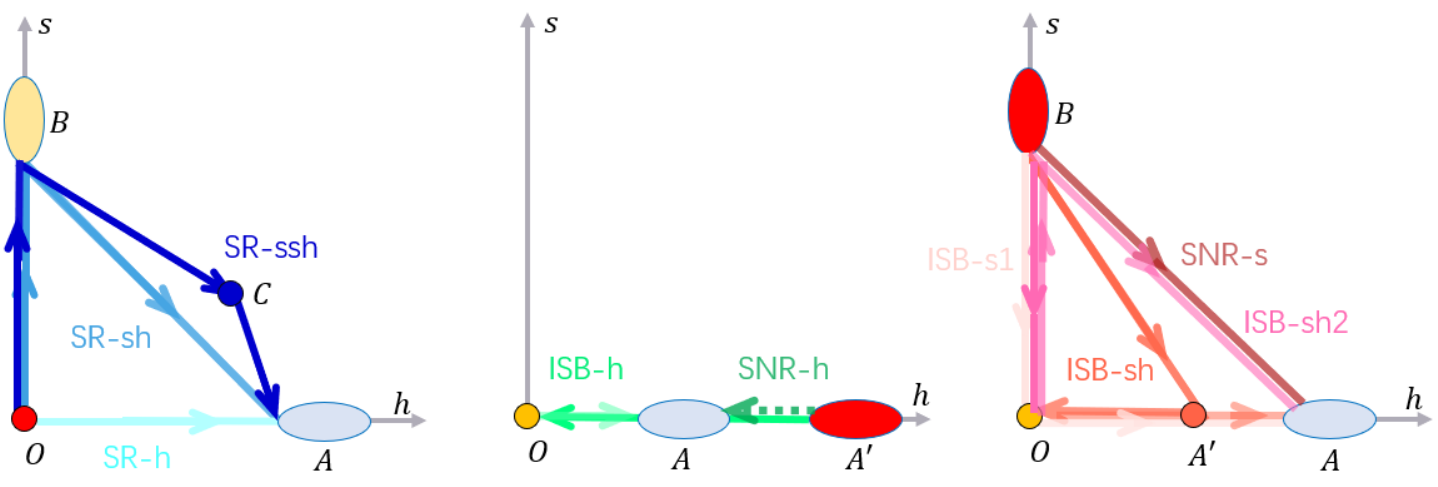}
	(a) \hspace{5cm} (b) \hspace{5cm} (c)
	\caption{A schematic diagram illustrating the thermal histories of the vacuum phase, starting from three different high-temperature vacuum phases: (a) the origin, (b) an EW-broken vacuum and (c) a $\mathbb{Z}_2$-broken vacuum. In each case, the arrow indicates the evolution of the vacuum as the temperature decreases. 
}
	\label{fig:SNRmode}
\end{figure}
%%%%%%%%%%%%%%%%%%%%%%%%%%%%%%%%

\begin{figure}[t]
\centering
\includegraphics[width=0.6\textwidth]{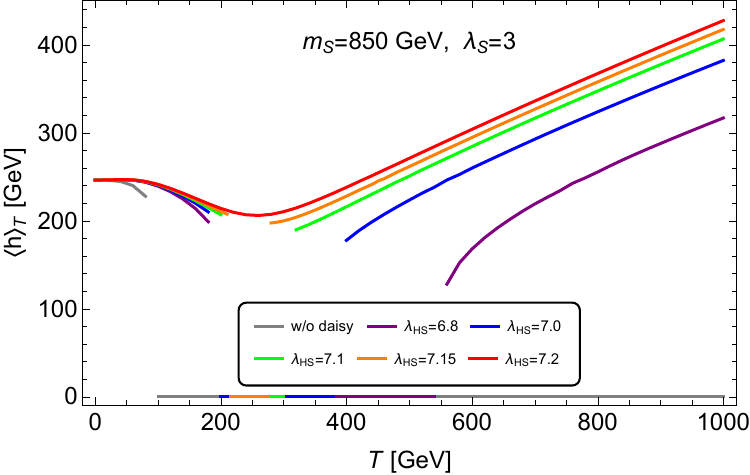}
\caption{Phase diagrams of the EW-broken vacuum scenario are presented for different values of $\lamhs$. The SNR-h EWPT is realized for a large coupling, specifically for $\lamhs=7.2$ (red line). In contrast, smaller values of $\lamhs$ (represented by other color lines) leads to an ISB-h EWPT.
}
	\label{fig:snr}
\end{figure}

Since we restrict the analysis to scenarios where the $\mathbb{Z}_2$ symmetry carried by the $S$ field remains unbroken at $T=0$, the EWPT must end up with an absolutely stable EW vacuum with $\langle s \rangle =0$ (point A in Fig.~\ref{fig:SNRmode}).
Possible thermal histories of the vacuum evolution are depicted in Fig.~\ref{fig:SNRmode}. 
Let us first consider the scenario of the EW-broken vacuum (Fig.~\ref{fig:SNRmode}(b)). In this scenario, the PTs begin from a vacuum phase with a non-zero Higgs condensate $\vTh =v'\ne 0$ (point $A'$), and subsequently proceed only within the $\vs_T=0$ plane. This implies that the $\ztwo$ symmetry remains unbroken throughout the entire process. The thermal evolution of the Higgs condensate $\vT$ for different values of $\lamhs$ is illustrated in Fig.~\ref{fig:snr}.  
As the temperature $T$ decreases, $\vT$ may either jumps back to the origin, resulting in an ISB
\begin{itemize}
	\item ISB-h: $A'(v',0) \to O(0,0) \to A(v,0)$
\end{itemize}
where the broken EW symmetry is restored, or it may remain non-zero due to enhanced couplings, thereby preventing a first-order PT at the EW scale
\begin{itemize}
	\item SNR-h: $A'(v',0) \to A(v,0)$
\end{itemize}
This latter scenario is called the SNR, a concept first suggested in Ref.~\cite{Weinberg:1974hy} and  further investigated in recent years within various BSM models~\cite{Baldes:2018nel,Meade:2018saz,Matsedonskyi:2020mlz,Carena:2021onl}. 
Notably, we find that SNR is closely tied to the contribution of daisy resummation.
When the daisy contribution is excluded, SNR no longer occurs, as illustrated by the gray line in Fig.~\ref{fig:snr}.

%%%%%%%%%%%%%%%%%%%%%%%%%%%%%%%%
%%%%%%%%%%%%%%%%%%%%%%%%%%%%%%%%

\begin{figure}[t]
	\centering
        \subfloat[SNR-s]{\includegraphics[width=0.46\textwidth]{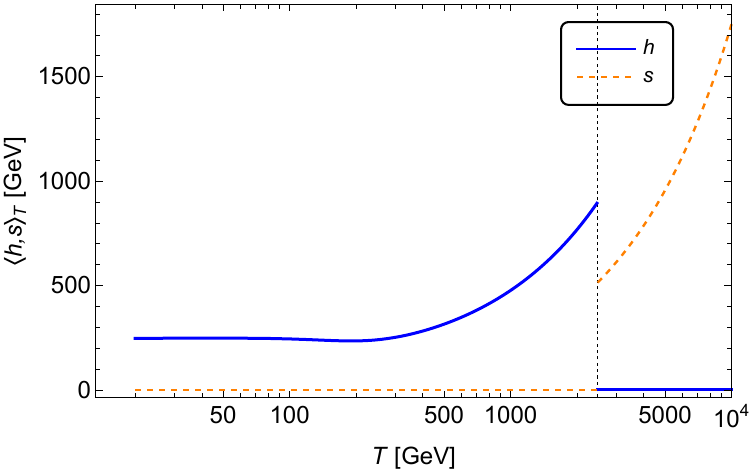}}
        	\subfloat[ISB-s1]{\includegraphics[width=0.45\textwidth]{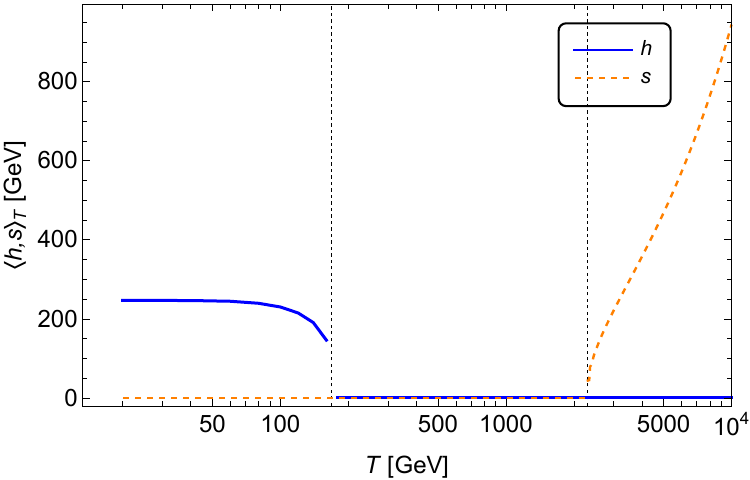}}\\
        \subfloat[ISB-sh]{\includegraphics[width=0.46\textwidth]{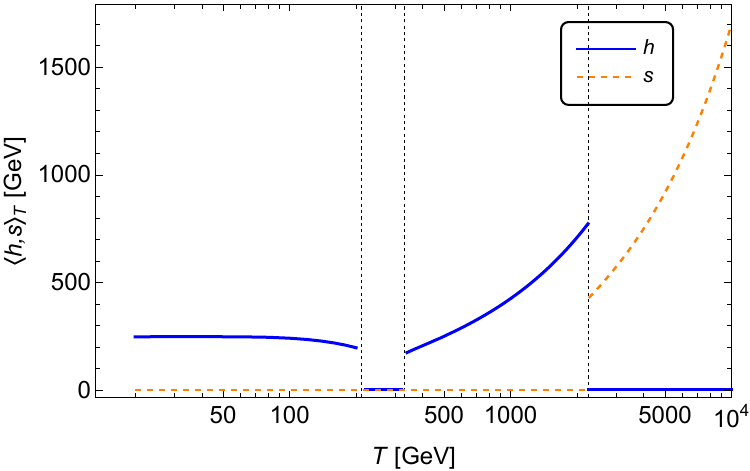}}
	\subfloat[ISB-sh2]{\includegraphics[width=0.45\textwidth]{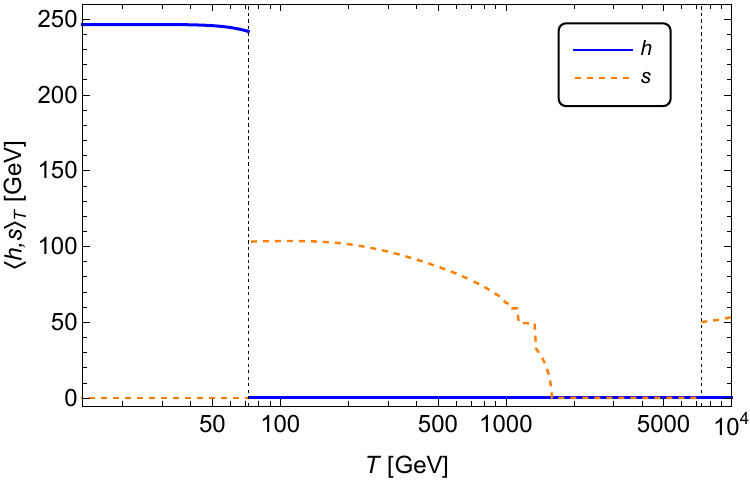}}
	\caption{Possible EWPT patterns originating from the $\mathbb{Z}_2$-broken vacuum scenario at $T_{\rm high} \gtrsim 3~{\rm TeV}$.}
	\label{fig:phase_diagram_SNR}
\end{figure}

In the scenario of the $\ztwo$-broken vacuum (Fig.~\ref{fig:SNRmode}(c)), prior to the EWPT, the universe resided in a vacuum where the $s-$field has a non-zero condensate $\vTs=w\ne 0$ (point $B$). Both the $h$ and $s$ condensates participate in the EWPT, leading to a proper EW vacuum and the restoration of the $\ztwo$-symmetry at $T=0$ through various possible paths.
Similar to the scenario of the EW-broken vacuum, this scenario includes both one-step EWPT
\begin{itemize}
	\item SNR-s: $B(0,w) \to A(v,0)$
\end{itemize}
and those consisting of multiple steps, characterized by complicated thermal histories of the vacuum phase
\begin{itemize}
	\item ISB-s1: $B(0,w) \to O(0,0) \to A(v,0)$
	\item ISB-sh: $B(0,w) \to A'(v',0) \to O(0,0) \to A(v,0)$
	\item ISB-sh2: $B(0,w) \to O(0,0) \to B(0,w) \to A(v,0)$.
\end{itemize}
While each of these PTs may involve potential instances of ISB, they are primarily categorized based on alternating periods of symmetry restoration and non-restoration. The phase diagrams for each EWPT pattern are shown in Fig.~\ref{fig:phase_diagram_SNR}. Upon examining Fig.~\ref{fig:phase_diagram_SNR} (a), (b) and (c), we observe that at temperatures exceeding approximately 3~TeV, the $s$-field consistently remains in a non-zero vacuum state. As the temperature decreases, this state transitions either to point $A$ (or $A'$) or to point O via either a first-order or second-order EWPT. In contrast, Fig.~\ref{fig:phase_diagram_SNR} (d) reveals that a higher temperature is required for the $s$-field direction to manifest as a non-zero vacuum.

%%%%%%%%%%%%%%%%%%%%%%%%%%%%%%%%
%%%%%%%%%%%%%%%%%%%%%%%%%%%%%%%%

On the other hand, within the bound, there exists PTs starting from the vacuum phase (point O), where both the EW and $\ztwo$ symmetries were preserved at sufficiently high temperatures. 
\begin{itemize}
	\item SR-h: $O(0,0) \to A(v,0)$
	\item SR-sh: $O(0,0) \to B(0,w) \to A(v,0)$
	\item SR-ssh: $O(0,0) \to B(0,w) \to C(v',w') \to A(v,0)$
\end{itemize}
The SR-h EWPT is a one-step PT that proceeds only in the $h$-field, while the SR-sh and SR-ssh PTs involve both the $h$ and $s$ fields participating in the EWPT. 

\begin{figure}[t]
\centering
\includegraphics[width=0.76\textwidth]{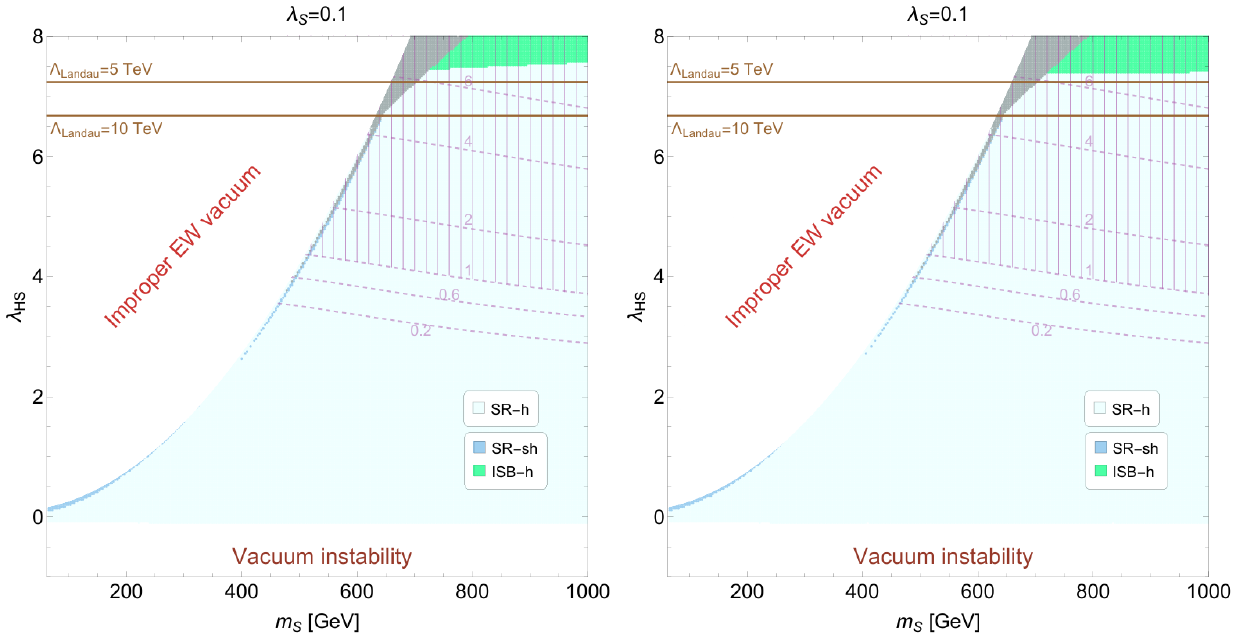}\\[5pt]
\includegraphics[width=0.76\textwidth]{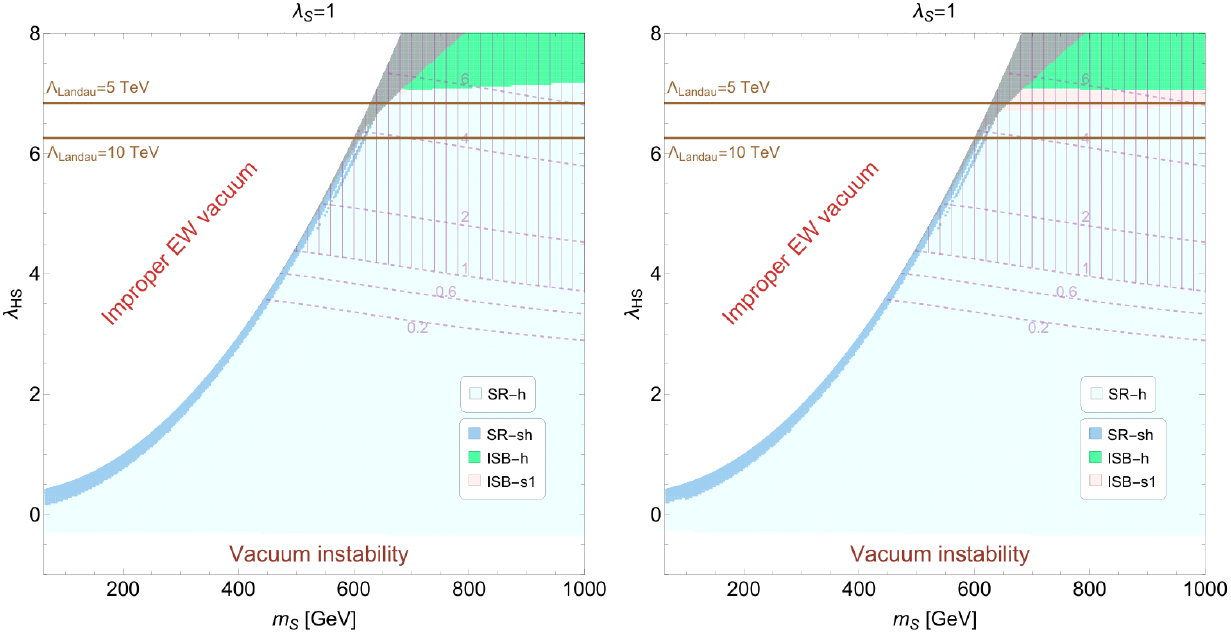}\\[5pt]
\includegraphics[width=0.76\textwidth]{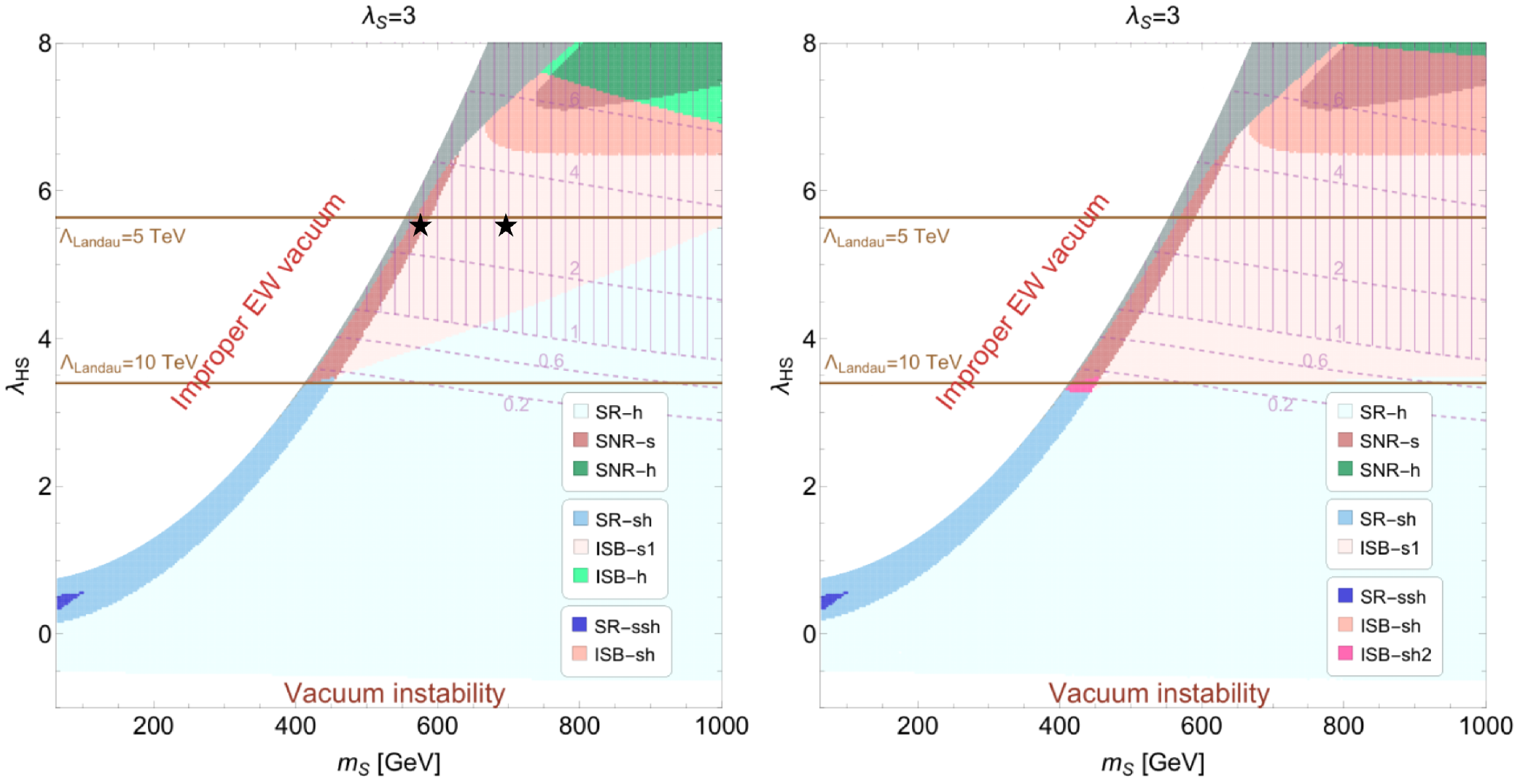}
	\caption{Possible PTs for different values of $\lams$ are shown. The identification of the EWPT pattern is sensitive to the choice of $T_{\rm high}$, which is set to $2~{\rm TeV}$ (left) and $10~{\rm TeV}$ (right). An improper EW vacuum develops at $T=0$ based on the tree-level (top-left region) and the one-loop (gray-shaded region) effective potentials. 
The purple contours show the ratio of the one-loop correction to the triple Higgs coupling $\lambda_{hhh}$ relative to its tree-level value. The purple hatched region marks the parameter space where this ratio exceeds one, indicating dominance of  the one-loop contribution over the tree-level term. The brown lines indicate the scale of the Landau pole.}
	\label{fig:thigheq2}
\end{figure}

For the points with an absolute stable EW vacuum at zero temperature (see details in Appendix~\ref{sec:vac})\footnote{In fact, a long-lived metastable EW vacuum remains consistent with current cosmological observations (see Sec.~\ref{ssec:proew} for further details).}, we use the \textsf{PhaseTracer} program~\cite{Athron:2020sbe} to determine the transition pattern, the order of the transitions (for each step of multi-step PTs), 
and the critical temperature $T_c$ for each first-order PT.
Striking examples for three different values of $\lams$ are provided in Fig.~\ref{fig:thigheq2}.
It is important to note that the identification of the EWPT pattern can sometimes depend on the choice of the high-temperature cutoff, $T_{\rm high}$, used in the analysis. 
To illustrate this sensitivity in the full parameter space, we show in Fig.~\ref{fig:thigheq2} the EWPT patterns that are possible in this model, using $T_\text{high}=2$~TeV (left panel) and $T_\text{high}=10$~TeV (right panel) as a comparison. 
Clearly, when $T_\text{high}$ increases from 2~TeV to 10~TeV, a significant portion of the SR-h region transforms into the ISB-s1 pattern. For $\lams=3$, the EW-broken vacuum scenario (which includes both the SNR-h and ISB-h patterns) in the top right corner shifts to the $\ztwo$-broken vacuum scenario, corresponding to the SNR-s and ISB-sh patterns, respectively. Additionally, the ISB-sh2 pattern appears in a small region of the SR-sh case. 
These changes are primarily due to the fact that as the temperature $T$ decreases, the vacuum PTs from the $\ztwo$-broken vacuum (phase B) to the EW-broken one (phase A) for large values of the $\lamhs$ coupling, or to the origin (phase O) for relatively weak coupling. This is illustrated in Fig.~\ref{fig:phase_diagram_SNR}, which shows the evolution of the vacuum phase. 
For example, in Fig.~\ref{fig:phase_diagram_SNR}(a), if we choose $T_{\rm high}\gtrsim 3~{\rm TeV}$, the universe would have initially resided in a $\ztwo$-broken vacuum, from which the EWPT starts, corresponding to the SNR-s pattern. In contrast, for a lower value of $T_{\rm high}$ (as typically considered in most analysis), an SNR-h EWPT will occur.

%%%%%%%%%%%%%%%%%%%%%%%%%%%%%%%%	
%%%%%%%%%%%%%%%%%%%%%%%%%%%%%%%%

In general, for $\lams\gtrsim 1$, the $\mathbb{Z}_2$-broken vacuum scenario typically arises for moderate values of $\lambda_{HS}$, while the EW-broken vacuum scenario is only realized for extremely large values of $\lambda_{HS}$. 
For sufficiently high $T_\text{high}$, the value of $\lamhs$ above which the $\mathbb{Z}_2$-broken vacuum scenario occurs becomes independent of $m_S$. In fact, this value coincides with the bound derived in Eq.~\eqref{eq:lamSsnrbound_s} (or Fig.~\ref{fig:ldpole}) for any given value of $\lams$. However, the bound that indicates the onset of the EW-broken vacuum scenario is inconsistent with the one given in Eq.~\eqref{eq:lamSsnrbound_h}. 
Above this bound for $\lamhs$, the effective potential in the $h$-field direction may possess a local minimum, but not necessarily a global one.
As a result, this bound does not guarantee the emergence of a stable EW-broken vacuum.
On the other hand, once $\lams \lesssim 1$, the $\mathbb{Z}_2$-broken vacuum phase no longer exists, regardless of the value of $T_{\rm high}$. In this case, when $\lamhs$ exceeds the bound in \eq{eq:lamSsnrbound_h}, the EW-broken vacuum phase appears at high temperatures, leading to an ISB-h EWPT. 

However, it is important to note that the results regarding the SNR and ISB PTs may not be entirely reliable. This is because the analysis above is based on the one-loop effective potential at finite temperature, which is known to suffer from several theoretical issues~\cite{Croon:2020cgk,Papaefstathiou:2020iag,Gould:2021oba,Athron:2022jyi,Ekstedt:2024etx}. As shown in Fig.~\ref{fig:thigheq2}, these EWPT patterns are typically generated by a large $\lamhs$, which falls within the non-perturbative regime, as discussed in previous analysis~\cite{Curtin:2014jma}. 
In such cases, loop-level effects can become significant. For example, the triple Higgs coupling $\lambda_{hhh}$ receive substantial radiative corrections at the one-loop level. In Fig.~\ref{fig:thigheq2}, we highlight the region where the one-loop contribution to $\lambda_{hhh}$ exceeds the corresponding tree-level value, indicating a potential breakdown of perturbativity. Nevertheless, we have explicitly verified that the $\mathbb{Z}_2$-broken vacuum persists at sufficiently hight temperature, with a slightly enhanced non-zero thermal vev $\langle s \rangle_T$ after incorporating two-loop effects, driven by the significant rise of the potential in the large $s$-field region.\footnote{This result is consistent with recent findings~\cite{Bahl:2024ykv,Bittar:2025lcr}. Two benchmark points (marked by stars in Fig.~\ref{fig:thigheq2}) from Pattern II and Pattern IV, characterized by relatively large $\lambda_S$ and $\lambda_{HS}$ couplings, are presented in Appendix~\ref{app:2loop} to illustrate the stability of vacuum configuration at the two-loop level . A more exhaustive analysis of higher-loop effects is left for future work.} 

Additionally, these transitions are associated with the appearance of a Landau pole $\Lambda$ at a few TeVs, which raises concerns from a theoretical perspective. 
For instance, if the model is free from a Landau pole below 5~TeV (i.e., if $\Lambda \gtrsim 5$~TeV), then PTs starting from an EW-broken vacuum become impossible, meaning that the EW symmetry remains unbroken prior to the EWPT.  
Consequently, the possible PTs associated with the EWSB in this model would either begin from the origin or from a $\ztwo$-broken vacuum, leading to the various EWPT patterns summarized in Table~\ref{tab:PTsum}. 
It can be observed that the smaller the value of $\lams$ is, the fewer EWPT patterns are possible.
Moreover, if one imposes the constraint that $\Lambda \gtrsim 10$~TeV, then only the SR-h and SR-sh are viable, provided that the three-step PTs are excluded. 
Furthermore, as noted in Ref.~\cite{Carena:2019une}, models exhibiting spontaneous $\mathbb{Z}_2$ symmetry breaking also encompass scenarios where $\mathbb{Z}_2$ symmetry is not restored, even within parameter spaces characterized by small couplings. This observation suggests that the Landau pole issue is likely not a significant concern in these cases.

%%%%%%%%%%%%%%%%%%%%%%%%%%%%%%%%
\begin{table}[t]
	\centering
	\vspace{-5pt}
	\renewcommand{\tablename}{Table}
	\caption{Possible EWPT patterns are shown assuming $\Lambda \gtrsim 5$~TeV. A single tick ($\surd$) indicates the presence of the corresponding EWPT, while a double tick ($\surd\surd$) signifies that the EWPT persists even when $\Lambda$ increases to $\gtrsim 10$~TeV.}
\begin{tabular}{c|l|l|ccc}
\toprule
Vacuum at $T_{\rm high}$ & EWPT patterns & Dynamics & $\lams=0.1$  & $\lams=1$ & $\lams=3$\\
\hline
\multirow{3}{*}{The origin} & SR-h (I) & one-step: $O \to A$ & $\surd\surd$ & $\surd\surd$ & $\surd\surd$ \\
&  SR-sh (III) & two-step: $O \to B \to A$ &  $\surd\surd$ & $\surd\surd$ & $\surd\surd$ \\
& SR-ssh & three-step: $O \to B \to C \to A$ & & &  $\surd\surd$\\
\hline
\multirow{3}{*}{$\ztwo$-broken vacuum}  & SNR-s (II) & one-step: $B \to A$ & & & $\surd$\\
& ISB-s1 (IV) & two-step:  $B \to O \to A$  & & $\surd$ & $\surd$\\
& ISB-sh2 & three-step: $B \to O \to B' \to A$ & & & $\surd\surd$\\	
\bottomrule
		\end{tabular}%
	\label{tab:PTsum}%
\end{table}
%%%%%%%%%%%%%%%%%%%%%%%%%%%%%%%%

%%%%%%%%%%%%%%%%%%%%%%%%%%%%%%%%
\section{Dynamics of the EWPTs}
\label{sec:dyn}

%%%%%%%%%%%%%%%%%%%%%%%
\subsection{Four patterns of EWPTs}
\label{ssec:PT}
%%%%%%%%%%%%%%%%%%%%%%%

\begin{figure}[t]
	\begin{center}	
	\includegraphics[width=0.92\textwidth]{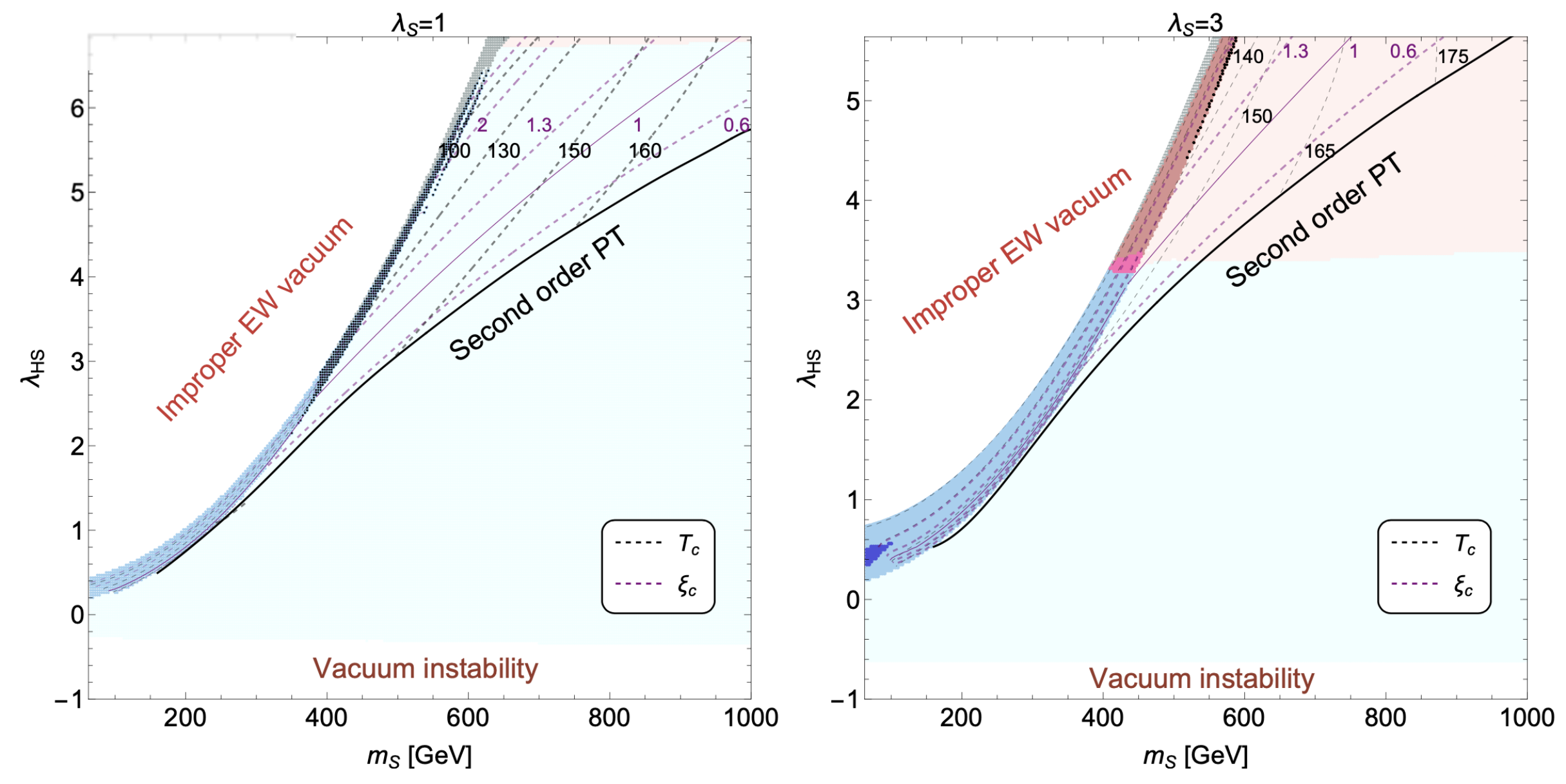}
	\end{center}
	\vspace{-5mm}
	\caption{The same as Fig.~\ref{fig:thigheq2}. Additionally, black dots denote Pattern III-1 (left) and Pattern IV-1 (right), both of which involve two successive first-order PTs. Excluding these two EWPT Patterns, the values of $T_c$ and $\xi_c$ for the first-order PT in other Patterns are indicated by the black and purple contours, respectively. Below the solid curve, the EWPT is second order, and no bubbles are nucleated.	
} 
\label{fig:PT_pattern}
\end{figure}

Among the possible EWPT patterns listed in Table~\ref{tab:PTsum}, there are a variety of possibilities for phase changes,  including SB and ISB for either symmetry. Interestingly, the same phase change can occur in different patterns of EWPT. The results are summarized in Table~\ref{tab:PTsum2}. 
All of these patterns consist of, at least, one step of the PT occurring at $T \lesssim v_{\rm EW}$ during which the SB occurs with the EW symmetry. 
Since the SR-ssh and ISB-sh2 PTs involve three-step processes and have complex thermal history of the vacuum phase, we will not analyze them in this work. Instead, we focus on the remaining four EWPT patterns: SR-h, SNR-s, SR-sh and ISB-s1, which are denoted as Pattern I, Pattern II, Pattern III and Pattern IV, respectively. Among these, Pattern~II PTs are always of first order. For Pattern~III PTs, we find that the second step (III\normalsize{\textcircled{\scriptsize{2}}}) is always first order, while the first step (III\normalsize{\textcircled{\scriptsize{1}}}) can be either first or second order, denoted as Pattern III-1 and Pattern III-2, respectively. The situation for Pattern~IV EWPT is more complicated. If the second step (IV\normalsize{\textcircled{\scriptsize{2}}}) is of second order, then the first step (IV\normalsize{\textcircled{\scriptsize{1}}}) must also be of second order. If IV\normalsize{\textcircled{\scriptsize{2}}} is first order, then IV\normalsize{\textcircled{\scriptsize{1}}} can be either first or second order, represented by Pattern IV-1 and Pattern IV-2, respectively. 
Thus, both Pattern III-1 and Pattern IV-1 consist of two first-order PTs and are marked with black dots in Fig.~\ref{fig:PT_pattern}, where $T_{\rm high}=10$~TeV is chosen. The details of the two-step PTs will be discussed in Sec.~\ref{ssec:twostep}.

%%%%%%%%%%%%%%%%%%%%%%%%%%%%%%%%
\begin{table}[t]
	\centering
	\vspace{-5pt}
	\renewcommand{\tablename}{Table}
	\caption{The properties related to EWPT dynamics for each phase change. P refers to unbroken symmetry, SB denotes symmetry breaking, and ISB represents inverse symmetry breaking. Note that for Pattern IV, the second step must be first-order if the first step is first-order.}
\begin{tabular}{c|c|c|l|c|c}
\toprule
Phase changes & EW & $\mathbb{Z}_2$ & EWPT processes  & First-order & $T_c$\\
\hline
$O \to A$ & SB & P & I & Y/N & $<300$ \\
& & & IV\normalsize{\textcircled{\scriptsize{2}}}& Y/N & $<300$\\
\hline
$O \to B$ & P & SB & III\normalsize{\textcircled{\scriptsize{1}}}& Y/N & $<300$\\
\hline
$B \to O$ & P & ISB & IV\normalsize{\textcircled{\scriptsize{1}}}& Y/N & $<400$\\
\hline
$B \to A$ & SB & ISB & II & Y & $<300$\\
 & & & III\normalsize{\textcircled{\scriptsize{2}}}& Y & $<300$\\
\bottomrule
		\end{tabular}%
	\label{tab:PTsum2}%
\end{table}
%%%%%%%%%%%%%%%%%%%%%%%%%%%%%%%%

For the case of $\lams=3$, both Pattern~II and Pattern~III EWPTs occur in a very narrow region of parameter space, located to the left of Pattern~I and Pattern~IV regions. 
These patterns are separated by the bound given in \eq{eq:lamSsnrbound_s}, which results from their different vacuum phases that existed prior to the EWPT.
Since this bound on $\lamhs$ increases as $\lams$ decreases, Pattern~II is eliminated in the small $\lams$ case. 
As expected, Pattern III-1 requires a relatively large $\lamhs$ coupling and, consequently, significant thermal effects to generate a potential barrier during the first-step of the PT. However, we find that Pattern III-1 is not viable for $\lams=3$. This is because a sufficiently large $\lams$ induces substantial thermal effects, resulting in a $\mathbb{Z}_2$-broken vacuum that instead follows Pattern~II to complete the EWPT.

%%%%%%%%%%%%%%%%%%%%%%%%
%%%%%%%%%%%%%%%%%%%%%%%

Excluding the two successive first-order PTs, Pattern III-1 and Pattern IV-1 (discussed in Sec.~\ref{ssec:twostep}), Fig.~\ref{fig:PT_pattern} presents the contours of the critical temperature $T_c$ and the PT strength $\xi_c\equiv v_c/T_c$ for the first-order PTs in the remaining Patterns. A solid purple line marks the boundary where $\xi_c=1$, above which a `naive' strong first-order PT occurs, requiring a relatively large coupling $\lamhs \gtrsim 2$. 
In the available parameter space, $T_c$ ranges from $120\gev$ to $180\gev$.
Notably, it is possible to keep both $T_c$ and $\xi_c$ invariant by simultaneously decreasing the values of $\ms$ and $\lamhs$. This is because a lower $T_c$ in the EWPT can be achieved either by decreasing $\ms$ or by increasing $\lamhs$, with these two opposite effects counteracting each other to maintain the invariance of $T_c$. 
In fact, each $T_c$ contour originates from the island of the EW-broken vacuum scenario (visible in the top right corner of the $\lams=3$ case in Fig.~\ref{fig:thigheq2}), near which the dependence of $T_c$ on $\ms$ and $\lamhs$ becomes anomalous. 
It is also worth noting that the contours of $T_c$ are smoothly continuous across regions where the EWPTs proceed through different patterns.  

In addition, the vacuum structure at zero temperature (see details in Appendix~\ref{sec:vac}) appears to play an important role in determining the pattern of the EWPT. 
By comparing Fig.~\ref{fig:PT_pattern_0} and Fig.~\ref{fig:PT_pattern}, we observe that the parameter region associated with the Type~C vacuum structure favors EWPTs of both Pattern~I and Pattern~IV (with the latter appearing only in the case of $\lams=3$). 
In this scenario, a PT from point $O$ to point $A$ is inevitable at low temperatures. This is because the Type~C vacuum structure lacks stationary points of $V_0(h,s)$ in the $s$-field direction, forcing the transition to occur along the $h$-field direction. 
In contrast, the vacuum structures of Type~A and Type~B typically give rise to the EWPTs of Pattern~II and Pattern~III, where both the $h$ and $s$ fields are involved. 
This observation suggests that EWPTs involving both fields generally require the presence of stationary points of $V_0(h,s)$ in the $s$-field direction, as in Type~A and Type~B vacua (c.f.~Table~\ref{tab:vac}). 
As the temperature increases, the stationary point characterized by $\langle s \rangle \neq 0$ evolves into a local minimum (and possibly a global minimum) of the effective potential. This process leads to the emergence of Pattern~II and Pattern~III. The primary distinction between these two Patterns lies in whether the breaking of $\mathbb{Z}_2$ symmetry occurs in the high-temperature vacuum. This outcome depends on the coupling strength and determines whether a false vacuum forms at the origin (point $O$) or at point $B$.

%%%%%%%%%%%%%%%%%%%%%%%
\subsection{Settling into the EW vacuum at zero temperature}
\label{ssec:proew}
%%%%%%%%%%%%%%%%%%%%%%%

\begin{figure}[t]
	\begin{center}
      \includegraphics[width=0.8\textwidth]{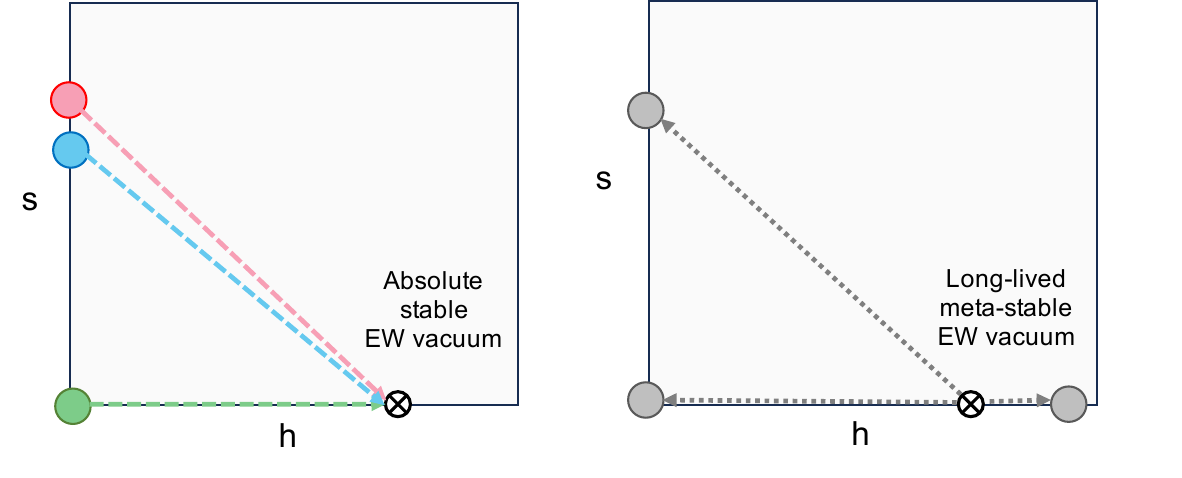}
	\end{center}
	\vspace{-5mm}
	\caption{Two distinct thermal histories of the EW vacuum are illustrated. The black cross indicates the desired EW-broken vacuum at zero temperature. Before the EWPT, the possible vacuum phases associated with different Patterns are represented by the colored circles.}
	\label{fig:EWvac}
\end{figure}

To ensure the realization of EWSB, we have so far assumed that the EW-broken vacuum is the global minimum of the zero-temperature effective potential, and the PTs from the EW-restored vacuum to it is successful. This picture of the thermal evolution is illustrated in Fig.~\ref{fig:EWvac} (left graph), but it may not always hold true when the full thermal history is considered. Two potentially important issues have been discussed. 

%%%%%%%%%%%%%%%%%%%%%%%%
%%%%%%%%%%%%%%%%%%%%%%%

First, the transition may not occur even if the EW-broken vacuum has the lowest free energy of the potential at zero temperature.
In this situation, our universe would be trapped in the false vacuum that existed prior to the EWPT, which would be unable to account for EWSB and thus would be incompatible with the experimental measurements at the LHC. This phenomenon has recently been termed ``vacuum trapping''~\cite{Biekotter:2022kgf}, and its phenomenology has also been studied in the Next-to-2HDM (N2HDM)~\cite{Biekotter:2021ysx} and in supersymmetric models~\cite{Baum:2020vfl}. We will examine this constraint in the next section. 

In contrast, another scenario is illustrated in Fig.~\ref{fig:EWvac} (right graph), where the EW-broken vacuum that emerges at high temperature is metastable at zero temperature.  
Suppose there exists a global minimum (indicated by the gray circles) with lower free energy at $T=0$, in this case, the EW-broken vacuum could transition into this lower-energy vacuum via quantum tunneling. If the transition time exceeds the age of the universe, the metastability of the EW-broken vacuum would still be compatible with experimental results. 
The vacuum of Type~E, as listed in Table~\ref{tab:vac}, could be a candidate for realizing this possibility within the framework of this model. However, we find that in this case, the EW-broken vacuum fails to evolve into a stable vacuum throughout the thermal history, thereby ruling out this scenario within this model.

%%%%%%%%%%%%%%%%%%%%%%%
\subsection{Successful bubble nucleation}
\label{ssec:Bn}
%%%%%%%%%%%%%%%%%%%%%%%

Quantum tunneling to the true vacuum state proceeds through the nucleation of bubbles of the true vacuum phase within the surrounding medium of the false vacuum. The onset of nucleation critically depends on two factors: the size of the bubbles and the number of bubbles that are nucleated. Assessing these factors requires the details of the bubble profiles, denoted as $\vec{\phi}(r)$. We will postpone discussing this issue until Sec.~\ref{ssec:bub_pro} and will focus here on the second factor. 

The nucleation rate for bubbles of the true vacuum (or the decay rate of the false vacuum) per unit time per unit volume at a finite temperature $T$ of the Universe, denoted as $\Gamma(T)$, can be expressed in the semi-classical approximation as follows~\cite{Coleman:1977py,Linde:1981zj}:
\beq
\label{eq:nucl_rate}
\Gamma(T)\approx A(T)e^{-S_E(T)}
\eeq
where $S_E(T)$ is the Euclidean tunneling action 
and the dimensionful pre-factor $A(T)$ arises from the functional determinants, which are difficult to compute except in the limits, $T\ll T_{\rm dif}$ and $T\gg T_{\rm dif}$, where $T_{\rm dif}$ is the temperature corresponding to the typical size inverse $R_0^{-1}$ of the O(4)-symmetric bubble at $T=0$. In this model, we find $R_0 \simeq \mathcal{O}(10^{-1})~{\rm GeV}^{-1}$.

In both low- and high-temperature limits, the pre-factor $A(T)$ admits a simple form at leading order~\cite{Coleman:1977py,Linde:1981zj},
\begin{equation} 
\begin{split}
	A(T)\simeq \left \{
	\begin{array}{ll}
	T_{\rm dif}^4\big(\frac{S_4(T)}{2\pi}\big)^2, & T \ll T_{\rm dif}	\\[3mm]
	T^4\big(\frac{S_3(T)}{2\pi T}\big)^{3/2}, & T \gg T_{\rm dif} 
	\end{array}
	\right.
\end{split}
\end{equation}
Here the factor $T^4$ is introduced for dimensional analysis. The leading contribution to $S_E(T)$ can be evaluated as follows
\begin{equation}
\begin{split}
	S_E(T) \simeq 	\left \{
	\begin{array}{ll}
		S_4(T)=2 \pi^2 \int^{+\infty}_{0}r^3{\rm d}r~\left[\frac{1}{2}\left(\frac{\partial \vec{\phi}}{\partial r}\right)^{2} + V_{\rm eff}(\vec{\phi};T)\right],&  T \ll T_{\rm dif}\\[5mm]
		S_3(T)/T, \,\, S_3(T)=4\pi\int^{\infty}_0 dr \ r^2\Big[\frac{1}{2}(\frac{d\vec{\phi}}{dr})^2+V_\text{eff}(\vec{\phi},T)\Big],  & T \gg T_{\rm dif}.\\
	\end{array}
	\right.
\end{split}
\label{eq:action}
\end{equation}
where $S_3(T)$ is often termed the three-dimensional Euclidean action. 

Close to $T_c$ the quantity $S_3/T$ is substantially large, indicating an exponential suppression on the nucleation rate $\Gamma$. Hence, at this moment the bubbles of the true vacuum are unlikely to be nucleated. 
In the cosmological sense, the nucleation of bubbles becomes efficient when 
the first bubble is nucleated in the casual Hubble volume\footnote{Ref.~\cite{Athron:2022mmm} suggests that a first-order PT can occur without nucleating $\mathcal{O}(1)$ bubbles per unit Hubble volume. However, we leave this possibility for future work and do not consider it here.}
\begin{align}
\label{Tn_def}
	N(T_n)=\int^{t_n}_{t_c}dt\ \frac{\Gamma(t)}{H(t)^3}=\int^{T_c}_{T_n}\frac{dT}{T}\frac{\Gamma(T)}{H(T)^4}=1 
\end{align}
Here the differential form ${dT}/{dt}=-H(T)T$ is used and $T_n$ is defined as the nucleation temperature at which the PT begins with nucleating a handful of vacuum bubbles in the entire Hubble volume.
Additionally, we relate the age of the universe to the cosmic temperature through the Hubble parameter $H(T)$,
\begin{align}
	H(T)=\sqrt{\frac{\rho_\text{rad}(T)+\rho_\text{vac}(T)}{3M^2_\text{pl}}},
\end{align}
where $\rho_\text{vac}(T)\simeq V_\text{eff}(\vec{\phi}_{\rm F},T)-V_\text{eff}(\vec{\phi}_{\rm T},T)$ represents the difference in the free energy densities between the true and false vacua, and
$\rho_\text{rad}(T)=\frac{\pi^2}{30} g_* T^4$ is the radiation energy density.
In the model studied here, the effective number of the relativistic degrees of freedom is $g_*\approx106.75$. For the numerical analysis, we take the reduced Planck mass $M_\text{pl}=2.4\times 10^{18}$ GeV.

\begin{figure}[t]
\centering
\includegraphics[width=0.92\textwidth]{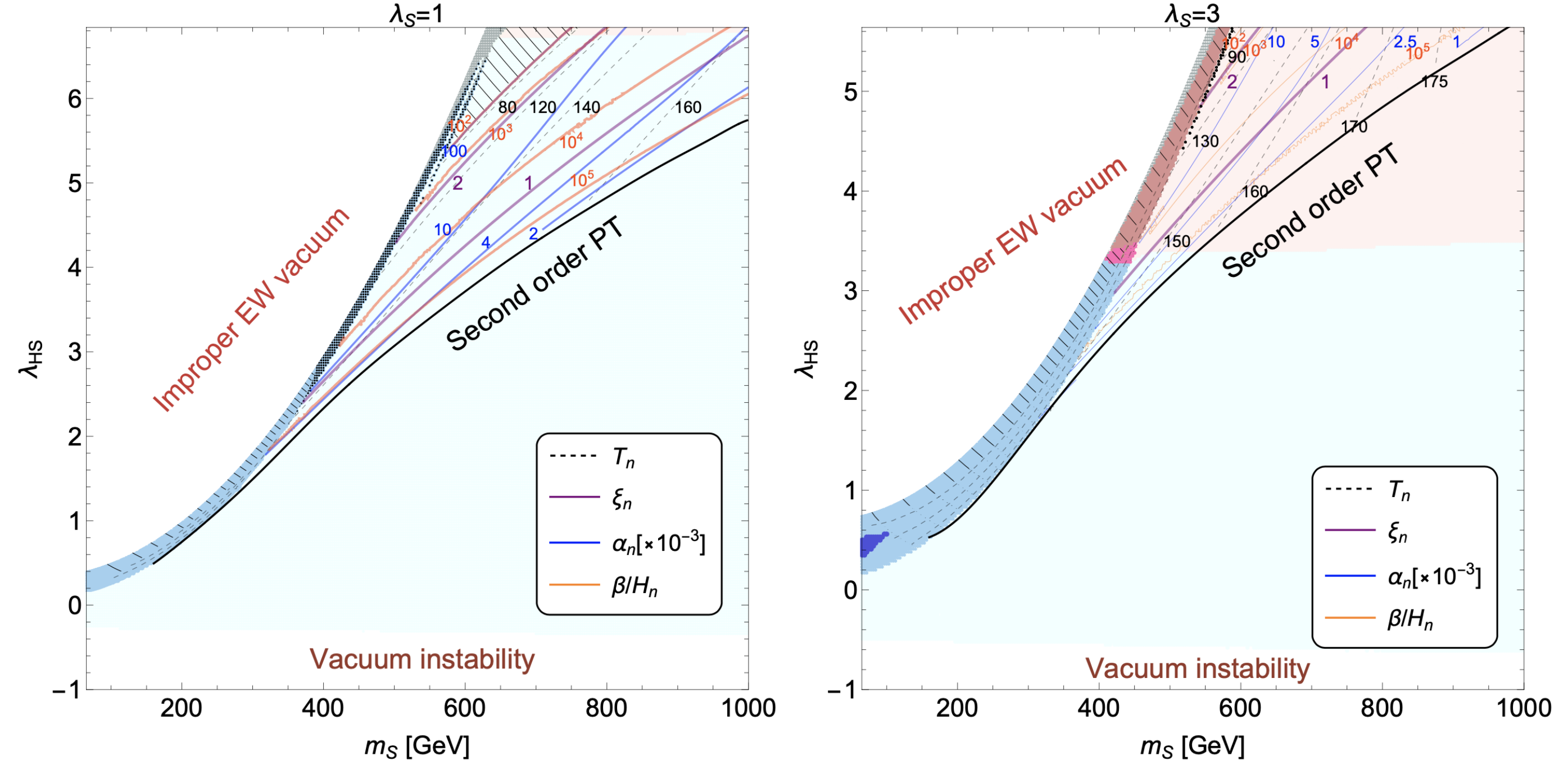}
\caption{The same as Fig.~\ref{fig:thigheq2}. The region with backslashes indicates the failure of bubble nucleation. Contours of $T_n$, $\xi_n$, $\alpha_n$, and $\beta/H_n$ are also shown.}
\label{fig:sucnucl}
\end{figure}

For the points that undergo first-order PTs, we evaluate the Euclidean bounce action, as given by \eq{eq:action},  for temperatures $T\lesssim T_c$. The nucleation temperature $T_n$ is then determined using the \textsf{SimpleBounce}~\cite{Sato:2019wpo}, which employs the technique of gradient flow equations to numerically solve the bounce solution.\footnote{Recently, a new method has been proposed to compute the tunneling action by introducing the tunneling potential~\cite{Espinosa:2018hue}. While computationally fast, this method remains at the cutting edge of research for the case of multi-field PTs.}
The results are shown in Fig.~\ref{fig:sucnucl}. 
First of all, nucleating the bubbles of the true vacuum (in the statistical sense) is not guaranteed in any pattern of the EWPTs, even if a strong first-order PT with $\xi_c \geq 1$ is predicted. 
In fact, the failure (indicated by backslashes) occurs in the large $\lamhs$ region, which corresponds to the largest PT strength $\xi_c$ predicted by this model.
Therefore, the constraint for successful nucleation, or the elimination of the vacuum trapping~\cite{Biekotter:2022kgf}, places an upper bound on $\xi_c \lesssim 1.8$ independent of EWPT pattern (similar to what is observed in the two-Higgs-doublet models~\cite{Biekotter:2022kgf}), as well as $\xi_n \lesssim 3.2$. These constraints are independent of the EWPT patterns, whether involving single-field or double-field transitions, and regardless of whether the transition occurs in one step or two steps. 
This, in turns, implies that the commonly used criterion $\xi_c\gtrsim 1$---often employed to characterize a successful strong first-order PT---is insufficient. 

%%%%%%%%%%%%%%%%%%%%%%%%%%%%%%%%
%%%%%%%%%%%%%%%%%%%%%%%%%%%%%%%%

\begin{figure}[t]
	\centering
	\includegraphics[width=0.45\textwidth]{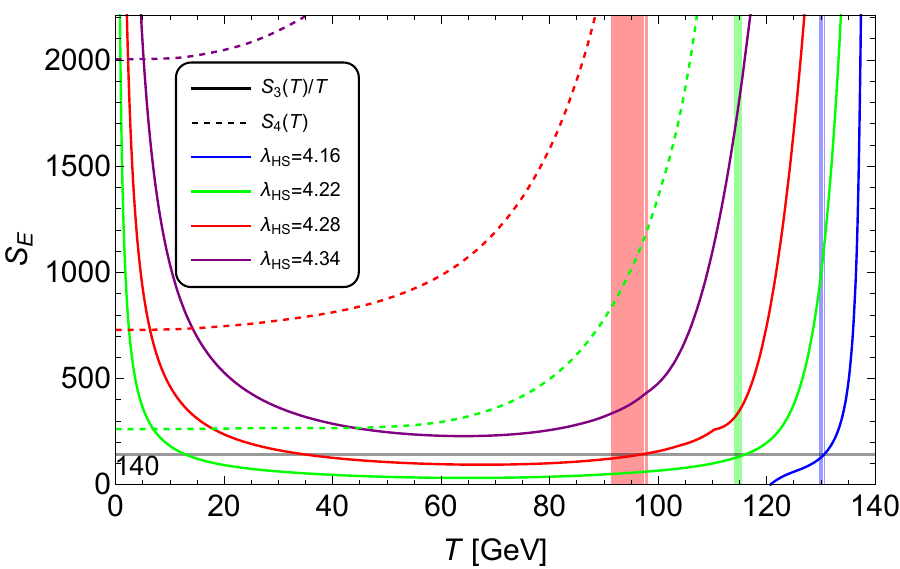}
	\includegraphics[width=0.43\textwidth]{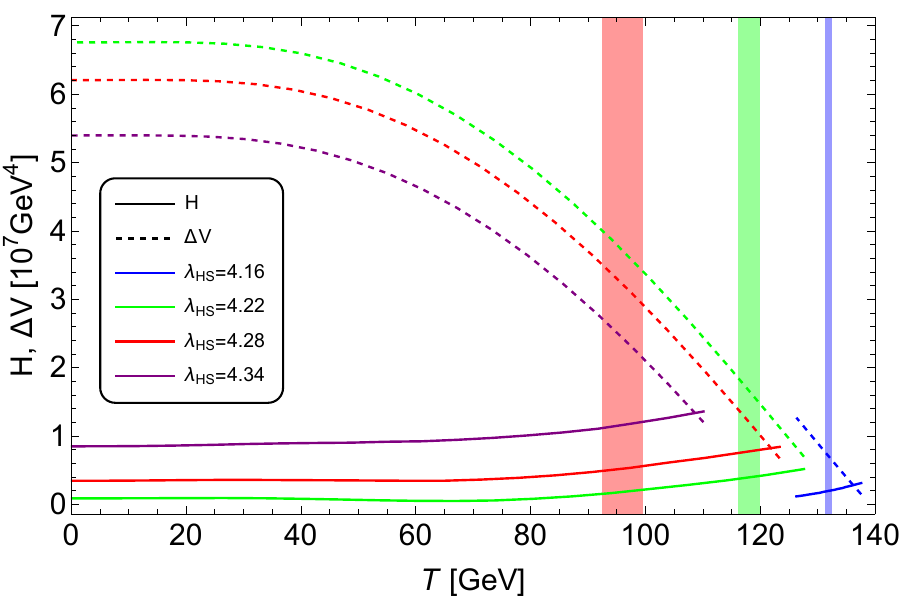}
	\caption{The action $S_3/T$ (left) and the barrier
height $H$ and the vacuum depth of the potential $\Delta V$ (right) are shown as functions of temperature $T$ for several examples of Pattern~II EWPT with $\lams=3$ and $m_S=500\gev$. Each curve is computed starting from the critical temperature $T_c$.  The shaded area represents the temperature range from the nucleation temperature $T_n$ to the percolation temperature $T_p$ defined in Sec.~\ref{ssec:perc}.}
	\label{fig:S3T}
\end{figure}

It might be interesting to understand what prevents the nucleation and whether this physical property is determined by the zero-temperature potential. Taking the example of the Pattern~II EWPT, where hybrid $h$-$s$ bubbles\footnote{The bubbles nucleated in Pattern III also consist of two fields. However, the domain walls formed in the first step can induce bubble nucleation in the vicinity of the domain wall sites, thereby making the additional contribution to the nucleation rate~\cite{Blasi:2022woz,Agrawal:2023cgp}. These calculations, which involve the dynamical evolution of the DW networks, are beyond the scope of this work.} are nucleated, we analyze how the bounce action changes with $T$ for different values of $\lamhs$, as illustrated in Fig.~\ref{fig:S3T}.
Regardless of the value of $\lamhs$, the ratio $S_3/T$ exhibits an exponential decrease near the critical temperature $T_c$. 
This occurs because that, at the onset of the PTs, the true vacuum develops a deeper minimum, and the barrier between it and the false vacuum becomes lower. Both of these effects reduce the difficulty of nucleating bubbles, leading to a significant decrease in $S_3$. In contrast, at low temperatures, the barrier height $H$ and the vacuum depth of the potential $\Delta V$ become almost stable, resulting in little change in $S_3$. Instead, the decreasing temperature causes $S_3/T$ to increase.    
From Fig.~\ref{fig:S3T}, it can be observed that $S_3/T$ decreases to a vanishingly small value just before the barrier disappears. However, for strong couplings, loop corrections become increasingly important. As a result, before $S_3/T$ rises, it does not decrease sufficiently to the order necessary for successful bubble nucleation, leading to the absence of $T_n$. Consequently, the PT fails to occur, and the universe remains trapped in the false vacuum. 
Therefore, we conclude that regions where vacuum trapping occurs are characterized by
a shallow depth between the the EW-broken vacuum and the EW-preserved vacuum in the zero-temperature effective potential.
This behavior is analogous to the case of the single-field bubble, as we have verified.

 %%%%%%%%%%%%%%%%%%%%%%%%%%%%%%%%
%%%%%%%%%%%%%%%%%%%%%%%%%%%%%%%%

In the remaining parameter regions where bubble nucleation is successful, we show the value of $T_n$ in the dashed contours, with a solid line indicating $\xi_n\equiv v_n/T_n=1$, above which a relatively large coupling $\lamhs \gtrsim 2$ is required. Similar to the $T_c$ distribution, the $T_n$ contours are also smoothly continuous,  even when the EWPTs proceed with different patterns. This continuity is somewhat surprising to us for the following reasons. Determining $T_n$ involves solving \eq{eq:eom}, which is non-linear, with the boundary conditions given by the false vacuum, which differs across the four Patterns.
It is observed that $T_n$ is roughly between $90~{\rm GeV}$ and $150~{\rm GeV}$ in both Pattern~II and Pattern~III\normalsize{\textcircled{\scriptsize{2}}}, while the EWPTs in Pattern~I and Pattern~IV\normalsize{\textcircled{\scriptsize{2}}} can reach as high as $180~{\rm GeV}$.

%%%%%%%%%%%%%%%%%%%%%%%
\subsection{Percolation of EWPTs}
\label{ssec:perc}
%%%%%%%%%%%%%%%%%%%%%%%

For successful nucleation, below $T_n$ an increasing number of bubbles emerge, expand, and collide. 
As these bubbles grow, they progressively absorb regions that were initially in the old state of the metastable vacuum, leading to the coalescence of true vacuum bubbles. This percolation process ultimately results in the dominance of the true vacuum phase and the fragmentation and dissipation of the old vacuum phase~\cite{shante1971introduction}.

In percolation theory, the percolation temperature $T_p$ is defined as the temperature at which the probability of finding the false vacuum within one Hubble volume, or equivalently, the volume fraction of the false vacuum, is $\mathcal{P}(T_p)=e^{-I(T_p)}\simeq 0.7$\footnote{The applicability of this numerical result to exotic bubble configurations, such as the co-existence of two distinct types of bubbles discussed in the next subsection, has not yet been confirmed.}~\cite{Rintoul:1997tze,lorenz2001precise,lin2018continuum,li2020numerical}.  
Here, $I(T)$ represents the total volume in spheres (with appropriate multiple counting of overlaps) per unit volume of space at a given temperature $T$~\cite{Guth:1981uk}. The expression for it is given by: 
\begin{align}
	\label{Tp_cal}
	I(T)=\frac{4\pi}{3}\int^{Tc}_TdT'\frac{\Gamma(T')}{H(T')T'^4}\Big(\int^{T'}_TdT''\frac{v_w(T'')}{H(T'')}\Big)^3,
\end{align}
where $v_w$ is the bubble wall velocity, which generally varies over time. 
Recent large-scale simulations~\cite{Athron:2022mmm} have demonstrated that by the end of the percolation stage, bubble collisions become highly frequent energetic, accelerating the coalescence of bubbles into a universe-spanning cluster and driving the rapid completion of the PT. Consequently, $T_p$ is commonly defined as the temperature at which the PT is effectively complete. This definition generally holds, except in the case of PTs occurring during inflation or in supercooled (or super-slow) PTs~\cite{Turner:1992tz,Ellis:2018mja,Athron:2022mmm}, where cosmic expansion may impede the percolation process. 
In such cases, it is often necessary to apply a criterion formula~\cite{Turner:1992tz} 
\begin{align}
	\frac{1}{\mathcal{V}_{\rm false}}\frac{d\mathcal{V}_{\rm false}}{dt}=3H(t)-\frac{dI(t)}{dt}=H(T)\left(3+T\frac{dI(T)}{dT}\right)
	<0
	\label{eq:checkTp}
\end{align}
to determine whether the false vacuum volume begins to shrink (at least at $T_p$) as the temperature decreases. We numerically verified Eq.~\eqref{eq:checkTp} within the parameter region where bubble nucleation is possible.

%%%%%%%%%%%%%%%%%%%%
\begin{figure}[t]
\centering
\includegraphics[width=0.45\textwidth]{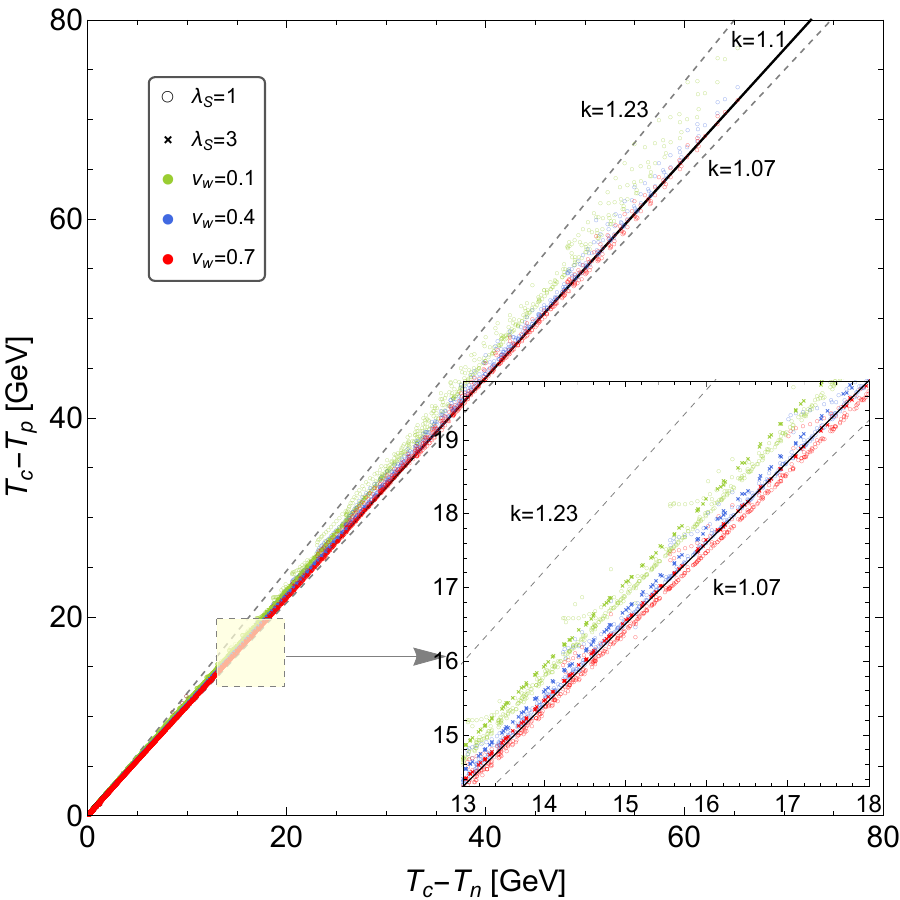}
\includegraphics[width=0.45\textwidth]{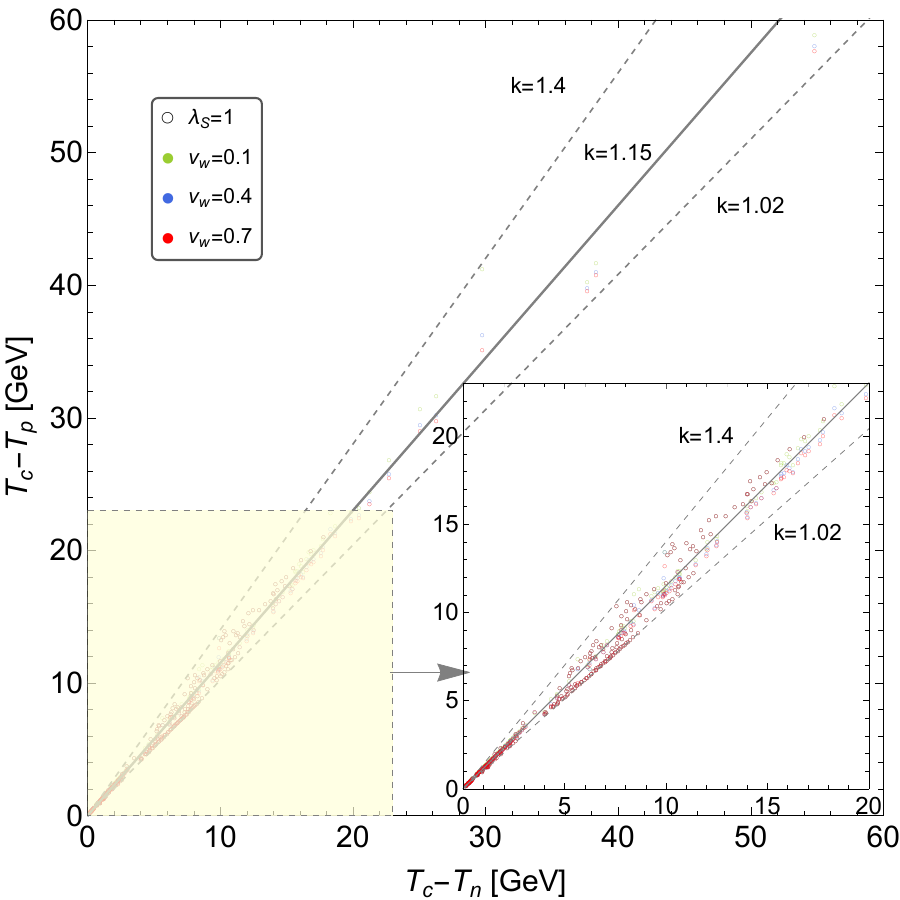}
\caption{The numerical correlation between the differences $T_c-T_n$ and $T_c-T_p$ is shown for $h$ bubbles from $O\to A$ transition (left) and for $s$ bubbles from $O\to B$ transition (right).
Three distinct values of the wall velocity $v_w$ are used, each indicated by a different color. 
The numerical relation suggested as a useful approximation for estimating $T_p$ is depicted by the solid line.}
\label{fig:Tpnum}
\end{figure}
%%%%%%%%%%%%%%%%%%%%

\begin{figure}[t]
	\centering
    \includegraphics[width=0.47\textwidth]{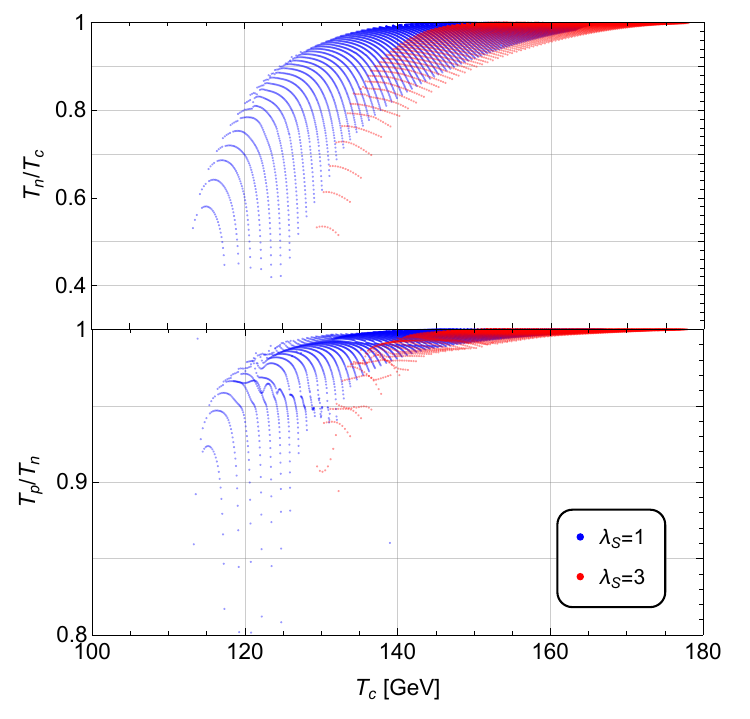}
    \includegraphics[width=0.47\textwidth]{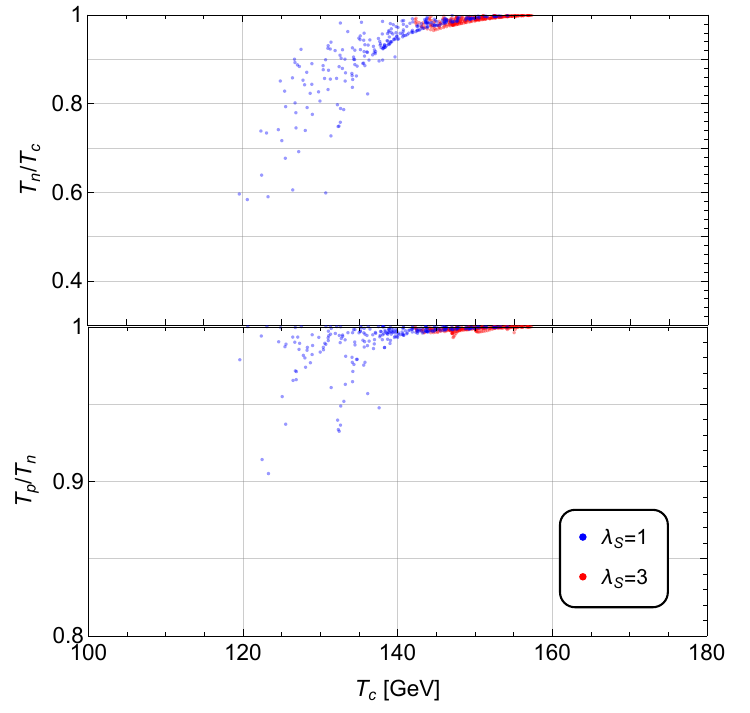}
	\caption{The plot illustrates the variations of the ratios $T_n/T_c$ and $T_p/T_n$ as functions of changes in $T_c$ for the first-order PT processes, $O \to A$ (left) and $B \to A$ (right). In each case two different values of $\lams$ are shown for comparison, with $v_w$ set to 0.8. 
} 
	\label{fig:Tpnum2}
\end{figure}

Fig.~\ref{fig:Tpnum} demonstrates the dependence of the $T_p$ prediction on the bubble velocity $v_w$ and the coupling $\lams$. In principle, $v_w$ is a quantity determined by the model parameters~\cite{Kozaczuk:2015owa,Friedlander:2020tnq,Lewicki:2021pgr} but its calculation requires a high level of complexity and computational capacity, which is beyond the scope of this work. For illustrative purposes, we treat $v_w$ as a constant parameter and select a set of values that correspond to the deflagration and detonation bubbles~\cite{Athron:2023xlk}.
It turns out that the ratio $(T_c-T_p)/(T_c-T_n)$ for the phase change $O \to A$ roughly ranges from 1.07 (1.09) to 1.23 (1.15) for $\lams=1$ ($\lams=3$), while for the phase change $O \to B$, the ratio spans a broader range, from 1.02 to 1.4. 
A ratio greater than 1 implies that $(T_c-T_p)$ is smaller for smaller values of $(T_c-T_n)$. An extremely fast first-order PT, with highly degenerate values of $T_c, T_n$ and $T_p$, is also possible and tend to occur at the highest $T_c$ for each phase change process, as shown in Fig.~\ref{fig:Tpnum2}, where $v_w=0.8$ is used. For the EWPT with a lower $T_c$, the lower bounds of $T_n/T_c$ and $T_p/T_c$ exhibit a larger departure from unity, causing slower PTs with a large difference between $T_n$ and $T_p$. 

It can be also observed in Fig.~\ref{fig:Tpnum} that this ratio approaches the lower bound for large $v_w$, which is favorable for generating large gravitational wave signals. 
Thus, we suggest the numerical relation $T_c-T_p=1.1(T_c-T_n)$, which is equivalent to $T_c-T_n=0.1(T_n-T_p)$, as a useful approximation for estimating $T_p$ without the need for running the real-time dynamical evolution, provided that  $T_c$ and $T_n$ are known.
However, due to the lack of a rigorous derivation, it remains unclear to us whether this relation has strong model dependence.

%%%%%%%%%%%%%%%%%%%%%%%%%%%%%%%%
%%%%%%%%%%%%%%%%%%%%%%%%%%%%%%%%

%%%%%%%%%%%%%%%%%%%%%%%%%%%%%%%%
\section{Properties of nucleated bubbles}
\label{sec:bubbles}
%%%%%%%%%%%%%%%%%%%%%%%%%%%%%%%%

In the EWPT of all the Patterns achievable in the model, there are four distinct phase changes, which are summarized in Table~\ref{tab:PTsum2}. 
Generally, for each phase change involving different symmetry breaking, a specific type of bubble with a distinct field configuration can be nucleated. 
Since the $s$ field condensate remains zero during the phase change $O \to A$ (which occurs In Pattern~I and the second step of Pattern~IV), the nucleated bubbles in these cases are constituted solely by the $h$ field. We refer to these as the {\it $h$ bubbles}. A similar situation occurs in the phase change $O \leftrightarrow B$ (which occurs in the first step of Pattern~III and Pattern~IV), where the $h$ field remains zero. In these cases, the nucleated bubbles are constituted solely by the $s$ field. We refer to the bubbles nucleated from $O \to B$ as the {\it $s$ bubbles} and those from $B \to O$ as {\it inverse $s$ bubbles}, due to the inverse symmetry breaking occurring in the $\ztwo$-symmetry in the latter case. Additionally, during the phase change $B \to A$ (present in Pattern~II EWPT and the second step of Pattern~III), both the $h$ and $s$ fields participate in the PT. The nucleated bubbles in these cases are constituted by both the $h$ and $s$ fields, and we term these the {hybrid $h$-$s$ bubbles}.  

%%%%%%%%%%%%%%%%%%
\subsection{Bubble profile analysis}
%%%%%%%%%%%%%%%%%%
\label{ssec:bub_pro}

\begin{figure}[t]
	\hspace{-2mm}
	\includegraphics[width=0.53\textwidth]{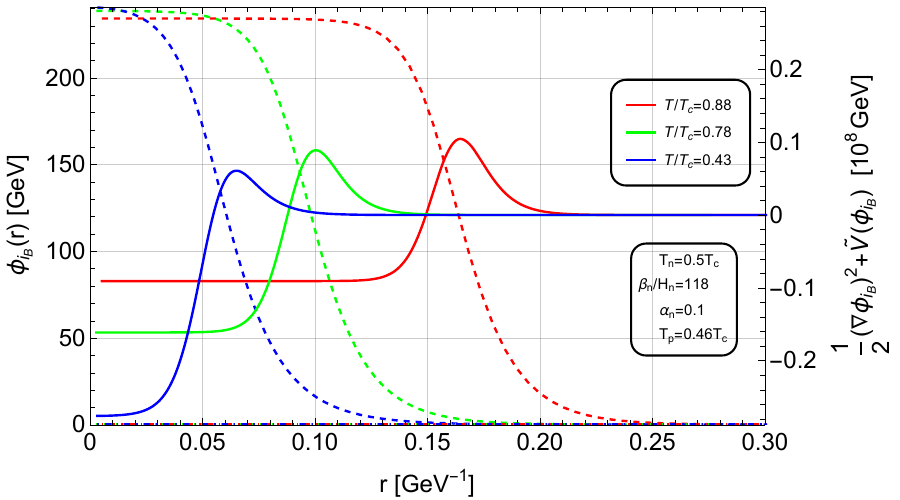}
	\includegraphics[width=0.53\textwidth]{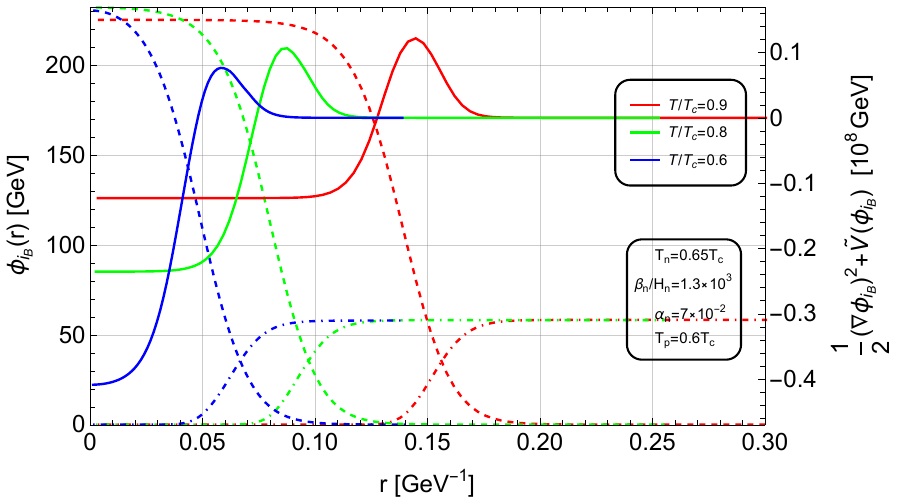}
	\\[8pt]
	\includegraphics[width=0.53\textwidth]{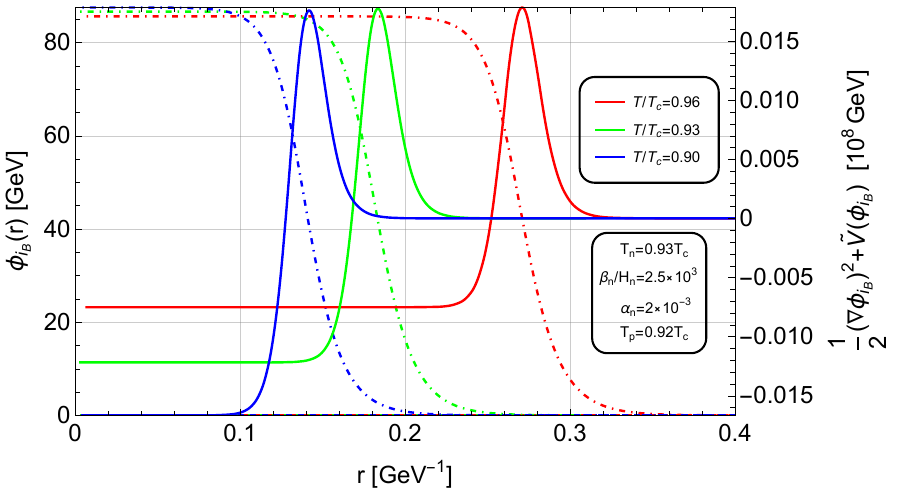}
	\includegraphics[width=0.53\textwidth]{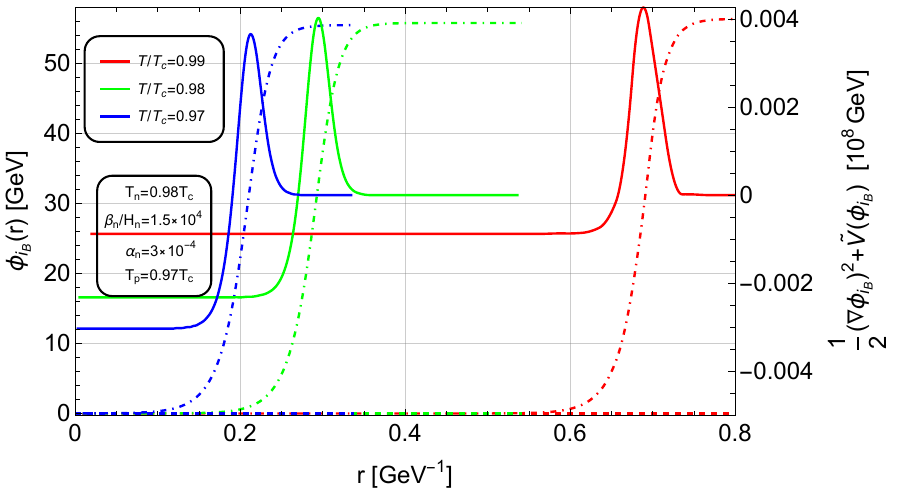}
	\caption{The bounce solution (dashed lines) and the distribution of the total energy density (solid lines) for the bubbles nucleated from the four Patterns are shown: pure $h$-bubble from Pattern~I (top left), hybrid $h$-$s$ bubble from Pattern~II (top right), pure $s$-bubble from Pattern~III (bottom left) and inverse $s$-bubble from Pattern~IV (bottom right). The bounce solution $\phi_b$ use the left y-axis while the total energy density is shown in the right y-axis.}
	\label{fig:multifield_Bubble_confi}
\end{figure}

In any case, the vacuum bubble is characterized by two important length scales: the bubble size $R_b$ and the bubble wall thickness $L_w$. These are directly associated with the wall region where $d\vec{\phi}(r)/dr$ is large, with $\vec{\phi}(r)$ being the “bounce” solution of O(3)-symmetric bubbles that satisfy the classical Euclidean field equation,
\begin{align}
	\frac{d^2 \vec{\phi}(r)}{d r^2}+\frac{2}{r}\frac{d \vec{\phi}(r)}{d r}&=\vec{\nabla} V_\text{eff}(\vec{\phi}),
	\label{eq:eom}
\end{align}
together with the boundary constraints
\begin{align}
	\frac{d\vec{\phi}}{d r}\Big|_{r=0}=0, \ \ \lim_{r \rightarrow\infty}\vec{\phi}(r)=\vec{\phi}_F\,.
\end{align}
Here the notation $\vec{\phi}$ represents the complete list of fields that make up the vacuum bubble, and $\vec{\phi}_F$ denotes the field configuration at the false vacuum. For example, $\vec{\phi}=h$ for $h$ bubbles, $\vec{\phi}=s$ for (inverse) $s$ bubbles, while $\vec{\phi}=\{h,s\}$ becomes two dimensional for hybrid $h$-$s$ bubbles. The bounce solution can be derived either analytically or by using computational packages, especially when the form of $V_\text{eff}$ is complicate or when the equations couple multiple fields. 
In Fig.~\ref{fig:multifield_Bubble_confi} we present the four bubble profiles corresponding to each pattern.
Once the bounce solution is obtained, these two scales can be determined . 
Numerically, $R_b$ is defined as the $r$-value at which the field value reaches half of its value at the bubble center, i.e., 
\beq
\label{eq:Rb}
\phi(R_b)={\phi(0)+\phi(\infty) \over 2}.
\eeq
The wall thickness $L_w$ is given by the difference between the radii $r_+$ and $r_-$, where the bubble wall lies in the region between $r=r_-$ and $r=r_+$. For thin-wall bubbles, $r=r_\pm$ corresponds to the radial locations where the field value satisfies $\phi(r_\pm)=\phi(0)\left [1\pm \tanh(1/2)\right]/2$~\cite{Cutting:2020nla}.

%%%%%%%%%%%%%%%%%%%%%%%
%%%%%%%%%%%%%%%%%%%%%%%

The approach described above does not work for multi-field bubbles that are nucleated in the Pattern~II PTs. The reason is that there are multiple field configurations that describe the bubble, making it unclear which one should be used to determine the bubble scales. 
To address this issue, it will be useful to develop a formalism that identifies the typical scales of the bubble without relying on the specific bubble profile.
Recall that vacuum bubbles are topological objects where the energy is strongly localized.  
Therefore, we suggest the bubble radius $R_b$ can be characterized by the extremum of the energy density, i.e., 
\beq
\label{eq:emax}
\mathcal{E}(R_b) = \mathcal{E}_{\rm max},
\eeq 
where the energy density $\mathcal{E}(r)$ is given by
\beq
\mathcal{E}(r)=\frac{1}{2}(\nabla\phi_i(r))^2+\tilde V(\phi(r)), 
\eeq
where $\tilde V(\phi(r))\equiv V_{\rm eff}(\vec{\phi}(r))-V_{\rm eff}(\vec{\phi}_{F})$ is the effective potential for $\vec{\phi}(r)$, relative to the one in the false vacuum. However, there is no analogous method for determining the wall thickness $L_w$. Based on the observation in Fig.~\ref{fig:multifield_Bubble_confi} that the distribution of the energy density resembles the Breit–Wigner shape of a resonance, we define $L_w$ as the full width at half maximum of this distribution 
\beq
L_w= R_2-R_1,
\eeq
with $R_1$ and $R_2$ being the $r$-values ($R_1<R_2$) at which the energy density reaches the half sum of the peak density and the asymptotic energy density $\mathcal{E}_{\infty}$, far from the bubble wall, 
\beq
\label{eq:asymenergy}
\mathcal{E}(R_1)=\mathcal{E}(R_2)=\frac{\mathcal{E}_{\rm max}+\mathcal{E}_{\infty}}{2}
\eeq
where $\mathcal{E}_{\infty}$ coincides with the difference between the free energy densities of the false and true vacuum phases, $\mathcal{E}_{\infty}=\Delta V \equiv V_{\rm eff}(\vec{\phi}_F)-V_{\rm eff}(\vec{\phi}_T)$.

As shown in Fig.~\ref{fig:multifield_Bubble_confi}, for a single-field bubble, the total energy density reaches its maximum at approximately half of the field values, both inside and outside the bubble. However, for dual-field bubbles, the location of the maximum energy does not correspond to the center value of either field configuration. 
The comparison between two approaches for determining the bubble properties is summarized in Table~\ref{tab:bub}, where we quantify the numerical error as
\beq
\epsilon(X)={X(\mathcal{E})-X(\phi)\over X(\mathcal{E})}, 
\eeq 
where $X(\phi)$ and $X(\mathcal{E})$ denote the quantities $X=R_b, L_w$ obtained from the bubble profiles and the energy distribution, respectively. 
Clearly, for single-field bubbles, the results for both $R_b$ and $L_w$ obtained by the two approaches are in perfect agreement at the beginning of the first-order PT during which $T$ is close to $T_c$ and thin-wall bubbles are typically nucleated. 
As $T$ decreases, however, the results from the two approaches begin to diverge, and the difference becomes more pronounced. When the discrepancy occurs, the results based on the energy distribution predict a larger value for $R_b$ and a smaller value for $L_w$. Furthermore, for a two-field bubble we observe that $R_1$ is close to $R_b$ as defined in Eq.~\eqref{eq:Rb}, while $R_2$ corresponds to the value of $r$ at the intersection of the two field configurations, where $h(R_2) \simeq s(R_2)$. 
It is unclear whether this is merely a coincidence or if it suggests some deeper implication related to bubble dynamics. 

%%%%%%%%%%%%%%%%%%%%%%%%%%%%%%%%
\begin{table}[t]
	\centering
	\vspace{-5pt}
	\renewcommand{\tablename}{Table}
	\caption{A comparison between the traditional $\phi$ approach and the newly proposed $\mathcal{E}$ approach for determining the properties of four types of bubbles such as the bubble radius $R_b$ and the wall thickness $L_w$. In addition, the critical radius $R_c$ is also included.}
	\begin{tabular}{c|c|c|cc|cc|cc|c}
		\toprule
		\multirow{2}{*}{Bubbles}  & PT & \multirow{2}{*}{$T/T_c$} & \multicolumn{2}{c|}{$\phi$ approach} &  \multicolumn{2}{c|}{$\mathcal{E}$ approach}  & \multicolumn{2}{c|}{Errors [\%]}  & \multirow{2}{*}{$R_c$}\\
		 & Patterns & & $R_b(\phi)$ &  $L_w(\phi)$ &  $R_b(\mathcal{E})$  & $L_w(\mathcal{E})$ & $\epsilon(R_b)$  & $\epsilon(L_w)$ &\\ 
		\hline
		\multirow{3}{*}{$h$-bubbles} &  I & 0.88 & 0.165 & 0.024 & 0.165 & 0.024 & 0 & 0 & 0.105\cr
		& IV\normalsize{\textcircled{\scriptsize{2}}}&  0.78 & 0.098  &  0.024 & 0.100 & 0.022 & 2.0 & -8.3 & 0.047\cr
		& & 0.43 & 0.061  & 0.026 & 0.065 & 0.020 & 6.2 & -23.1 & 0.009\cr
		\hline
		\multirow{3}{*}{$h$-$s$ bubbles} & II & 0.90 &-  &- & 0.144 & 0.024 & - & - & 0.141\cr
		& III\normalsize{\textcircled{\scriptsize{2}}}  & 0.80 & -  &  - & 0.086  & 0.020 & - & - & 0.079\cr
		& & 0.60 & -  & - & 0.058 & 0.018 & - & - & 0.045\cr
		\hline
		\multirow{3}{*}{$s$ bubbles} & III\normalsize{\textcircled{\scriptsize{1}}} &  0.96 & 0.271 & 0.023 & 0.271   & 0.024 & 0 & 4.3 & 0.213\cr
		& & 0.93 & 0.183  & 0.023 & 0.183  & 0.023 & 0 & 0 & 0.123\cr
		& & 0.90 & 0.141  & 0.022 & 0.142  & 0.021 & 0.7 & -4.5 & 0.085\cr
		 \hline
		Inverse & IV\normalsize{\textcircled{\scriptsize{1}}} &  0.99 & 0.688  & 0.031 & 0.689  & 0.033 & 0.1 & 6.5 & 0.720\cr
		  $s$ bubbles &  & 0.98 & 0.290 & 0.031 & 0.295 & 0.031 & 1.7 & 0 & 0.279\cr
		 & & 0.97 & 0.206 & 0.031 & 0.213 & 0.030 & 3.3 & -3.2 & 0.209\cr
		\bottomrule
	\end{tabular}
	\label{tab:bub}
\end{table}
%%%%%%%%%%%%%%%%%%%%%%%%%%%%%%%%

%%%%%%%%%%%%%%%%%%%%%%%
\begin{figure}[t]
	\includegraphics[width=0.5\textwidth]{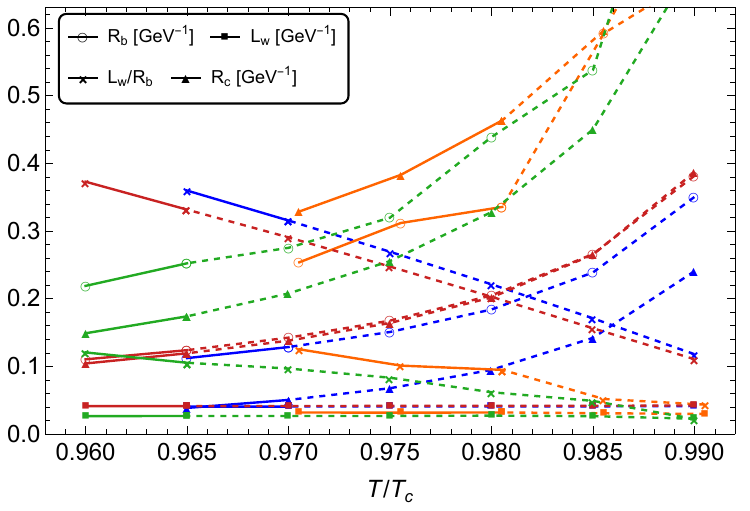}
	\hspace{2mm}
	\includegraphics[width=0.5\textwidth]{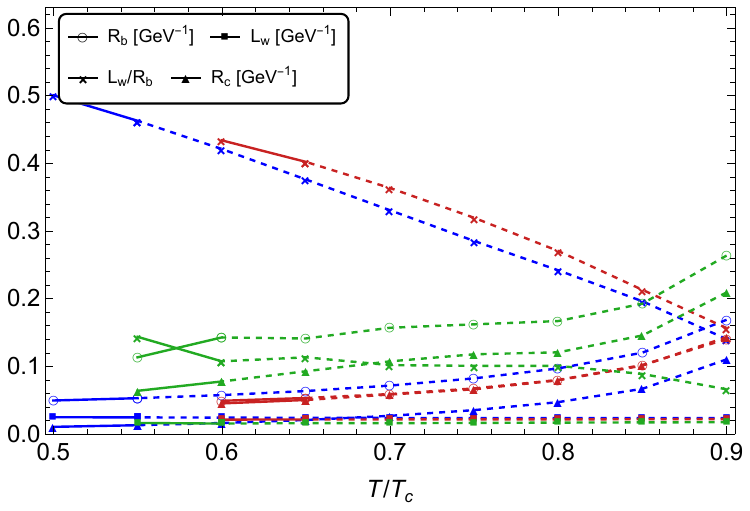}	
	\caption{The evolution of bubble properties with decreasing temperature $T$ is shown for an $h$-bubble (blue), a hybrid $h$-$s$ bubble (red), a $s$-bubble (green) and an inverse $s$-bubble (orange) from both the fast FOPTs (left panel) and relatively slow ones (right panel). Results above $T_n/T_c$ are depicted in dashed lines.}
	\label{fig:multifield_Bubble}
\end{figure}
%%%%%%%%%%%%%%%%%%%%%%%

Since the traditional approach based on the bubble profile is not applicable to multi-field bubbles, and the numerical errors in the two approaches for single-field bubbles is sufficiently small, we adopt the new method to analyze the bubble properties such as $R_b$, $L_w$ and their ratio. The results are shown in Fig.~\ref{fig:multifield_Bubble}. 
As the value of $\beta/H$ from the first step of Pattern~IV FOPTs exceeds $\mathcal{O}(10^2)$, all inverse $s$-bubbles arise from ultra-fast FOPTs. For the other three types of bubble, we distinguish between those generated by fast FOPTs (left panel) and those formed by relatively slow FOPTs (right panel). For any bubble, we present results ranging from $T_n/T_c$ to $T_p/T_c$. 
In general, bubbles formed during fast FOPTs tend to be larger than those formed during slow FOPTs. This can be understood as follows: at $r \simeq R_b$, where the bubble energy is maximized as described by \eq{eq:emax}, the field value reaches an inflection point, so $d^2\phi/dr^2 \simeq 0$ and $\Delta V_{\rm eff}(\phi) \simeq 0$, effectively making $R_b^{-1} d\phi/dr$ a constant. Since $d\phi/dr$ is relatively larger for fast PTs, the bubble radius $R_b$ is correspondingly larger.
Furthermore, for bubbles nucleated at a lower $T$, whether from fast or slow FOPTs, the radius $R_b$ decreases while the thickness $L_w$ remains nearly constant. Consequently, the ratio $L_w/R_b$ increases, reducing the accuracy of the thin-wall approximation. 

Another key issue related to bubble dynamics is the determination of the critical radius $R_c$, below which a newly formed bubble cannot survive. To understand this, we consider a vacuum bubble of size $R$, which has the free energy relative to the free energy of the false vacuum given by 
\beq
F(R)=4\pi R^2\sigma - \frac{4\pi}{3} R^3 \Delta V
\eeq
where $\Delta V$  is the difference in energy densities between the false and true vacua (as defined below \eq{eq:asymenergy}), and $\sigma$ is the surface energy density of the bubble wall, which is given by:
\beq
\sigma=\int^{\infty}_{0} \!\! dr \,\, {d\vec{\phi} \over dr} \cdot {d\vec{\phi} \over dr}.
\eeq
Here we observe that for sufficiently small sizes, the free energy of the bubble decreases as $R$ decreases, but for sufficiently large $R$ it changes oppositely. This implies that small nucleated bubbles will spontaneously collapse due to surface tension,  whereas large bubbles will expand, causing the universe to transition to the true vacuum phase. 
Thus, the minimum size of a bubble that can avoid collapse is determined by the condition $d F/d R=0$, 
which gives the radius of the critical bubble, 
\beq
 R_c=\frac{2\sigma}{\Delta V}
\eeq

For single-field bubbles, $\sigma$ can be obtained by solving the equation of motion while neglecting the damping term. In this case, one has
\beq
\frac{d\phi}{dr}=-\sqrt{2 \tilde{V}(\phi)},
\eeq
giving rise to
\beq
\label{eq:mu_single-field}
\sigma=\int^{\phi_0}_{\phi_F}\!\!d\phi \,\sqrt{2 \tilde{V}(\phi)}.
\eeq
where $\phi_0$ represents the field value inside the bubble, which closely approaches $\phi_T$ in the thin-wall limit. 
However, an analytical solution for two-field bubbles is not possible because, even under the thin-wall approximation, the damping terms in the coupled equations of motion cannot be simultaneously neglected due to their interdependence.  
The numerical result for the critical radius, $R_c$, for these bubbles is shown in Fig.~\ref{fig:multifield_Bubble} and Table~\ref{tab:bub}.
It is important to note that not all nucleated bubbles exceed the critical radius. 
An exception occurs for the inverse $s$-bubbles from Pattern~IV\normalsize{\textcircled{\scriptsize{1}}}, 
where the difference in vacuum energy between the interior and exterior of the nucleated bubble becomes negligible compared to the gradient energy of the bubble wall, causing the bubble to collapse due to surface tension. 
Additionally, for the hybrid $h$-$s$ bubbles from Pattern~II, the radius of the nucleated bubbles is nearly equal to critical radius.

%%%%%%%%%%%%%%%%%%
\subsection{Assessment of the thin-wall approximation}
\label{sec:est_tw}
%%%%%%%%%%%%%%%%%%
The thin-wall approximation is a crucial concept in nucleation theory, as it typically allows for semi-analytic solutions of bounce solutions or bounce actions. It is commonly employed as an initial condition in numerical simulations~\cite{Jinno:2016vai,Jinno:2017fby}. In general, there are two popular definitions of the thin-wall approximation, which are as follows: (i) the difference in energy between the false and true vacuum, $\Delta V$, is small compared to the height of the barrier $H$, and (ii) the thickness of the nucleated bubble, $L_w$, is much smaller than the radius, $R_b$. Furthermore, under the first definition, it is possible to derive a simple, approximate analytic expression for the Euclidean action $S_E$~\cite{Linde:1981zj},
\beq
S^\text{TW}_3={16\pi \over 3} (\Delta V)^{-2}\sigma^3
\eeq
where $\sigma$ is the surface energy density of the bubble wall, as given in \eq{eq:mu_single-field}.

As mentioned in Ref.~\cite{Cline:2021iff}, the action obtained using the full one-loop potential may differ significantly from the one derived using the thin-wall approximation, specifically $S_4^\text{tw}$ and $S_3^\text{tw}/T$. For a quantitative assessment, we define the relative error for the action $\epsilon(S_3/T)$ and the nucleation temperature $T_n$, which can serve as a new way to define the thin-wall approximation:
\beq
\epsilon(X)=\frac{X-X^\text{TW}}{X}
\eeq
where $X=S_3/T$ and $T_n$. 

%%%%%%%%%%%%%%%%%%%%%%%%%%%%%%%%
\begin{figure}[t]
\centering
\includegraphics[width=0.5\textwidth]{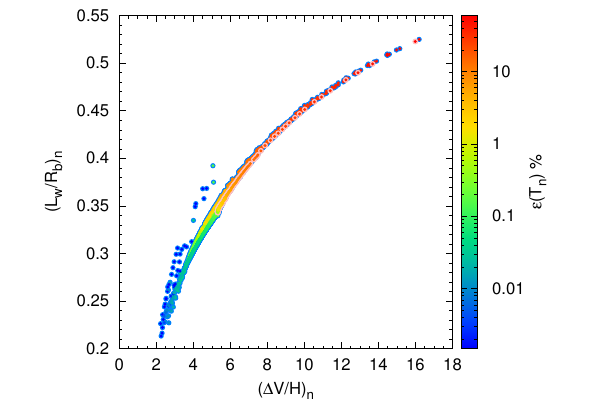}
\hspace{-9mm}
\includegraphics[width=0.33\textheight]{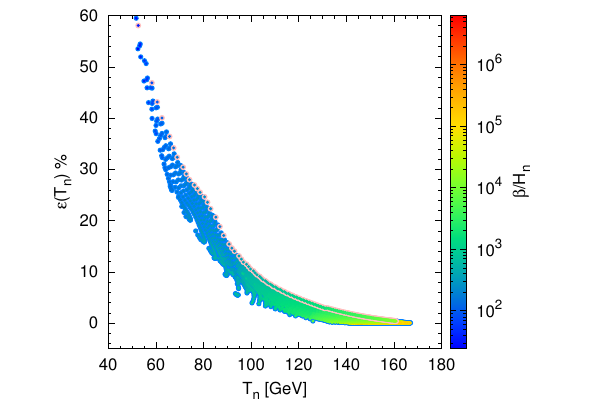}
\caption{The applicability of the thin-wall approximation, measured by $\epsilon(T_n)$, is analyzed for the case of $\lams=1$. The left plot demonstrates the consistency of the thin-wall approximation using three different descriptions. The right plot reveals that larger values of $\epsilon(T_n)$ are associated with smaller values of $\beta/H_n$, indicating slow PTs. 
In both plots, points enclosed in light blue and dark blue circles correspond to Pattern~I PTs and Pattern~III-2 PTs, respectively, and those enclosed in pink circles represent Pattern~IV PTs.}
	\label{fig:epsTn_para}
\end{figure}
%%%%%%%%%%%%%%%%%%%%%%%%%%%%%%%%

To clarify the relevance of these three definitions of the thin-wall approximation, 
we denote the ratio $\Delta V/H$ at a temperature of $T_n$ as $(\Delta V/H)_n$. In the following discussion, we choose $\epsilon(T_n)$ as the quantity to assess the applicability of the thin-wall approximation, rather than using $\epsilon(S_3/T)$, based on the following considerations: $T_n$ is a crucial physical quantity of interest in gravitational waves, and the determination of the action is aimed at obtaining the nucleation temperature $T_n$. 

Taking the example of $\lams = 1$, we examine in Fig.~\ref{fig:epsTn_para} the applicability of the thin-wall approximation, measured by  $\epsilon (T_n)$. 
The left panel illustrates a positive correlation between $(\Delta V/H)_n$ and $(L_w/R_b)_n$, suggesting that as the ratio $(\Delta V/H)_n$ increases, the relative thickness of the bubble wall correspondingly grows. Additionally, $\epsilon (T_n)$ rises with both parameters, suggesting that $\epsilon (T_n)$, $(\Delta V/H)_n$ and $(L_w/R_b)_n$ are closely related and any of these quantities may serve as reliable indicators for assessing the validity of the thin-wall approximation. 
The right panel illustrates the relationship between $\epsilon(T_n)$, $T_n$, and $\beta/H_n$. $\epsilon (T_n)$ increases as $T_n$ decreases, indicating that the thin-wall approximation becomes less valid at lower nucleation temperatures. 
Moreover, a clear relationship is observed between $\epsilon (T_n)$ and $\beta/H_n$, where lower values of $\beta/H_n$ correspond to higher $\epsilon (T_n)$. This suggests that the thin-wall approximation is better suited for faster PTs with higher $\beta/H_n$. 
For instance, if a $10\%$ error in $\epsilon (T_n)$ is considered acceptable, corresponding limits on the first-order PT parameters can be established: $T_n\gtrsim100$ GeV and $\beta/H_n\gtrsim1000$, which is approximately equivalent to $T_n-T_p\lesssim4~{\rm GeV}$.

%%%%%%%%%%%%%%%%%%
\subsection{Bubble nucleation modes}
%%%%%%%%%%%%%%%%%%

The manner in which bubbles are nucleated is of great importance in the numerical simulation of bubble collisions, as a deep understanding of how bubbles form can significantly influence simulation outcomes.
Nucleation processes during PTs can occur through various modes, each characterized by distinct features. 

The modes of bubble nucleation can be classified in multiple ways based on the nature of bubble formation. 
Each mode provides valuable insights into how time and the underlying physical processes affect bubble formation, ultimately enhancing the accuracy of numerical simulations of bubble collisions and dynamics.  
As shown in \eq{eq:nucl_rate}, the nucleation rate $\Gamma$---the probability of nucleation events---is governed by the tunneling action $S_E(t)$. Consequently, a crucial concept in understanding these modes is the tunneling action,
with each mode reflecting different behaviors as temperature declines, thereby influencing the PT dynamics.

To systematically categorize and analyze the various modes of bubble nucleation, one effective approach involves applying a Taylor expansion to the tunneling action $S_E(t)$,
\beq
\label{eq:Sexp}
S_E(t)= S_E(t_*)+\frac{dS_E(t)}{dt}\Big|_{t_*}(t-t_*)+\frac{1}{2}\frac{d^2S_E(t)}{dt^2}\Big|_{t_*}(t-t_*)^2+\cdots
\eeq 
where $t_*$ is a reference time around which the expansion is performed. The derivatives in each term are evaluated at $t=t_*$.

Three primary nucleation modes~\cite{Megevand:2016lpr}---constant, exponential and simultaneous---serve as approximation approaches for $S_E(t)$, allowing the derivation of semi-analytical results. 

(i) Constant nucleation: This mode is applicable when all higher-order terms in $S_E(t)$ are negligible, leaving only the zeroth-order term, which is is time-independent. 
As a result, the nucleation rate $\Gamma$ remains constant over time. 

(ii) Exponential nucleation: In this mode, $S_E(t)$ is approximated using a linear expansion, corresponding to a first-order approximation. The first-order term in the expansion plays a significant role, highlighting its relevance in the nucleation process.

(iii) Simultaneous nucleation: This mode employs a Gaussian approximation that expands $S_E(t)$ to quadratic order, while the the linear term contributes negligibly.

The conditions\footnote{A rigorous derivation is provided in Appendix~\ref{app:Nm}.} under which each mode is realized are
\beq
\left\{
\begin{aligned}
\frac{1}{2} \left|\frac{\beta_n}{H_n}\right|\left(\frac{T_n^2}{T_p^2}-1\right)\leq\epsilon_0,& \,\, {\rm constant~nucleation}\\
\frac{1}{2} \left|\frac{\beta_p}{H_p}\right|\left(1-\frac{T_p^2}{T_n^2}\right)>\epsilon_0\,\, {\rm and}\,\, \frac{1}{8}\left| \frac{3\beta_n}{H_n} +\frac{\beta'^2_n}{H^2_n}\right| \left(\frac{T_n^2}{T_p^2}-1\right)^2\leq\epsilon_1,& \,\, {\rm exponential~nucleation}\\
\frac{1}{8}\frac{\beta_m^{'2}}{H_m^2}\left(1-\frac{T_m^2}{T_n^2}\right)^2>\epsilon_0
\,\, {\rm and}\,\, \frac{1}{16}\left| \left(\frac{5\beta_n}{H_n}+\frac{3\beta^{'2}_n}{H_n^2}\right)\right|\left(\frac{T_n^2}{T_m^2}-1\right)^3\leq\epsilon_2,& \,\, {\rm simultaneous~nucleation}
\end{aligned}
\right. 
\label{eq:expnucl_sat}
\eeq
where $\beta^{'2}=H(T)^2T^2\frac{d^2S}{dT^2}$ and $T_m$ is the temperature corresponding to the time $t_m$ at which the extremum of $S_E(t)$ occurs. The subscripts $n$, $m$ and $p$ indicate that the quantity is evaluated at the temperatures $T_n$, $T_m$ and $T_p$, respectively. The parameters $\epsilon_0, \epsilon_1, \epsilon_2$ are the error terms introduced to assess the applicability and range of each nucleation mode.

%%%%%%%%%%%%%%%%%%%%%%%%%%%%%%%%
 \begin{figure}[t]
 	\centering
 	\includegraphics[width=0.88\textwidth]{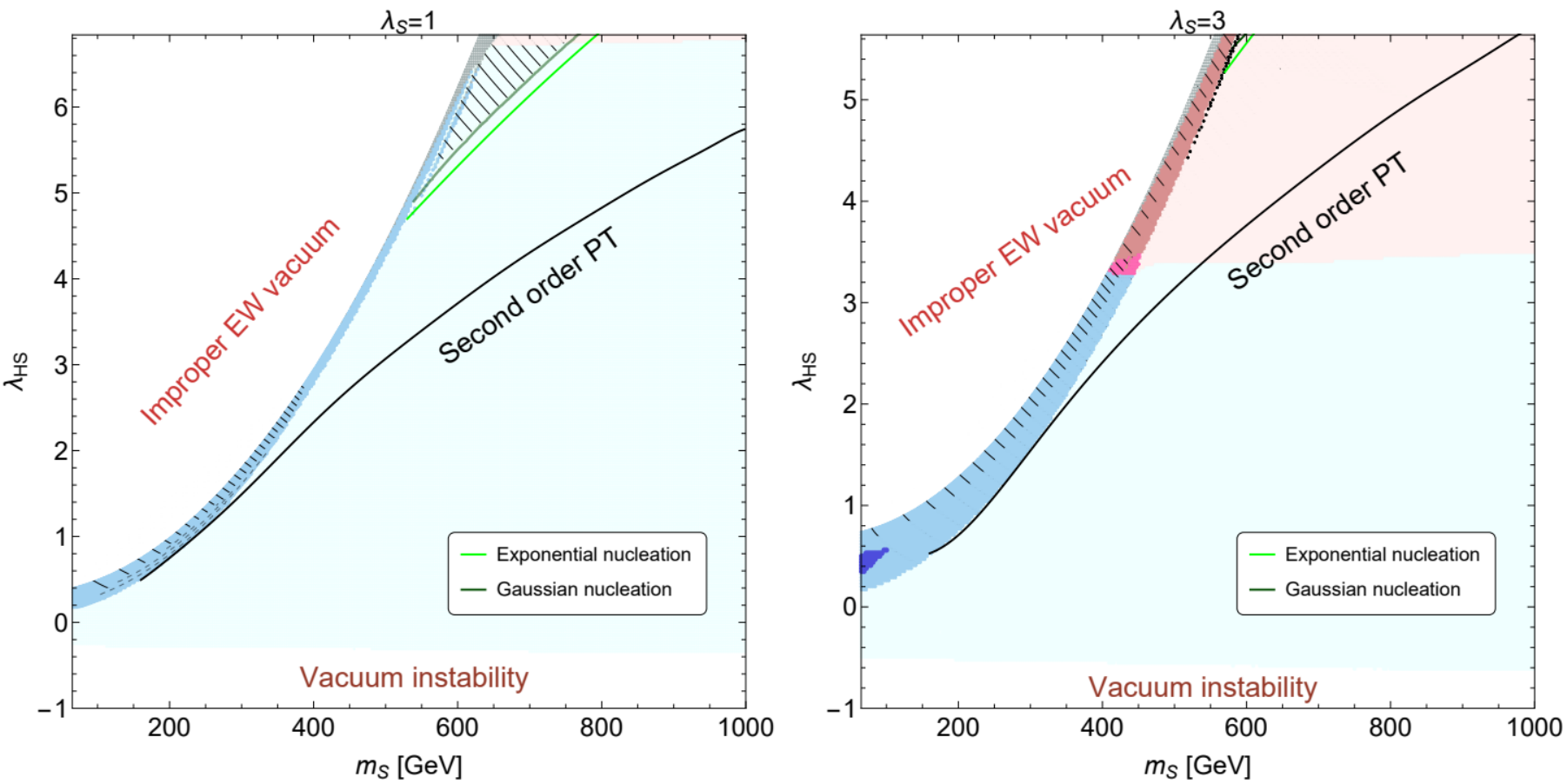}
 	\caption{The distribution of nucleation modes in the parameter space for $\lams=1$ (left) and $\lams=3$ (right) is illustrated. For the $O\to A$ PT, exponential nucleation predominates, occurring in the region below the light green curve. In contrast, Gaussian nucleation happens in the very narrow region above the dark green curve.}
 	\label{fig:nuclmode}
 \end{figure}
 %%%%%%%%%%%%%%%%%%%%%%%%%%%%%%%%
 
Consider the phase change $O \to A$ as an example, in Fig.~\ref{fig:nuclmode} we highlight the
boundaries within the parameter region where different nucleation modes dominate, assuming $\epsilon_0 \sim \epsilon_1 \sim \epsilon_2 \sim 1$. 
Both exponential and Gaussian nucleation processes are present across the parameter region.  
It is evident that exponential nucleation is the predominant mode for bubble formation, 
indicating that, in most scenarios, the nucleation rate grows exponentially,  and the formation of phase $A$ bubbles follows an exponential probability distribution.
However, a very narrow region (above the dark green curve) exists within the parameter space where Gaussian nucleation can take place. This region is located just below the boundary where the nucleation rate becomes suppressed and unsuccessful bubble nucleation is resulted. 
This suggests that under certain conditions---—likely involving a moderately suppressed rate---the formation of bubbles occurs nearly simultaneously across the system, deviating from the typical exponential behavior.
While simplistic, this analysis provides valuable insights into how the bubble formation shifts with variations in the parameter space, offering a clearer understanding of the interplay between two nucleation modes commonly employed in numerical simulation.

%%%%%%%%%%%%%%%%%%%%%%%
\section{Exotic bubbles in consecutive first-order PTs}
\label{ssec:twostep}
%%%%%%%%%%%%%%%%%%%%%%%

\begin{figure}[t]
\centering
\includegraphics[width=0.48\textwidth]{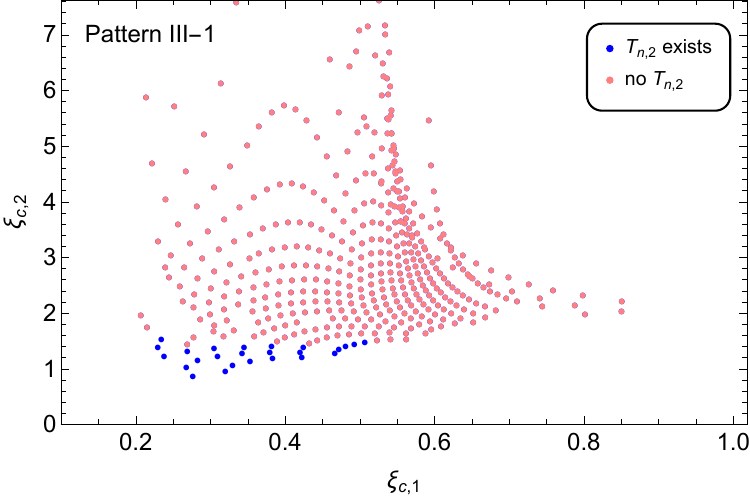}
\includegraphics[width=0.48\textwidth]{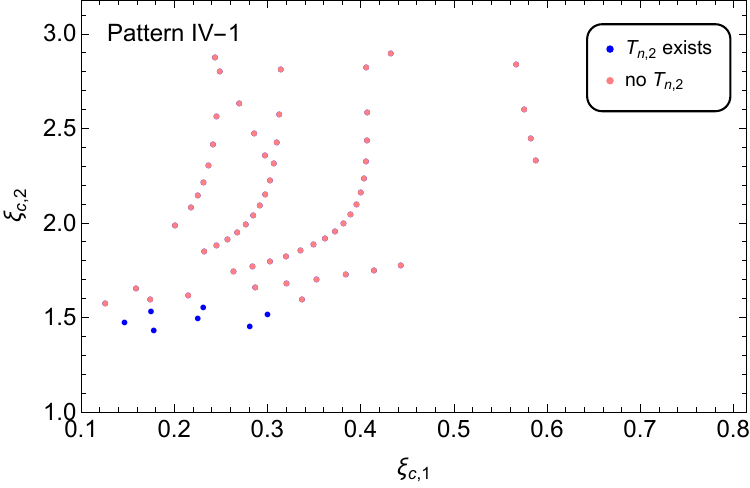}
\caption{The PT strengths for the first and second steps, denoted by $\xi_{c,1}$ and $\xi_{c,2}$, respectively, are shown for the case of two consecutive first-order PTs: Pattern~III-1 (left) and Pattern~IV-1 (right). The points are categorized based on whether bubble nucleation occurs during the second step of the PT. Blue points represent successful bubble nucleation in the second step, while red points indicate failure to nucleate.}
\label{fig:twoPTs1}
\end{figure}

In this section we present a concise analysis of two consecutive first-order PTs, specifically Pattern~III-1 and Pattern~IV-1. Pattern~III-1 corresponds to a small value of $\lams=1$, while Pattern~IV-1 occurs at a large value of $\lams=3$, as illustrated in Fig.~\ref{fig:PT_pattern}. Detailed results are shown in Fig.~\ref{fig:twoPTs1}. 
In both Patterns, we observe that the second-step transition is significantly strong, whereas the first-step transition is rather weak. 
As a result, bubble nucleation in the first step is always successful provided the bubbles exceed the critical size. However, nucleation in the second step is not guaranteed. 
In Fig.~\ref{fig:twoPTs1} we classify the points based on the occurrence of bubble nucleation in the second step. Our analysis reveals that successful nucleation in the second step is associated with a critical PT strength $\xi_{c,2} \lesssim 1.8$, consistent with the findings discussed in Sec.~\ref{ssec:Bn}.

In general, the dynamics of successive FOPTs are considerably more complex. During the thermal history of Pattern~III-1 (Pattern~IV-1) EWPT, there exist three possible phase changes: $O (B) \to B (O)$, $B (O)\to A$ and also $O (B) \to A$ (see Fig.~\ref{fig:SNRmode}). 
The presence of these phase changes allows for the formation of diverse transition sequences, leading to more exotic bubble configurations in addition to the sequential formation of two-step bubbles and the vacuum trapping~\cite{Kurup:2017dzf,Kozaczuk:2019pet,Baum:2020vfl,Biekotter:2021ysx,Biekotter:2022kgf} of the intermediate $B (O)$ vacuum phase. 
For instance, if multiple nucleation processes occur within the same temperature range, there is a potential for the co-existence of two types of bubbles or the emergence of nested bubbles~\cite{Aguirre:2007an,Morais:2019fnm,Croon:2018new}. 

%%%%%%%%%%%%%%%%%%%%%%%
%%%%%%%%%%%%%%%%%%%%%%%

%%%%%%%%%%%%%%%%
\subsection{Bubble configurations in Pattern~III-1 PTs}
%%%%%%%%%%%%%%%%

\begin{figure}[t]
\centering
\includegraphics[width=0.5\textwidth]{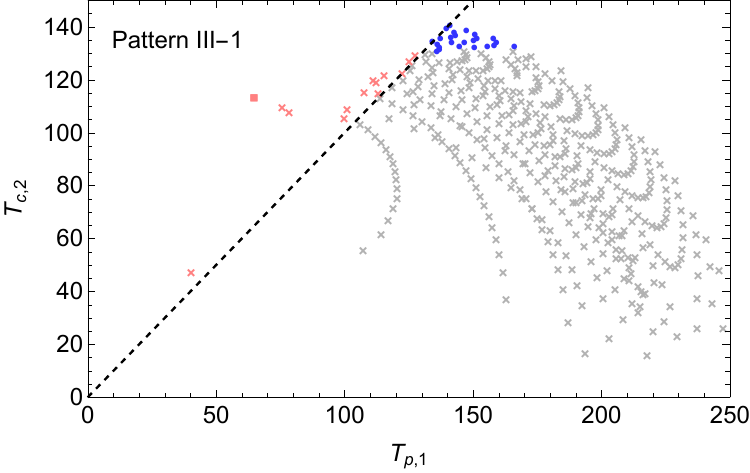}
\caption{Temperature scales relevant to the formation of exotic bubble configurations in Pattern~III-1 EWPT. Blue circles denotes successful EWPTs via sequential bubble nucleation. The red box highlights the coexistence of both $A$ and $B$ bubbles. In contrast, gray and red crosses indicate scenarios leading to vacuum trapping, distinguished by $T_{p,1}>T_{c,2}$ (gray crosses) or $T_{p,1}<T_{c,2}$ (red crosses).
}
\label{fig:III-1PTs2}
\end{figure}

To explore the details of Pattern~III-1 EWPT, we present in Fig.~\ref{fig:III-1PTs2} the relevant temperature scales for all points: $T_{p,1}$ and $T_{c,2}$, which indicates the completion of the first-step PT ($O \to B$) and the onset of the second-step PT ($B \to A$), respectively. 
For points where bubble nucleation is successful in the second step (indicated by blue points), we find that $T_{p,1} > T_{c,2}$. This implies that the second-step PT occurs only after the first-step PT has fully completed, thereby preventing the formation of nested bubbles. 
In these cases, the initial $O$ vacuum first transitions to the intermediate $B$ vacuum through the nucleation of $B$ bubbles, followed by a subsequent transition to the final $A$ vacuum through the nucleation of $A$ bubbles. Point III-1a, listed in Table~\ref{tab:ExoBub}, serves as a benchmark point for this scenario. 
Conversely, for points where nucleation fails in the second step, the intermediate $B$ vacuum phase is typically trapped. Point III-1b and III-1c given in Table~\ref{tab:ExoBub} exhibit this behavior, with the distinction being that either $T_{p,1}$ or $T_{c,2}$ is higher. 
However, an important exception arises when the nucleation temperature $T_n (O \to A)$ exists and $T_n (O \to A)>T_{p,1}$, corresponding to Point III-1d and highlighted by a red box in Fig.~\ref{fig:III-1PTs2}. In this case, as the universe cools below $T_n (O \to A)$, both $A$ and $B$ bubbles can nucleate within the initial $O$ vacuum phase. This scenario represents the simultaneous occurrence of two FOPTs: $O \to B$ and $O \to A$. 

%%%%%%%%%%%%%%%%%%%%%%%
%%%%%%%%%%%%%%%%%%%%%%%

%%%%%%%%%%%%%%%%
\subsection{Bubble configurations in Pattern~IV-1 PTs}
%%%%%%%%%%%%%%%%

%%%%%%%%%%%%%%%%%%%%%%%
\begin{figure}[t]
\centering
\includegraphics[width=0.5\textwidth]{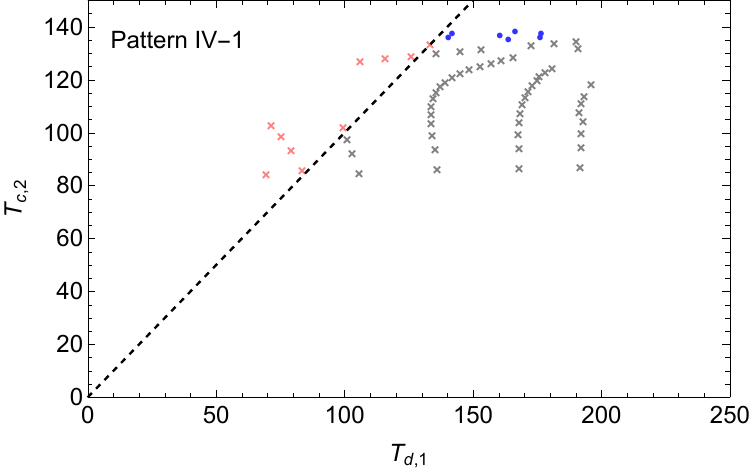}
\caption{Temperature scales relevant to exotic bubble configurations from Pattern~IV-1 EWPT. Blue circles denote successful EWPTs via bubble nucleation after a smooth first-step transition. Gray and red crosses indicate vacuum trapping, distinguished by $T_{d,1}>T_{c,2}$ (gray crosses) or $T_{d,1}<T_{c,2}$ (red crosses).
}
\label{fig:IV-1PTs2}
\end{figure}
%%%%%%%%%%%%%%%%%%%%%%%

In contrast, the analysis of PT dynamics for this Pattern is less intricate. This simplicity arises because the inverse $s$-bubbles formed within this pattern are generally smaller than the critical bubbles, as clearly demonstrated in Fig.~\ref{fig:multifield_Bubble}. Consequently, the transition from the $B$ vacuum state to the $O$ vacuum state cannot proceed via bubble nucleation, potentially leaving the universe trapped in the initial $B$ vacuum state. 
Nevertheless, this description may not fully capture the PT dynamics if the potential barrier between $B$ and $O$ vacuum states disappears or if a direct transition from the $B$ vacuum to the $A$ vacuum occurs at low temperatures. Our analysis reveals that the latter scenario is impossible due to an insufficient nucleation rate, even as the free energy of the $A$ vacuum gradually increases, causing it to become degenerate with the $B$ vacuum before the potential barrier between $B$ and $O$ vanishes (as exemplified by Point IV - 1c in Table~\ref{tab:ExoBub}. 
Therefore, during this Pattern of EWPT, the universe first undergoes a smooth transition from the initial $B$ vacuum to the $O$ vacuum state when the potential barrier between $B$ and $O$ disappears at $T_{d,1}$. We find that $T_{d,1}$ can be either higher or lower than $T_{c,2}$, the critical temperature at which the $A$ and $O$ vacuum states become degenerate while remaining separated by a potential barrier in between. This relationship is illustrated in Fig.~\ref{fig:IV-1PTs2}. In either case, the ultimate fate of PT dynamics depends entirely on whether the second-step transition---from the $O$ vacuum to the desired $A$ vacuum state---can successfully occur. For cases where bubble nucleation succeeds in the second step (indicated by blue points), we find that $T_{d,1} > T_{c,2}$. This allows the universe to transition to the final $A$ vacuum through the nucleation of $A$ bubbles following the initial smooth transition. Point IV-1a, listed in Table~\ref{tab:ExoBub}, serves as a benchmark point for this phenomenologically viable scenario. 
Conversely, if the second-step transition fails, the universe remains trapped in the intermediate $O$ vacuum, where the EW symmetry remains unbroken. These cases are represented by Point III-1b and III-1c in Table~\ref{tab:ExoBub}, differing only in the relative ordering of $T_{d,1}$ and $T_{c,2}$.  Despite this difference, both cases result in the universe being confined to an EW-unbroken vacuum state, persisting until zero temperature. Such scenarios are therefore phenomenologically excluded.

%%%%%%%%%%%%%%%%%%%%%%%%%%%%%%%%
\begin{table}[t]
 \centering
 \vspace{-5pt}
 \renewcommand{\tablename}{Table}
 \caption{Exotic bubble configurations and their associated dynamics from the consecutive first-order PTs: Pattern~III-1 (upper division) and Pattern~IV-1 (lower division).}
\begin{footnotesize}
\begin{tabular}{c|c|ccc}
\toprule
\multicolumn{5}{c}{Pattern~III-1 ($\lams=1$)}\\
\hline
Parameters & III-1a & III-1b & III-1c & III-1d\\ 
\hline
$m_S$ & 441 & 500 & 475 & 569\\
$\lamhs$ & 3.35 & 4.28 & 3.86 & 5.30\\
\hline
$T_c (O\to B)$ & 138.8 & 145.1 & 130.8 & 122.6\\
$T_n (O\to B)$ & 136.2 & 140.0 & 125.9 & 67.7\\
$T_p (O\to B)$ & 135.9 & 139.3 & 125.5 & 64.9\\
$T_c (B\to A)$ & 130.7 & 119.2 & 127.6 & 112.7\\
$T_n (B\to A)$ & 78.2 & - & - & -\\
$T_c (O\to A)$ & 131.8 & 125.6 & 123.6 & 114.3\\
$T_n (O\to A)$ & - & - & - & 66.3\\
\hline
PT dynamics &  $O \to B \to A$ &  $O \to B$ (trapped) &  $O \to B$ (trapped) &  $O \to B$ \& $O\to A $ simultaneously\\
\hline
Bubble & Sequential $B$ and $A$ bubbles & $B$ bubbles only & $B$ bubbles only & Co-existence of $A$\\
configurations& in $O$ and $B$ vac, respectively &&& and $B$ bubbles in $O$ vac\\
\bottomrule
\toprule
\multicolumn{5}{c}{Pattern~IV-1 ($\lams=3$)}\\
\hline
Parameters & IV-1a & IV-1b & IV-1c \\ 
\hline
$m_S$ & 519 & 556 & 550\\
$\lamhs$ & 4.43 & 5.06 & 4.97 \\
\hline
$T_c (B\to O)$ & 167.3 & 174.5 & 143.9\\
$T_d (B\to O)$ & 166.3 & 172.9 & 133.6\\
$T_c (B\to A)$ & - & - & 134.4 &\\
$T_c (O\to A)$ & 138.3 & 133.5 & 133.9 &\\
$T_n (O\to A)$ & 131.0 & - & - & \\
\hline
PT dynamics &  $B \to O \to A$ &  $B \to O$ (trapped) &  $B \to O$ (trapped)  &   \\
\hline
Bubble & $A$ bubbles in $O$ vac & no bubbles & no bubbles &  \\
configurations&  &&&  \\
\bottomrule
  \end{tabular}
\end{footnotesize}
 \label{tab:ExoBub}
\end{table}
%%%%%%%%%%%%%%%%%%%%%%%%%%%%%%%%

%%%%%%%%%%%%%%%%%%%%%%%%%%%%%%%%
\section{GW signals: observational signatures}
\label{sec:gwparas}
%%%%%%%%%%%%%%%%%%%%%%%%%%%%%%%%

As we understand it, the production of GWs from FOPTs begin with bubble collisions, which are typically completed by the percolation temperature $T_p$. 
Let $T_*$ represent the temperature at which GW production occurs. In general, this gives the relation, $T_n > T_* > T_p$. For fast PTs, the approximation $T_* \simeq T_n$ is sufficiently accurate for practical purposes. 

In addition to $T_n$, two key parameters are essential for computing the produced GW power spectrum. 
The first parameter quantifies the strength of the PT, defined as the ratio of the vacuum energy density to the background radiation energy density of the universe. We adopt the definition from~\cite{Giese:2020rtr}, which is expressed as
\begin{align}
	\alpha_n={\rho_{\rm vac}(T_n)\over\rho_{\rm rad}(T_n)}=\frac{1}{\rho_\text{rad}(T_n)}\Big( \Delta V(T) - \frac{T}{4} \frac{\partial}{\partial T}\Delta V(T)\Big)\Big|_{T=T_n} ,
\end{align}
where $\Delta V(T) \equiv V_{\rm eff}(\vec{\phi}_F, T)-V_{\rm eff}(\vec{\phi}_T, T)$, as defined in Sec.~\ref{ssec:bub_pro}. 
The second parameter is the ratio $\beta/H$, which is typically expressed as
\begin{align}
	\beta/H_n=T\frac{d}{dT}\Big(\frac{S_3}{T}\Big)\Big|_{T=T_n} ,
\end{align}
where the subscript $n$ indicates that the value is evaluated at $T_n$. 
The quantity $\beta^{-1}$ is related to the coefficients of the first-order terms in the Taylor expansion of the tunneling action (see Appendix~\ref{app:Nm}). 
Assuming an exponential bubble nucleation rate, $\beta^{-1}$, can be interpreted as the duration of the PT. 
Strictly speaking, both of these parameters should be evaluated at $T_*$, the temperature corresponding to the time of GW production. 

%%%%%%%%%%%%%%%%%%%%%%%%%%%%%%%%
\begin{figure}[t]
\centering
\includegraphics[width=0.5\textwidth]{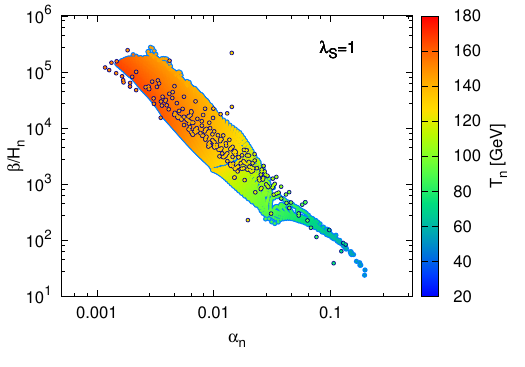}
\hspace{-3mm}
\includegraphics[width=0.5\textwidth]{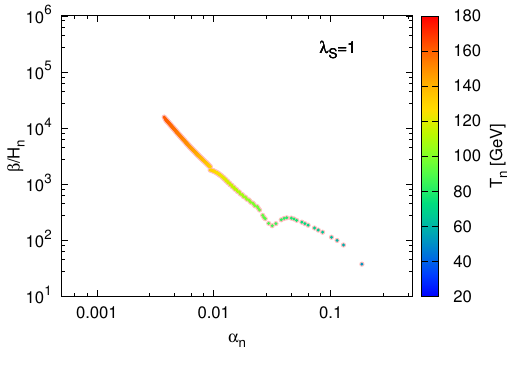}\\
\includegraphics[width=0.5\textwidth]{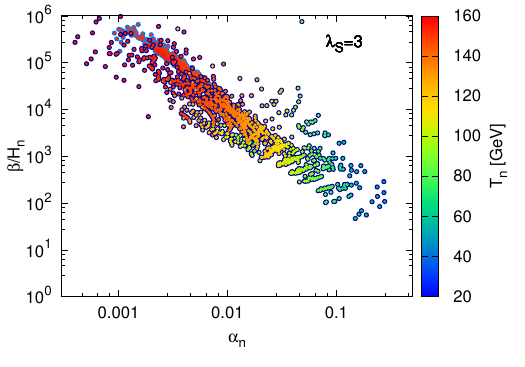}
\hspace{-3mm}
\includegraphics[width=0.5\textwidth]{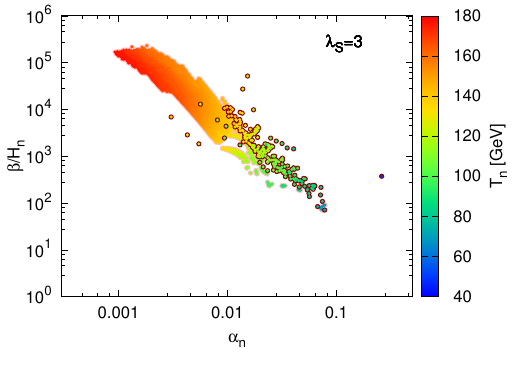}
\caption{Key PT parameters ($\alpha_n$, $\beta/H_n$, and $T_n$) relevant for predicting GW signals are presented in the parameter space for $\lams=1$ (upper panel) and $\lams=3$ (lower panel). For multi-step PTs (Pattern~III-2 and Pattern~IV-2), the parameters are evaluated for the step in which EWSB occurs. 
The left panel illustrate scenarios where the high-temperature vacuum resides at the origin. Like Fig.~\ref{fig:epsTn_para}, points enclosed in light blue and dark blue circles correspond to Pattern~I PTs and Pattern~III-2 PTs, respectively. 
Scenarios with a $\mathbb{Z}_2$-broken vacuum are depicted in the right panel, where points enclosed in dark-red and pink circles correspond to Pattern~II PTs and Pattern~IV-2 PTs, respectively.
}
\label{fig:alpha_beta}
\end{figure}
%%%%%%%%%%%%%%%%%%%%%%%%%%%%%%%%

Fig.~\ref{fig:alpha_beta} illustrates the strong correlation among the three key parameters that determine the production of gravitational waves. We observe that both $\beta/H_n$ and $T_n$ decrease as $\alpha_n$ increases. Furthermore, there is no point in the parameter space where both $\alpha_n$ and $\beta/H_n$ take high or lower values simultaneously. This pattern reflects underlying constraints on the dynamics of the first-order EWPT.
This behavior can be well understood through Fig.~\ref{fig:sucnucl}, which shows the contours of $\alpha_n$ and $\beta/H_n$ in the parameter space that gives rise to a first-order EWPT. 
A key feature in these plots is that $\alpha_n$ and $\beta/H_n$ change oppositely in magnitude as the model parameters, $\ms$ and $\lamhs$, are varied in any direction. This inverse relationship is a direct consequence of the potential structure of the model.
For instance, increasing $\lamhs$ typically increases the height of the potential barrier between the two degenerate vacuum states at $T_c$. As the barrier height grows, the tunneling rate of the false vacuum decay is suppressed, delaying the onset of the nucleation.
This delay occurs because nucleation only begins when the barrier weakens sufficiently at lower  $T_n$. Consequently, the difference between $T_c$ and $T_n$ increases (which corresponds to a smaller value of $\beta/H_n$), and the deeper potential in the true vacuum generates more latent heat, thereby increasing $\alpha_n$.
This interplay between $\alpha_n$ and $\beta/H_n$ leads to characteristic behaviors in EWPTs within this model. 
Strong EWPTs tend to occur later in the evolution of the universe and take longer to complete. In contrast, weaker EWPTs occur earlier and can proceed with near-instantaneous nucleation in extreme cases.

Additionally, as shown in Fig.~\ref{fig:sucnucl}, the distribution of the $\alpha_n$ and $\beta/H_n$ contours is nearly degenerate. This suggests that $\alpha_n$ and $\beta/H_n$ are not fully independent parameters; instead, they are strongly correlated, as highlighted in Fig.~\ref{fig:alpha_beta}. In this context, the typical approach where $\alpha_n$, $\beta/H_n$ and $T_n$ are treated as free parameters is no longer perfectly suitable for analyzing the GW signals within this model.
On the other hand, one might wonder whether it is possible to distinguish the processes through which EWSB occurred. For example, if the four Patterns discussed earlier exhibit distinctly different values for $\alpha_n$, $\beta/H_n$ and $T_n$, the resulting GW power spectrum would peak in different frequency bands, potentially. This could allow us to probe the details of the EWPT through GW detectors. 
To this end, we attempt to distinguish the four Patterns 
in Fig.~\ref{fig:alpha_beta}, although the results are not as conclusive as we had anticipated. 
For instance, Pattern~I shows the largest $\alpha_n$. Most of the points, however, are similar, with the only notable difference being the lack of a specific characteristic in Pattern~II.

The key difference lies in the fact that an ultra-fast EWPT, with $\beta/H_n \gtrsim 10^4$, cannot be achieved in Pattern~II. This implies that Pattern~II would generate a GW power spectrum that is peaked at a relatively high frequency, $f_{\rm  p}\gtrsim 1~{\rm Hz}$, with a small amplitude---well below the sensitivity of ongoing GW detection projects like TianQin, Taiji and LISA. As a result, distinguishing Pattern~II from the other Patterns will be very challenging in the near future.
It remains difficult to determine whether or not the EWPTs begin with a $\mathbb{Z}_2$-broken vacuum phase.
However, it may not be entirely hopeless to differentiate between Pattern~I and Pattern~III, even though these Patterns exhibit nearly identical ranges for the relevant parameters. This is because the mechanism underlying the generation of GWs in the two-step PT process might differ significantly from those in the one-step EWPT, particularly in the presence of domain walls during the transition~\cite{Blasi:2022woz,Agrawal:2023cgp,Wei:2024qpy}. We plan to explore this issue in future work.

%%%%%%%%%%%%%%%%%%%%%%%%%%%%%%%%
\section{Conclusions and future directions}\label{sec:concl}
%%%%%%%%%%%%%%%%%%%%%%%%%%%%%%%%

The study of EWPT dynamics has evolved into a sophisticated and multifaceted discipline, demanding precise modeling of the thermal history from the Big Bang to the electroweak scale, coupled with detailed analyses from bubble nucleation to percolation and the influence of topological defects. For new physics models, a comprehensive framework integrating these aspects this holistic approach is essential to uncover the full spectrum of EWPT phenomena and their cosmological implications.

In this work, we conduct an in-depth analysis of the real, $\mathbb{Z}_2$-odd scalar singlet model, focusing on its thermal history and the intricate bubble dynamics during PTs. 
We systematically explore the vacuum structures at temperature above the TeV scale and the diverse PT processes from high-temperature vacua to the EW vacuum at low temperatures. 
A key finding is that spontaneous symmetry breaking can occur in the high-temperature vacuum, typically requiring 
large values of the couplings $\lamhs$ and $\lams$. However, under the constraint that the model remains free of the Landau pole below a few TeV, the scenario of an EW-broken vacuum at high temperatures is ruled out. As a result, the high-temperature vacuum prior to EWPTs either preserves symmetry or breaks the $\mathbb{Z}_2$ symmetry. As the universe cools, these vacua evolve to the zero-temperature EW vacuum through various pathways, including one-step, two-step, or even three-step PTs, though multi-step transitions occur only in a small region of the parameter space. Notably, the existence of a $\mathbb{Z}_2$-broken vacuum phase prior to EWPTs gives rise to the unusual phenomenon of ISB. 

Excluding three-step PTs, we identify four distinct patterns of EWPTs, each characterized by unique symmetry-breaking mechanisms. As a result, the bubbles nucleated in these transitions exhibit distinct field configurations, including hybrid $h$-$s$ bubbles involving both fields and inverse $s$ bubbles arising from the ISB of $\mathbb{Z}_2$ symmetry. 
To address the complexities of analyzing these bubbles, we introduce a novel methodology based on energy-density distributions, allowing for precise determination of key properties such as bubble radius $R_b$ and wall thickness $L_w$. 
Our analysis reveals that bubbles nucleated at lower temperatures---whether from fast or slow FOPTs---tend to have smaller radii, while their wall thickness remains nearly constant. This results in an increased $L_w/R_b$ ratio, reducing the accuracy of the thin-wall approximation. 
Additionally, we find that bubbles formed during fast FOPTs are generally larger than those from slow FOPTs. 
However, inverse s-bubbles, which emerge from ultra-fast FOPTs associated with $\mathbb{Z}_2$ symmetry restoration, fail to exceed the critical radius and collapse instantaneously upon formation. This obscures the distinction between these transitions (when the potential barrier between degenerate vacua disappears) and corresponding second-order transitions. 
On the other hand, extremely strong FOPTs with $\xi_c\gtrsim 1.8$ in any pattern typically results in unsuccessful bubble nucleation, leading to a substantial reduction of Pattern-II EWPTs that originate from a $\mathbb{Z}_2$-broken vacuum. 
In both cases of unsuccessful bubble nucleation, the universe may remain trapped in a false vacuum,  preventing the generation of observable GW signals and making detection particularly challenging. 

Conversely, for FOPTs where bubble nucleation succeeds, our analysis reveals that strong EWPTs tend to occur later in the thermal evolution of the universe and take longer to complete, whereas weaker EWPTs occur earlier and, in extreme cases, proceed with near-instantaneous nucleation. By classifying nucleation modes, we find that exponential nucleation is the dominant mechanism for bubble formation, with the nucleation rate growing exponentially in most scenarios. In contrast, Gaussian nucleation occurs only in a very narrow region of the parameter space.
This classification offers key insights into how bubbles form across different patterns of PTs and provides important guidance for modeling bubble formation in numerical simulations.  
These advancements are crucial for refining our understanding of PT dynamics and accurately predicting the resulting GW signals.

Finally, we investigate successive FOPTs, which involve multiple phase changes occurring in sequence. These transitions demand meticulous analysis to fully unravel their dynamics and the exotic bubble configurations that emerge.  
By systematically assessing bubble nucleation at each PT step, we uncover complex dynamics, including sequential nucleation and the coexistence of bubbles from two distinct vacua. Such scenarios greatly increase the complexity of modeling bubble dynamics, highlighting the need for advanced theoretical frameworks and computational techniques to accurately capture and describe these intricate phenomena.

The rich EWPT processes identified in this work, particularly multi-step PTs, have profound implications for GW spectra, potentially generating distinctive signals that could be detected by future GW observatories. 
We emphasize that the presence of high-temperature $\mathbb{Z}_2$-broken vacua in Pattern~II and Pattern~IV prior to EWPTs leads to significant cosmological consequences and opens exciting opportunities for uncovering new physics. 
For instance, it may influence the relic abundance of dark matter observed today, as the singlet field---when acquiring a nonzero vev in the $\mathbb{Z}_2$-broken vacua--alters its interactions with the Higgs field or open new annihilation channels prior to freeze-out.\footnote{Refs.~\cite{Baker:2016xzo,Baker:2017zwx} investigated this scenario in extended frameworks where an additional scalar field triggers the EWPT, while an extra fermion, serving as dark matter, acquires its abundance through its own decay or freeze-in mechanisms.} Furthermore, if the singlet field interacts with the right-handed neutrinos (RHNs) $N_1,N_2$, it could enhance the CP asymmetry via the decay of the heavier RHN $N_2$, potentially generating a lepton asymmetry before EWPTs~\cite{LeDall:2014too,Alanne:2018brf}. On the other hand, spontaneous $\mathbb{Z}_2$ symmetry breaking in Pattern~III leads to the formation of domain walls, creating regions where the EW symmetry is restored. As these domain walls propagate through space, they could generate the observed baryon asymmetry~\cite{Brandenberger:1994mq,Schroder:2024gsi,Azzola:2024pzq}. Both of these scenarios could provide new mechanisms for successful baryogenesis and warrant further investigation. 

Overall, this study establishes a comprehensive framework for investigating complex PTs and the formation of exotic bubble configurations in a broad class of BSM models, extending well beyond the simple scalar singlet model explored here. 
While further validation through high-precision perturbative thermal field theory or non-perturbative methods is necessary, our findings provide crucial insights into the thermal history of the EW vacuum, bubble nucleation and the phenomenology of EWPTs, advancing our understanding of EWPTs in the context of BSM physics. 
Future studies will benefit from extending this framework through numerical simulations, exploring the interplay between PT dynamics and observable signatures, and investigating the broader implications for cosmology and particle physics.

Moreover, this work sets the stage for future investigations into multi-step PTs and their cosmological consequences, including potential GW signals and their role in shaping early-universe physics. 
It is also important to note that topological defects may form during multi-step PTs, significantly influencing the PT dynamics. A comprehensive exploration of these phenomena lies beyond the scope of this study but will be addressed in future studies. 

%%%%%%%%%%%%%%%%%%%%%%%%%
\subsubsection*{Acknowledgments}
We thank Tianjun Li for providing valuable suggestions regarding the Landau poles. This work is supported by the National Key Research and Development Program of China (Grant No. 2021YFC2203002) and in part by the GuangDong Major Project of Basic and Applied Basic Research (Grant No. 2019B030302001). Y. J. is also funded by the Guangzhou Basic and Applied Basic Research Foundation (No. 202102021092), the GuangDong Basic and Applied Basic Research Foundation (No. 2020A1515110150) and the Sun Yat-sen University Science Foundation. 

\appendix

\addcontentsline{toc}{section}{Appendices}

%%%%%%%%%%%%%%%%%%%%%%%%%%%%%%%%%
\section{Vacuum structure at zero temperature} \label{sec:vac}
%%%%%%%%%%%%%%%%%%%%%%%%%%%%%%%%%

The vacuum structure defines all possible vacua (including both true and false vacua) that may coexist for a given effective potential at any temperature. This structure can vary significantly depending on the specific set of free parameters of the potential. 
Nonetheless, the vacuum structure can provide an important indication about the possibility of different FOPTs experienced at high temperatures. 
At zero temperature, the vacuum structure can be constructed by analyzing the classical values of each physical degree of freedom associated with the $H$ and $S$ fields. 
For simplicity, we neglect the irrelevant Goldstone modes and focus on the vev of two neutral fields, $\langle H^\dagger H \rangle \equiv h^2/2$ and $ \langle S \rangle =s$, giving rise to the two-field potential at zero temperature,
\beq
	\label{eq:V0}
	V_0(h,s)=-\frac{1}{2}\mu^2_H h^2+\frac{1}{4 }\lambda_H h^4-\frac{1}{2}\mu_{S}^2s^2+\frac{1}{4}\lambda_Ss^4+\frac{1}{2}\lambda_{HS}h^2s^2
\eeq
The values of $h$ and $s$ at the potential minimum correspond to the homogenous background of the $H$ and $S$ fields. Therefore, constructing the vacuum structure involves finding the minima of the potential, \eq{eq:V0}.

%%%%%%%%%%%%%%%%%%%%%%%%%
\subsection{Tree-level analysis}

%%%%%%%%%%%%%%%%%%%%%%%
\begin{table}[t]
\centering
  \setlength\tabcolsep{8pt}
  \caption{The table summarizes the four potential stationary points in the tree-level potential at zero temperature: $(v,0)$, $(0,0)$, $(0,w)$ and $(v_h,w_s)$, where $v=\pm \mu_H/\sqrt{\lambda_H}$, $w=\pm \mu_S/\sqrt{\lambda_S}$. $v_h\ne v$ and $w_s\ne w$ are the non-trivial solutions determined by the minimization conditions, \eq{m.c.}. A \checkmark indicates a stationary point, while --- indicates that the point is stationary. The last four cases are excluded because $(v,0)$ is not a stationary point, precluding the possibility of EW vacuum. For the remaining cases, the properties of the stationary points are analyzed with the following possibilities: GMin: Global minimum, LMin: Local minimum, LMax: Local maximum, Sad: Saddle point. Neither Type~D nor Type~E can support a stable EW-broken vacuum, rendering them phenomenologically infeasible. 
} 
  \begin{tabular}{ c | c  c  c c | c}
   \hline
   \multirow{2}{*}{Type} & \multicolumn{4}{c|}{Potentially stationary points} & \multirow{2}{*}{Status} \\
   & $(v, 0)$ & $(0,0)$  & $(0, w)$  & $(v_h, w_s)$ & \\
   \hline
   A & \checkmark GMin & \checkmark LMax  & \checkmark LMin & \checkmark Sad & Viable \\
   \hline
   B & \checkmark GMin & \checkmark LMax & \checkmark Sad & --- & Viable\\
    \hline
   C & \checkmark GMin & \checkmark Sad & --- & ---  & Viable\\
   \hline
   D & \checkmark LMin & \checkmark Sad & --- & \checkmark Sad & No stable EW vacuum \\
   \hline
   E & \checkmark LMin  & \checkmark LMax  & \checkmark GMin & \checkmark Sad & Improper EW vacuum \\
   \hline
   F & ---  & \checkmark & \checkmark& \checkmark  & Excluded \\
   \hline
   G & --- & \checkmark & \checkmark & --- &  Excluded   \\
   \hline
   H & --- & \checkmark & --- & \checkmark  &  Excluded   \\
   \hline
   I & --- & \checkmark & --- & ---   &  Excluded  \\
   \hline
  \end{tabular}
\label{tab:vac}
\end{table}
%%%%%%%%%%%%%%%%%%%%%%%

The minima of the tree-level potential are determined by the minimization conditions,
\begin{align}
	\left\{
	\begin{aligned}
		\frac{\partial V_0}{\partial h}&=h( \lambda _H h^2+\lambda _{HS} s^2- \mu _H^2 )=0,  \\
		\frac{\partial V_0}{\partial s}&=s(\lambda _{HS} h^2  +\lambda _S s^2 - \mu_{S}^2)=0.  \\
	\end{aligned}
	\right.
	\label{m.c.}
\end{align}

Mathematically, Eq~\eqref{m.c.} has 9 possible stationary points, which could be either local extrema or saddle points. However, some of these points may be eliminated depending on the specific set of model parameters. Due to the symmetric structure of the potential in the $h-s$ field space, as given by \eq{eq:V0}, it is sufficient to focus on the four stationary points: one at the origin $(0,0)$, two along the direction of the either field, $(v,0)$ and $(0,w), $ and one in the first quadrant, $(v_h,w_s)$.
Based on the properties of these four points, we categorize the 9 different cases in Table~\ref{tab:vac}. 
Among these, we exclude the last four cases where the stationary point at $(v,0)$ is not possible, thus preventing a proper electroweak (EW) symmetry breaking.
Furthermore, Type~D and Type~E are physically unreasonable because the point $(v,0)$ is not the global minimum of the potential and, therefore, cannot serve as the stable EW vacuum. 
As a result, only three cases remain, under the requirement that 
$(v,0)$ is the global minimum.

\begin{figure}[t]
	\includegraphics[width=0.33\textwidth]{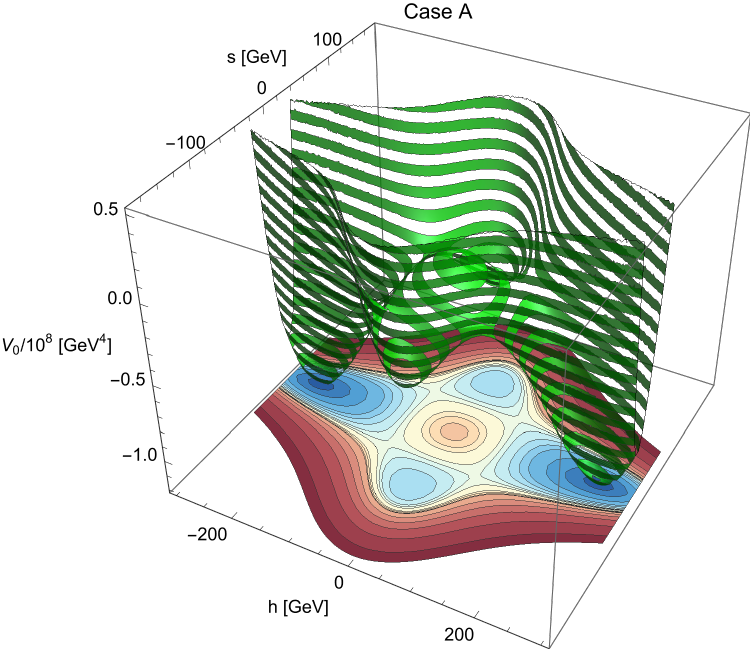}
	\includegraphics[width=0.33\textwidth]{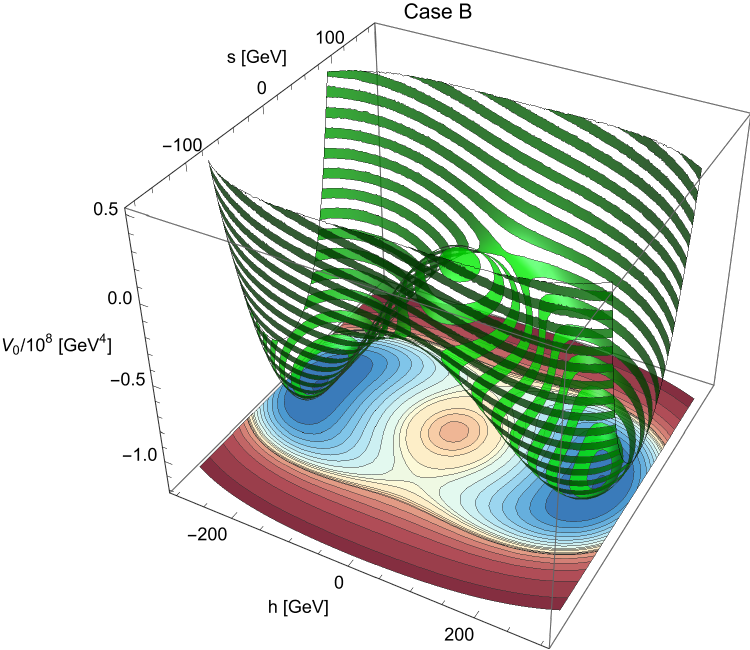}
	\includegraphics[width=0.33\textwidth]{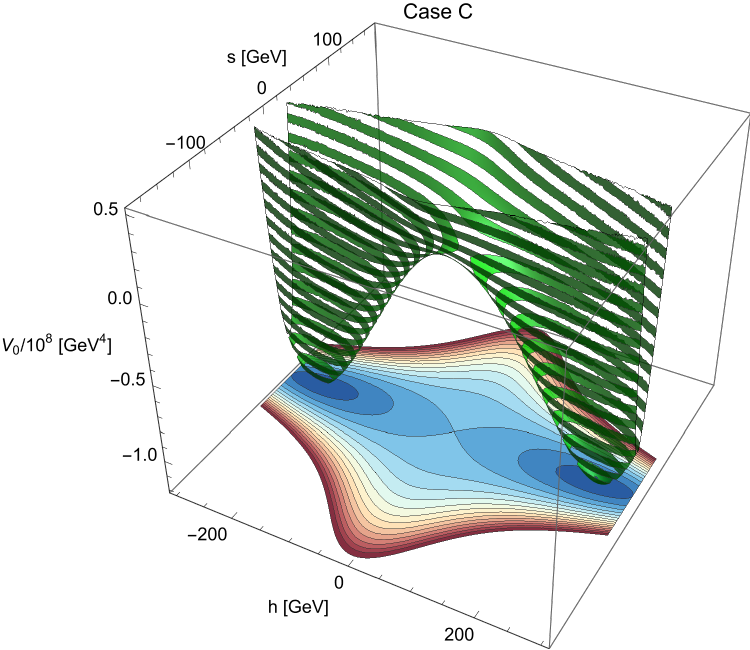}
	\caption{The three-dimensional contours and the two-dimensional projections of the tree-level potential, \eq{eq:V0}, for scenarios where the vacua of Type~A, B, and C (from left to right) are established at zero temperature. The horizontal axes represent the $h$ and $s$ field directions, while the vertical axis shows the value of $V_0/10^8$.} 
	\label{fig:vacpot}
\end{figure}

For the remaining Types (A, B, C), we analyze the property of the possible stationary points, as these will influence the dynamics of EWPT, as discussed in Sec~\ref{ssec:PT}. 
The contours of $V_0$, \eq{eq:V0}, for these cases are shown in Fig.~\ref{fig:vacpot}. 
In Type~C, the potential along the $s-$axis has a stationary point at the origin $(0,0)$, which is a saddle point. 
The situation is more intricate in Type~A and Type~B. In these cases, the origin $(0,0)$ can be either a local maximum or a saddle point.  The key difference lies in the behavior of the point $(0,w)$: 
in Type~A, it is a local minimum, whereas in Type~B, it is a saddle point. This distinction has significant implications for the PT dynamics.

%%%%%%%%%%%%%%%%%%%%%%%%%%%%%%%%
%%%%%%%%%%%%%%%%%%%%%%%%%%%%%%%%

\begin{figure}[t]
\centering
\includegraphics[width=0.98\textwidth]{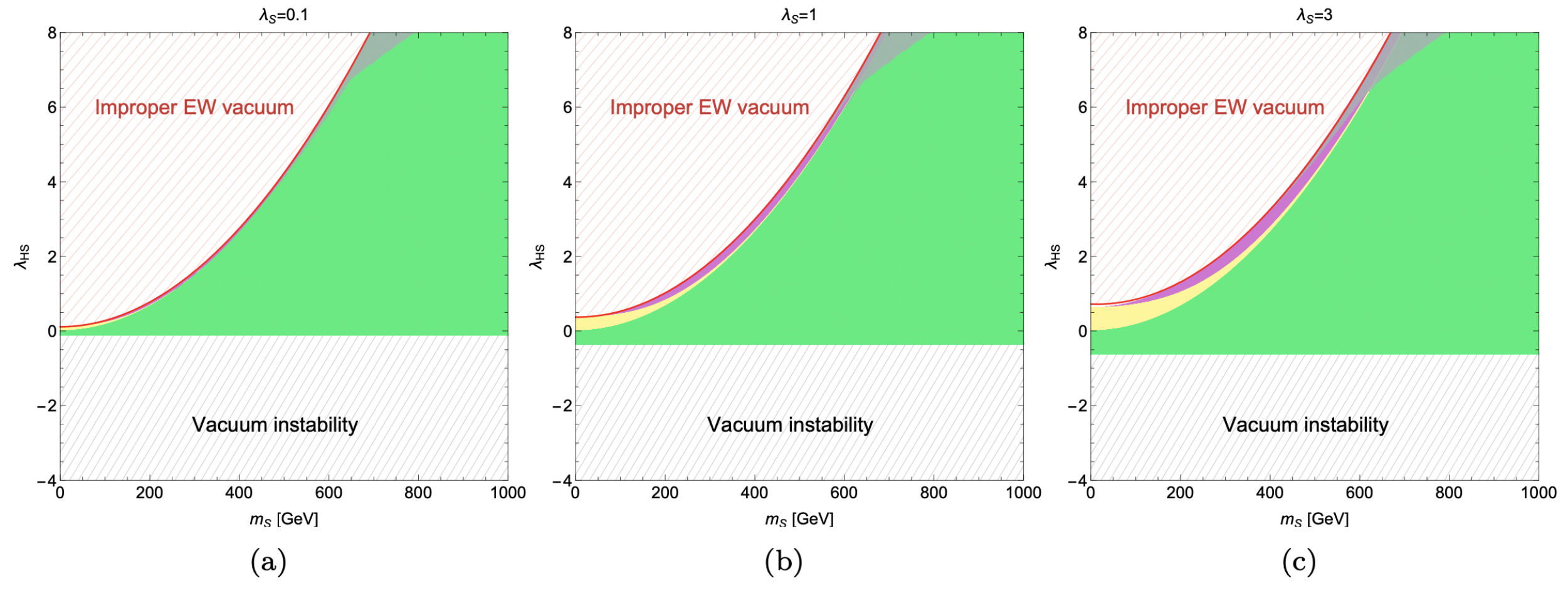}
\caption{The vacuum structures across the parameter space are illustrated for different values of $\lambda_S$: (a) $\lambda_S=0.1$, (b) $\lambda_S=1$, and (c) $\lambda_S=3$. The grey-hatched region at the bottom corresponds to Type~D vacuum, where the EW vacuum is unstable. An improper EW vacuum, classified as Type~E, is generated at tree level (red-hatched) and at one-loop level (gray-shaded). These cases fail to produce a stable EW vacuum at zero temperature and are therefore excluded.
The remaining parameter space supports a proper and stable EW vacuum with the following vacuum structures: Type~A (represented by the purple-shaded region), Type~B (by the yellow-shaded region) and Type~C (by the green-shaded region). 
}
\label{fig:PT_pattern_0}
\end{figure}

In Fig.~\ref{fig:PT_pattern_0} we show the three possible vacuum structures in the parameters space $(m_S,\lambda_{HS})$ for three different values of $\lams=0.1,1,3$. Once $\lams$ is fixed, the requirement for vacuum stability produces a lower bound on $\lamhs$. This bound is independent of the scalar mass $\ms$ and decreases slightly with increasing portal coupling $\lams$, thereby expanding the parameter space that allows for a proper and stable EW vacuum. 
For any value of $m_S$ above this lower bound, the potential always has a stable global minimum. As the coupling $\lamhs$ increases, the potential along the $s-$direction generates a barrier, and the vacuum structure transitions from Type~C to Type~B. If $\lamhs$ is sufficiently large, this barrier curls and forms a local minimum, leading to Type~A vacuum structure. 
However, $\lamhs$ cannot be arbitrarily large, or else the potential at the point $(v,0)$ will be lifted above the other minima, resulting in an improper EW vacuum. This sets an upper bound on $\lamhs$, which increases with $\ms$. 
On the other hand, for a small value of $\lams$, it becomes difficult to realize Type~A and Type~B vacuum structures. This is because the existence of Type~B (and Type~A) vacuum structure demands  the minimum value condition for vacuum: $(\lamhs v^2-\ms^2)^2<\lams\mu_{H}^4/\lamh $. As $\lams$ decreases, the relationship between $\lamhs$ and $\ms^2$ approaches a straight line. In other words, this region converges towards a parabolic curve in the $\ms-\lamhs$ plane.

%%%%%%%%%%%%%%%%%%%%%%%%%%%%%%%%
%%%%%%%%%%%%%%%%%%%%%%%%%%%%%%%%%

%%%%%%%%%%%%%%%%%%%%%%%%%
\subsection{Effects of loop corrections}

In addition to tree-level term given in \eqref{eq:V0}, the vacuum structure of the model is also influenced by loop corrections to the effective potential. 
The one-loop corrections, commonly known as the Coleman-Weinberg potential~\cite{Coleman:1973jx}, provide the dominant contribution. 
In the $\overline{\text{MS}}$ renormalization scheme, it takes the form
\beq
V_{\rm CW}(h,s)=\frac{1}{64\pi^2}\sum_i(-1)^{2s_i}n_i m_i^4(h,s)\left[\ln \frac{m_i^2(h,s)}{Q^2}-C_i\right]
\eeq
The sum over $i$ includes contributions from the top quark, $W^\pm$, $Z^0$ bosons, all Higgs bosons and Goldstone bosons. In the sum, $s_i$ and $n_i$ denote the spin and the numbers of degree of freedom for the $i$-th particle. The field-dependent squared masses $m_i^2(h,s)$ are functions of the two fields $h$ and $s$, and can be derived from the mass matrix as outlined in Appendix~\ref{sec:fielddepmass}. The renormalization scale $Q$ is fixed at $Q = v$. In the $\overline{\text{MS}}$ scheme used, 
$C_i = 5/6$ for gauge bosons and $C_i = 3/2$ for fermions and scalar bosons.

With the inclusion of $V_\text{CW}$, the EW vacuum $(v,0)$ at zero temperature is typically shifted. Consequently, a correction to the Higgs boson mass arises beyond the tree-level value $m_h^2$, which is given in Sec.~\ref{sec:model} and fixed at $125~{\rm GeV}$. To counteract this shift, a so-called ``counter-term" (CT) can be manually introduced, provided that the following relations are satisfied at the EW vacuum,
\beq
\label{eq:vevcond1}
\frac{\partial V_\text{CW}}{\partial h}\Big|_{(v,0)}=-\frac{\partial V_\text{CT}}{\partial h}\Big|_{(v,0)},\,\,
\frac{\partial^2 V_\text{CW}}{\partial h^2}\Big|_{(v,0)}=-\frac{\partial ^2V_\text{CT}}{\partial h^2}\Big|_{(v,0)},\,\,
\frac{\partial^2 V_\text{CW}}{\partial s^2}\Big|_{(v,0)}=-\frac{\partial ^2V_\text{CT}}{\partial s^2}\Big|_{(v,0)}.
\eeq

Respecting the invariance of gauge symmetry, the general form of the CT is parametrized as
\begin{align}
	V_\text{CT}(h,s)={1\over 2} \delta\mu_H^2 h^2+{1\over 4} \delta\lambda_H h^4+ {1\over 2} \delta\mu_S^2 s^2+{1\over 4} \delta\lams s^4+\frac{\delta\lambda_{HS}}{2}h^2s^2
\end{align} 
with 5 free coefficients to be determined. First, $\delta\mu_H^2$ and $\delta\lambda_H$ can be directly evaluated from the first two conditions in \eq{eq:vevcond1}. The remaining coefficients are determined by requiring the cancellation of shifts at other stationary points. 

For example, in Type~A and Type~B vacuum structures, the point $(0,w)$ is a local minimum and a saddle point, respectively. The coefficients $\delta\mu_S^2$ and $\delta\lambda_S$ can be simultaneously determined by applying the following conditions at this point
\beq
\label{eq:vevcond2}
\frac{\partial V_\text{CW}}{\partial s}\Big|_{(0,w)}=-\frac{\partial V_\text{CT}}{\partial s}\Big|_{(0,w)},\,\,\,\,
\frac{\partial^2 V_\text{CW}}{\partial s^2}\Big|_{(0,w)}=-\frac{\partial ^2V_\text{CT}}{\partial s^2}\Big|_{(0,w)}.
\eeq

In contrast, in Type~C vacuum structure, the point $(0,w)$ is not stationary, leaving $\delta \lambda_S$ unconstrained. In our analysis, we set $\delta \lambda_S=0$ and determine $\delta\mu_S^2$ by imposing the condition at the origin $(0,0)$
\beq
\frac{\partial^2 V_\text{CW}}{\partial s^2}\Big|_{(0,0)}=-\frac{\partial ^2V_\text{CT}}{\partial s^2}\Big|_{(0,0)}.\
\label{eq:vevcond3}
\eeq
Once $\delta\mu_S^2$ is determined, $\delta\lambda_{HS}$ can be derived from the last condition of \eq{eq:vevcond1}, completing the full solution for the CT term. It is worth noting that the construction of the CT is not unique.\footnote{For example, one of the conditions in \eq{eq:vevcond2} can be replaced by $\frac{\partial^2 V_\text{CW}}{\partial h^2}\Big|_{(0,w)}=-\frac{\partial ^2V_\text{CT}}{\partial h^2}\Big|_{(0,w)}$. We have verified that this alternative approach excludes a larger parameter space due to the requirement of a proper EW vacuum at zero temperature.}

While the constructed CT fully cancels the vacuum shift, it may fail to generate a sufficiently large potential to counteract the effect of $V_{\text{CW}}$ at the EW vacuum. If $V_{\text{CW}}$ raises the potential at the EW vacuum above that of other local minima, and the CT does not adequately lower the lifted potential, the proper EW vacuum established at tree level will be destabilized by loop corrections.
As illustrated in Fig.~\ref{fig:PT_pattern_0}, this destabilization effect becomes increasingly significant for larger values of $\lambda_S$.
 
%%%%%%%%%%%%%%%%%%%%%%%%%
\section{Field-dependent mass at finite temperature}
\label{sec:fielddepmass}
%%%%%%%%%%%%%%%%%%%%%%%%%

In this model, the masses of the scalar fields $H$ and $S$ are field-dependent, meaning their effective masses depend on the values of $h$ and $s$ at the potential minima. The field-dependent nature of the mass terms arises because the background values of the fields modify the potential, which in turn affects the mass terms.

The corresponding squared-mass matrix for the fields $h$ and $s$ is given by:
\begin{equation}
	\widehat{M}^2_{h,s}=	\left(\begin{array}{cc}   
		\frac{\partial^2 V_0}{\partial h^2} &\frac{\partial^2 V_0}{\partial h \partial s} \\  
		\frac{\partial^2 V_0}{\partial h \partial s} & \frac{\partial^2 V_0}{\partial s^2}  
	\end{array}
	\right) 
	=
	\left(\begin{array}{cc}   
		-\mu_H^2 +3\lambda_H h^2 +\lambda_{HS} s^2 &  2\lambda_{HS}hs \\  
		2\lambda_{HS}hs & -\mu_S ^2 +3\lambda_S s^2 +\lambda_{HS} h^2
	\end{array}
	\right) 
\end{equation}
where $\mu^2_H$ and $\mu^2_S$ are the mass parameters, and $\lamh$, $\lams$ and $\lamhs$ are the quartic couplings.
  
At the EW vacuum, which corresponds to $(v,0)$ at zero temperature (where $s=0$),  the off-diagonal term of the squared-mass matrix vanishes. This results in the physical masses for the Higgs boson and the new singlet particle, which are given by the eigenvalues of the above squared-mass matrix:
\begin{align}
m_h^2=m_h^2(v,0)&=3v^2\lambda_H-\mu_{H}^2\label{eq:hmss}, \\
m_S^2=m_s^2(v,0)&=\lambda_{HS}v^2-\mu_{S}^2\label{eq:smss}.
\end{align}

The field-dependent masses for other particles in the model, including the gauge bosons and fermions, are given by the following expressions:
\begin{align}
	m^2_{G^0,G^{\pm}}(h,s)&=-\mu_H^2 + \lambda_H h^2 +\lambda_{HS} s^2\\
	m^2_W(h,s)&=\frac{1}{4}g^2 h^2\\
	m^2_Z(h,s)&={1\over 4} (g^2+g'^2) h^2\\
	m^2_\gamma(h,s)&=0\\
	m^2_f(h,s)&=\frac{1}{2}y_f^2 h^2
\end{align}
The SM gauge couplings are defined at the $Z$-pole mass scale, $M_Z$, as:
\beq
\label{eq:inputsm}
	g'=\sqrt{4\pi \alpha_e(M_Z)},\,\, g=\alpha_e(M_Z)/\sin\theta_W,
\eeq
where the inputs are derived from precision electroweak data~\cite{particle2024review}: $ \alpha^{-1}_e(M_Z)=127.944$ and $M_Z = 91.1876~{\rm GeV}$, which yield $\sin^2\theta_W = 0.234$.

The Yukawa couplings are evaluated at the quark pole masses using the relation 
\beq
\label{eq:inputsm2}
y_f=\sqrt{2}m_f/v, 
\eeq
where $m_f$ is the measured quark mass. For example, the top quark mass is $172.5~{\rm GeV}$~\cite{ATLAS:2023eix}.

%%%%%%%%%%%%%%%%%%%%%%%
\section{Thermal corrections to the mass}
\label{app:thermal}
%%%%%%%%%%%%%%%%%%%%%%%%%

Thermal corrections to the mass of particles arise due to the interactions between particles and the thermal bath at finite temperatures. At high temperatures, these interactions modify the effective masses of particles, which can influence the PTs of the system. In the context of this model, these corrections are primarily due to the thermal distribution of particles, which can significantly affect the system's behavior.

The thermal self-energy corrections to the propagators for the fields are represented by the following matrix:
\beq
\begin{aligned}
	&\Pi_{h,s}=\left(\begin{array}{cc}   
		\Pi_{hh} & \Pi_{hs} \\  
		\Pi_{hs} & \Pi_{ss}  \\  
	\end{array}\right)
\end{aligned}
\eeq
where the individual components are given by:
\begin{align}
	&\Pi_{hh}=(\frac{3}{16}g^2+\frac{1}{16}g'^2+\frac{1}{2}\lambda_H+\frac{1}{4}y_t^2+\frac{1}{12}\lambda_{HS})T^2\label{eq:pih}\\
	&\Pi_{ss}=(\frac{1}{4}\lambda_S+\frac{1}{3}\lambda_{HS})T^2\label{eq:pis}\\
	&\Pi_{hs}\approx 0\\
	&\Pi_{G^0,G^{\pm}}=(\frac{3}{16}g^2+\frac{1}{16}g'^2+\frac{1}{2}\lambda_H+\frac{1}{4}y_t^2+\frac{1}{12}\lambda_{HS}) T^2
\end{align}

For the longitudinally polarized $W^\pm$ boson, the mass correction due to thermal effects is given by:
\begin{align}
	M^2_{W^{\pm}_{L}}=\frac{1}{4}g^2h^2+\frac{11}{6}g^2T^2
\end{align}
The masses of the longitudinal $Z$ boson and the photon ($\gamma_L$) are determined by diagonalizing the matrix that includes both the zero-temperature mass term and the thermal corrections:
\begin{align}
	\frac{1}{4}h^2
	\left(\begin{array}{cc}   
		g^2 & -gg' \\  
		-gg' & g'^2
	\end{array}
	\right) +
	\left(\begin{array}{cc}   
		\frac{11}{6}g^2T^2 & 0 \\  
		0 & \frac{11}{6}g'^2T^2
	\end{array}
	\right)
\end{align}

The eigenvalues of this matrix give the masses of the longitudinally polarized $Z$ boson and photon. They are:
\begin{equation}
	M^2_{Z_{L},\gamma_L}=\frac{1}{8}(g^2+g'^2)h^2+\frac{11}{12}(g^2+g'^2)T^2\pm\frac{1}{24}\Delta
\end{equation}
where
\begin{equation}
	\Delta^2=-176 g^2 g'^2 T^2 \left(3 h^2+11 T^2\right)+\left(g^2+g'^2\right)^2 \left(3 h^2+22 T^2\right)^2
\end{equation}

The thermal corrections to the masses of the $W^\pm$, $Z$ and photon fields are crucial for determining the behavior of the electroweak sector at high temperatures. These corrections shift the masses of the longitudinal components of the gauge bosons and alter the vacuum structure and EWPT patterns in the system. The key dependence on temperature comes from the thermal self-energy contributions, which introduce a temperature-dependent mass term that modifies the dynamics of the system, particularly near the critical temperature for electroweak symmetry breaking.

%%%%%%%%%%%%%%%%%%%%%%%%%%%%%%
\section{Two-loop contributions to the resummed effective potential} 
\label{app:2loop}
%%%%%%%%%%%%%%%%%%%%%%%%%

In this appendix, we provide the analytic expressions for the two-loop contributions from $h$ and $s$ fields to the effective potential, evaluated within the high-temperature approximation and incorporating thermal resummation following the Arnold–Espinosa scheme~\cite{Arnold:1992rz,Bahl:2024ykv}. 
As shown diagrammatically in Ref.~\cite{Bahl:2024ykv}, the two-loop contributions to the effective potential $\Delta V_{\rm eff, 2loop}^{\rm AE}$ include the figure-8 and sunset diagrams, along with one-loop mass and vertex counterterm diagrams, as well as thermal counterterm diagrams.
\beq
\Delta V_{\rm eff, 2-loop}^{\rm AE}=V_{\rm fig-8}+V_{\rm sunset}+V_{\rm c.t.}
\eeq

Since the scenario in which the EW symmetry is broken before the EWPT is excluded by the requirement of avoiding a Landau pole, we focus in this analysis on the vacuum structure along the $s$-field direction at high temperature. In particular, we investigate whether the inclusion of the two-loop contribution to the effective potential modifies the vacuum such that $s$ still acquires a nonzero expectation value, indicating breaking of $\mathbb{Z}_2$ symmetry. For this purpose, it is reasonable to set $h = 0$ to simplify the analysis. Below, we present the expressions for each contribution to the two-loop effective potential, evaluated in the high-temperature approximation and using the Arnold–Espinosa resummation scheme.

The finite contributions to the resummed two-loop effective potential, $\Delta V_{\rm eff, 2-loop}^{\rm AE}$, arising from the figure-8 and sunset diagrams are given by,
\beq
\begin{split}
V_{\rm fig-8} = & {3\lambda_H  \over 4}  I_0[m_h] I_0[m_h]+ {3\lambda_H \over 2} I_1[m_h] I_{-1}[m_h] \\
&+  {3\lambda_S \over 4} I_0[m_s] I_0[m_s] + {3\lambda_S \over 2} I_1[m_s] I_{-1}[m_s] \\
&+ {\lambda_{HS} \over 2} I_0[m_h] I_0[m_s] + {\lambda_{HS} \over 2} \left(I_1[m_h] I_{-1}[m_s]  + I_1[m_s] I_{-1}[m_h] \right),
\end{split}
\eeq
and
\beq
\begin{split}
V_{\rm sunset} & = -3 \lambda_S^2 s^2 H_0[m_s,m_s,m_s] - \lambda_{HS}^2 s^2 H_0[m_h,m_h,m_s]
\end{split}
\eeq
where we have expanded the Matsubara zero-mode resummed thermal functions of
\begin{align}
	\mathcal{I}[m]=\SumInt_{K}= \frac{1}{K^2+m^2} , \quad 
	\mathcal{H}[m_1,m_2,m_3]=\SumInt_{P,Q} \frac{1}{P^2+m_1^2}\frac{1}{Q^2+m_2^2}\frac{1}{(P+Q)^2+m_3^2}
\end{align}
as

\beq
\mathcal{I}_{\rm {zero-mode~resum}}[m]={1\over \eps} I_{-1} [m]+I_0[m] + \eps I_1[m] +\mathcal{O}(\eps^2)
\eeq
\beq
\mathcal{H}_{\rm {zero-mode~resum}}[m_1,m_2,m_3]={1\over \eps^2} H_{-2}+ {1\over \eps} H_{-1} +H_0+ \mathcal{O}(\eps)
\eeq
with~\cite{Laine:2017hdk,Ekstedt:2020qyp,Bahl:2024ykv} 
\begin{align}
I_{-1}[m] &= -{m^2 \over (4 \pi)^2},\\
I_0[m] & \simeq  {T^2 \over 12} - {M T \over 4 \pi} - {m^2 \over (4 \pi)^2} L_R +{2 m^4 \over (4 \pi)^4  T^2} \zeta(3), \\
I_1[m] & \simeq  {T^2 \over 6} \left({L_R \over 2} + \ln (2 \pi) - {\zeta'(2) \over \zeta(2)}\right) - {M T \over 2 \pi} \left[\ln \left({\bar{\mu} \over 2 M}\right) + 1\right]
- {2 m^2 \over (4 \pi)^2} \left[\left({L_R\over 2}\right)^2 - \gamma_E^2 - 2 \gamma_1 + {\pi^2\over 8}\right],\\
H_0[m_1,m_2,m_3] &=  {T^2 \over (4 \pi)^2} \left[ \ln \left( {\bar{\mu}\over M_1 + M_2 + M_3} \right) + {1\over 2}\right] \\&- 
 {T \over 64 \pi^3} \sum_i M_i \left[ 2\ln \left( {\bar{\mu}\over 2 M_i} \right) + L_R + 2\right] \\ &- 
 {1\over 256 \pi^4} (m_1^2 + m_2^2 + m_3^2) \left(L_R^2 + L_R - 2 \gamma_E^2 - 
    4 \gamma_1 + {\pi^2\over 4} + {3 \over 2}\right)\\&
    -  {3\zeta(2) \over 256 \pi^4} (M_1^2 + M_2^2 + M_3^2),
\end{align}
where $L_R\equiv 2 \left[ \ln(\bar{\mu}/T)+ \gamma_E - \ln(4 \pi)\right]$ with $\bar{\mu}$ being the MS renormalization scale, $\gamma_E\approx0.577$ is the Euler constant and $\gamma_1\approx -0.0728$ is the first Stieltjes constant and the Debye mass $M_i^2(T) \equiv m_i^2 +\Pi_i (T)$.

The total finite contribution to $\Delta V_{\rm eff, 2-loop}^{\rm AE}$ from the counter-term diagrams~\cite{Arnold:1992rz} is given by
\beq
\begin{split}
V_{\rm c.t.} & = \delta{m_h^2} I_1(m_h) + \delta{m_s^2} I_1(m_s)\\
& \quad\, + \delta_{\lams} s^2 I_1(m_s) + \delta_{\lamhs} s^2 I_1(m_h)\\
& \quad\, - {1\over 8\pi} \Pi_h T M_h - {1\over 8\pi} \Pi_s T M_s
\end{split}
\eeq
where the mass counter-terms are given by
\begin{align}
	\delta{m_h^2}=-{3\lambda_H\over 32\pi^2}\mu_H^2, \quad \delta{m_s^2}=-{3\lambda_S\over 32\pi^2}\mu_S^2
\end{align}
and the coupling counter-terms relevant for the resummation are:
\begin{align}
	\delta_{\lams}=\frac{27\lambda_S^2+3\lamhs^2}{32\pi^{2}}, \quad \delta_{\lamhs}=\frac{4\lambda_{HS}^2+3(\lamh+\lams)\lamhs}{32\pi^{2}}
\end{align}

So far, we have not specified how the field-dependent masses $m_i^2$, derived from the tree-level potential (see Eqs.~(\ref{eq:hmss}) and (\ref{eq:smss})), are replaced by the Debye masses $M_i^2(T)$ in the resummation procedure. A common approach uses the linear approximation $M_i^2(T)=m_i^2+c_i T^2$, where the thermal corrections $c_i T^2$, arising from the leading-order thermal self-energy, have been given in Eqs.~(\ref{eq:pih}) and (\ref{eq:pis}). While this is valid in the weak coupling and high-temperature limits, it may break down either near the temperature at which phase transition occurs or in strongly coupled scenarios.

To improve the accuracy of thermal mass resummation, we adopt the gap equation approach~\cite{Espinosa:1992gq,Espinosa:1992kf,Quiros:1992ez,Curtin:2016urg,Curtin:2022ovx,Bahl:2024ykv,Bittar:2025lcr}, in which $M_i^2(T)$ are computed self-consistently by evaluating the second derivative of the one-loop effective potential with respect to the corresponding field condensates. 
In our two-field model, this procedure yields a coupled system of equations in high-temperature limit~\cite{Bahl:2024ykv},
\beq
\begin{aligned}
		M_s^2&=m_s^2+c_sT^2-\frac{3\lambda_S}{4\pi}M_sT-\frac{\lamhs}{4\pi}M_hT 
		-\frac{3\lambda_S}{16\pi^2}L_RM_s^2-\frac{\lamhs}{16\pi^2}L_RM_h^2-\frac{L_R}{8\pi^2}(9\lambda_S^2+\lamhs^2)s^2,\\
			M_h^2&=m_h^2+c_hT^2-\frac{3\lambda_H}{4\pi}M_hT-\frac{\lamhs}{4\pi}M_s T
		-\frac{3\lambda_H}{16\pi^2}L_RM_h^2-\frac{\lamhs}{16\pi^2}L_RM_s^2-\frac{L_R}{8\pi^2}(9\lambda_H^2+\lamhs^2)h^2.
\end{aligned}
\eeq
We solve these equations numerically to obtain the self-consistent Debye masses $M_i(T)$, which are subsequently used in the resummed effective potential.

\begin{figure}[t]
\centering
\includegraphics[width=0.45\textwidth]{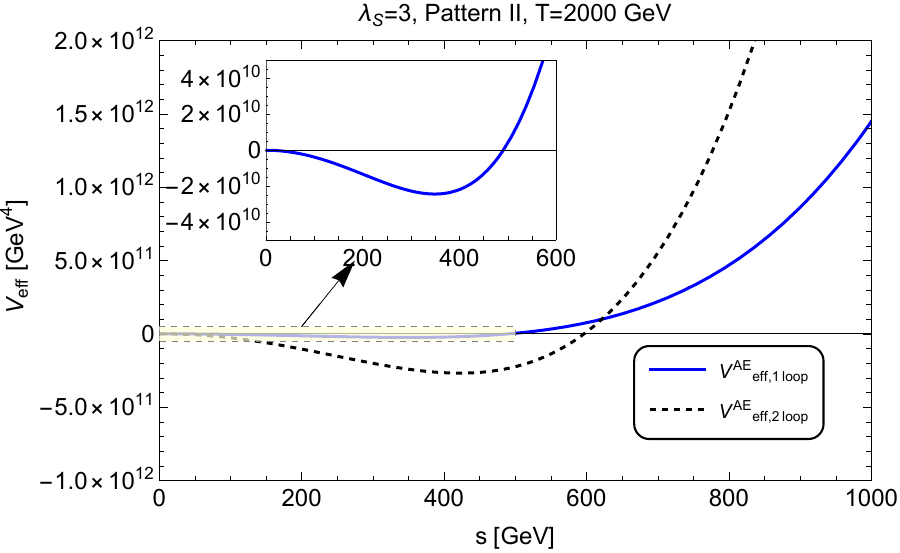}
\includegraphics[width=0.45\textwidth]{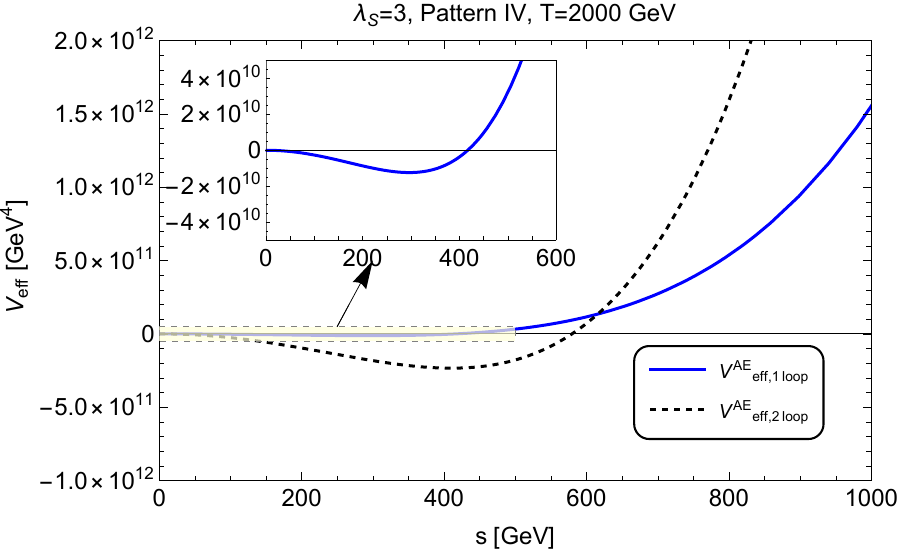}
\caption{
Comparison of one-loop and two-loop effective potentials. Shown are the one-loop effective potential $V_{\rm eff, 1-loop}^{\rm AE}$ and the two-loop result $V_{\rm eff, 2-loop}^{\rm AE}=V_{\rm eff, 1-loop}^{\rm AE}+\Delta V_{\rm eff, 2-loop}^{\rm AE}$, both evalauted at $T=2~{\rm TeV}$ along the $s$-field direction using the AE resummation scheme. Two benchmark points with relatively large couplings, $\lambda_S=3$ and $\lambda_{HS}=5.5$, are selected from Pattern II (left) and Pattern IV (right), as marked by stars in Fig.~\ref{fig:thigheq2}. In both cases, the minimum of the effective potentials remains at a non-zero thermal vev of $s$-field, confirming the stability of the $\mathbb{Z}_2$ broken vacuum. The two-loop contributions slightly enhance the thermal vev and significantly modify the shape of the effective potential compared to the one-loop result, particularly in the large-field region.}
	\label{fig:2loop_BPVeff}
\end{figure}

Based on the resummed two-loop effective potential, we compare the results with the one-loop potential for two benchmark points with relatively large couplings, $\lambda_S=3$ and $\lambda_{HS}=5.5$, selected from Pattern II and Pattern IV, respectively. The effective potentials along the $s$-field direction are evaluated at $T=2~{\rm TeV}$, with the comparison shown in Fig.~\ref{fig:2loop_BPVeff}. 
In both cases, the minimum of the effective potential remains at a non-zero thermal vev of the $s$-field condensate, indicating the qualitative stability of the $\mathbb{Z}_2$ broken vacuum. The inclusion of two-loop corrections slightly enhances the thermal vev and introduces noticeable modifications to the shape of the effective potential, particularly in the large-field region, suggesting that higher-order effects may play a non-negligible role away from the vacuum configuration.

%%%%%%%%%%%%%%%
%%%%%%%%%%%%%%%

%%%%%%%%%%%%%%%%%%%%%%%%%%%%%%
\section{Renormalization group equations} 
\label{app:RGEs}
%%%%%%%%%%%%%%%%%%%%%%%%%

RGEs describe the evolution of parameters (such as coupling constants) of a quantum field theory with respect to changes in the energy scale $\mu$. 
The one-loop RGEs for the quartic couplings $\lamh$, $\lamhs$ and $\lams$, along with the gauge and Yukawa couplings, are given by the following set of equations~\cite{Curtin:2014jma}:
\beq
\begin{aligned}
	16\pi^2\frac{d\lamh}{dt}&=\frac{3}{8}g'^4+\frac{9}{8}g^4+\frac{3}{4}g'^2g^2-6y_t^4+24\lamh^2+12y_t^2\lamh-3g'^2\lamh-9g^2\lamh+\frac{1}{2}\lamhs^2\\
	16\pi^2\frac{d\lamhs}{dt}&=\lamhs\Big(12\lamh+6\lams+4\lamhs+6y_t^2-\frac{3}{2}g'^2-\frac{9}{2}g^2\Big)\\
	16\pi^2\frac{d\lams}{dt}&=2\lamhs^2+18\lams^2\\
	16\pi^2\frac{dg'}{dt}&=\frac{41}{6}g'^3\\
	16\pi^2\frac{dg}{dt}&=-\frac{19}{6}g^3\\
	16\pi^2\frac{dg_s}{dt}&=-7g_s^3\\
	16\pi^2\frac{dy_t}{dt}&=y_t\Big(\frac{9}{2}y_t^2-\frac{17}{12}g'^2-\frac{9}{4}g^2-8g_s^2\Big)
\end{aligned}
\eeq
Here, $t\equiv\ln(\mu/m_z)$ represents the running scale, with $\mu$ being the renormalization scale. 

In our analysis these couplings are input at the $Z$-pole mass $M_Z$. 
In addition to two free parameters $\lamhs$ and $\lams$. the remaining parameters are fixed based on experimental values. These include: the Higgs quartic coupling $\lamh= 0.129$, the strong coupling $g_s=\sqrt{4\pi \alpha_s(M_Z)}$, where $ \alpha_s(M_Z)=0.1180$~\cite{particle2024review}, and other couplings as specified in \eqref{eq:inputsm} and \eqref{eq:inputsm2}.

To determine the energy scale of the Landau pole, we examine the coupling parameters $\lamhs$ and $\lams$. We then identify the energy scales at which these parameters diverge, taking the numerical divergence to be at $4\pi$. These energy scales are illustrated in Fig.~\ref{fig:ldpole}.
To understand the behavior of the theory at high energy scales, particularly the existence of a Landau pole, we examine the evolution of the coupling constants $\lamhs$ and $\lams$. The Landau pole occurs when a coupling diverges at some high energy scale. The point at which this divergence occurs can be numerically estimated by identifying the energy scale 
$\mu$ at which the coupling reaches a critical value, typically $4\pi$. The result is illustrated in Fig.~\ref{fig:ldpole}. This is often referred to as the energy scale where the theory becomes non-perturbative, and it is no longer valid to describe the interactions with a simple perturbative expansion.   

%%%%%%%%%%%%%%%%%%%%%%%%%%%%%
\section{Surface energy density of thin-wall bubbles}
\label{app:thin-wall}
%%%%%%%%%%%%%%%%%%%%%%%%%%%%%

For a bubble described by the field configuration $\vec{\phi}=(\phi_1,\phi_2,...)$, the free energy (relative to that of the false vacuum phase) is 
\beq
F[\vec{\phi}]=\int^\infty_0dr \ 4\pi r^2\Big[\sum_i\frac{1}{2}\big(\frac{d\phi_i(r)}{dr}\big)^2+\tilde{V}(\vec{\phi}(r))\Big]
\eeq
where $\tilde{V} (\vec{\phi}) \equiv V_\text{eff}(\vec{\phi})-V_\text{eff}(\vec{\phi}_F)$. 

%%%%%%%%%%%%%%%
\begin{figure}[t]
	\centering
	\includegraphics[width=0.55\textwidth]{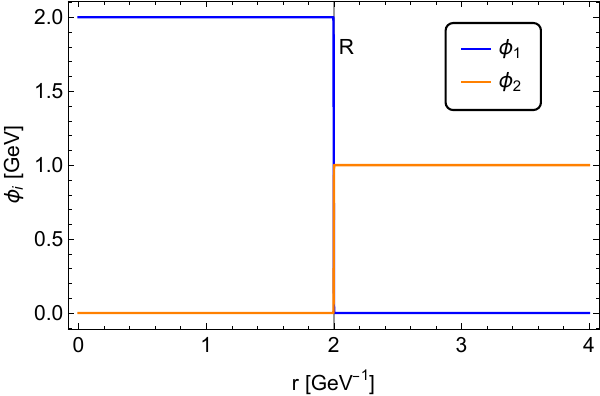}
	\caption{A schematic example of the field configuration for a thin-wall bubble consisting of two fields, $\phi_1$ and $\phi_2$ is presented. This example illustrates the spatial variation of the field values across the bubble wall, highlighting the sharp transition between the interior and exterior regions of the bubble. 
}
	\label{fig:tw}
\end{figure}
%%%%%%%%%%%%%%%

In general, finding an analytic solution to the bubble profile $\vec{\phi}$ is not possible. 
However, in the case of thin-wall bubbles, where the wall thickness is much smaller than the bubble radius, $L_w\ll R_b$ (an example of the two-field bubble profile is given in Fig.~\ref{fig:tw}), the calculation can be simplified by dividing the integral into three distinct regions: (i) inside the bubble $r<r_+\equiv R_b-L_w/2$, (ii) outside the bubble $r>r_-\equiv R_b+L_w/2$ and (iii) inside the wall $r_+ \leq r \leq r_-$. Inside the bubble, the field $\phi$ approaches its true vacuum value, $\phi_T$ and $\tilde{V}(\phi)=-\Delta V$, where $\Delta V= V_\text{eff}(\vec{\phi}_F)-V_\text{eff}(\vec{\phi}_T)$. Outside the bubble, the field approaches the false vacuum $\phi_F$, and $\tilde{V}(\phi)=0$. In both of these regions, $d\phi_i/dr$ is negligibly small and can be neglected, so the contribution from the integral outside the bubble is effectively zero. 
On the other hand, since the wall thickness $L_w$ is assumed to be very small, the factor $4\pi r^2$ inside the wall changes so slowly that can be approximated as a constant, $4\pi R^2_b$. Thus, the free energy becomes
\begin{align}
F[\vec{\phi}]&=\int^{r_+}_0 \! dr \ 4\pi r^2 (-\Delta V)
+4\pi R_b^2\int^{r_-}_{r_+}dr \left[\sum_i\frac{1}{2}\big(\frac{d\phi_i}{dr}\big)^2+\tilde{V}(\vec{\phi})\right]\\
&\simeq - \frac{4}{3}\pi R_b^3 \Delta V+4\pi R_b^2 \sigma
\end{align}
where $\sigma$ is the surface energy density of the bubble wall, defined as the energy stored per unit area of the wall surface:
\beq
\sigma=\int^{r_-}_{r_+}dr  \left[\sum_i\frac{1}{2}\big(\frac{d\phi_i}{dr}\big)^2+\tilde{V}(\vec{\phi})\right]
\eeq
The two terms in the integral are related to each other through the equation of motion for the extremum of the functional, as described in\eq{fig:tw}, which the field configuration obeys. 
\beq
	\frac{d^2\phi_i}{dr^2}=\frac{\partial \tilde{V}(\vec{\phi})}{\partial \phi_i} 
	\label{eq:tw}
\eeq
Compared with \eq{eq:eom}, the friction term proportional to $r^{-1}$ is absent for any field configuration inside the wall. While the form of the equations is relatively simple, it is not possible to solve for an individual $\phi_i$, unlike the single-field case. However, there is a differential relationship among all the fields that make up the bubbles.
\beq
	\sum_i\frac{1}{2}\Big(\frac{d\phi_i}{dr}\Big)^2=\tilde{V}(\vec{\phi})
\eeq
Given this relation, the surface energy density $\sigma$ in the general case becomes:
\beq
	\sigma=\int^{r_-}_{r_+} \!\!dr   \sum_i\big(\frac{d\phi_i}{dr}\big)^2 = \int^{\infty}_{0}\!\!dr  \sum_i\big(\frac{d\phi_i}{dr}\big)^2
\eeq
Note that the integral range has been formally extended from the bubble center to the infinity.

%%%%%%%%%%%%%%%%%%%%%%
\section{Modes of bubble nucleation}
\label{app:Nm}
%%%%%%%%%%%%%%%%%%%%%%

\begin{figure}[t]
	\centering
	\includegraphics[width=0.55\textwidth]{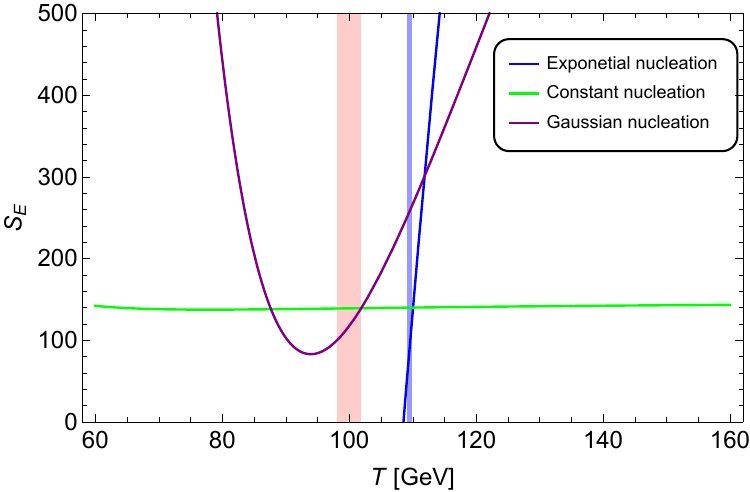}
	\caption{The typical curves of $S_E$  for various nucleation modes are illustrated. The blue shaded region indicates the temperature range where exponential nucleation occurs, while the pink shaded region highlights the temperature range within which other modes of bubble nucleation takes place.}
	\label{fig:nucl_mode}
\end{figure}

Depending on the order of expansion, three distinct analytical approximations for $S_E(t)$ typically emerge, each corresponding to a different nucleation mode~\cite{Megevand:2016lpr}. In Fig.~\ref{fig:nucl_mode} we illustrate the curves of $S_E$ for these various nucleation modes. 

%%%%%%%%%%%%
%%%%%%%%%%%%
 
(i) {\it Constant nucleation}. In this scenario, all higher-order terms in the expansion of $S_E$ are negligible except for the zeroth-order term. This results in:
\beq
S_E(t)=S_E(t_*)+R_0(t_*;t)
\eeq
where $R_0(t_*;t)$ is the zeroth-order remainder term, given in the Lagrange form by:
\beq
R_0(t_*;t)= \frac{dS_E(t)}{dt}\Big|_{t_\xi} (t-t_*)=-\beta_{\xi} (t-t_*)
\eeq
Here, $t_*<t_\xi<t$ and $\beta_\xi\equiv\beta(T_\xi)$.
 
Assuming that the PT occurs during the radiation-dominated (RD) period (as is assumed throughout this paper), the time difference 
$t-t_*=\left(H^{-1}(T)-H^{-1}(T_*)\right)/2$ and $\beta\equiv H(T)T\frac{dS_E(T)}{dT}$ decreases monotonically with decreasing temperature during the PT\footnote{Here, we focus on fast PTs where the assumption $T_n>T_p>T_m$, with $T_m$ being the temperature at which $S_E(T)$ extremizes and $\beta_m\equiv\beta(T_m)=0$ is reasonable. For slow PTs (e.g., supercooled PTs), scenarios where $T_p<T_n\sim T_m$ may arise, and the assumption of radiation dominance may break down. We exclude this scenario from our discussion.}.
Thus, we obtain the following estimate for $\left|R_0(t_*;t)\right|$:
\beq
\left|R_0(t_*;t)\right|<\left|\frac{\beta_*}{2}\left(\frac{1}{H}-\frac{1}{H_*}\right)\right|
=\frac{1}{2}\left|\frac{\beta_*}{H_*}\left(\frac{T_*^2}{T^2}-1\right)\right|
\eeq
where the relation $H(T) \sim T^2$ is used for the RD universe.

To ensure the error in this approximation is smaller than a given tolerance $\epsilon_0$, i.e., $\left|R_0(t_*;t)\right|=|S_E(t)-S_E(t_*)|\leq\epsilon_0$, we set $T_*=T_n$ and $T=T_p$. 
This gives the following condition:
\beq
\frac{1}{2}\left|\frac{\beta_n}{H_n}\right| \left(\frac{T_n^2}{T_p^2}-1\right) \leq\epsilon_0 
\label{eq:con}
\eeq

(ii) {\it Exponential nucleation}. In this case, the tunneling action $S_E(t)$ can be  approximated to first order as follows:
\beq
S_E(t)=S_E(t_*)+\frac{dS_E(t)}{dt}\Big|_{t_*}(t-t_*)+R_1(t_*;t) 
\eeq
where $R_1(t_*;t)$ represents the first-order remainder term, expressed as:
\beq
R_1(t_*;t)= \frac{1}{2} \frac{d^2S_E(t)}{dt^2}\Big|_{t_\xi} (t-t_*)^2
\eeq
where $t_*<t_\xi<t$. Introducing the additional parameter $\beta'^2(T)=H(T)^2T^2\frac{d^2S_E}{dT^2}$, we can expand this expression:
\beq
\left|R_1(t_*;t)\right|=\frac{1}{2} \left|3\beta_\xi H_\xi +\beta_\xi^{'2}\right| (t-t_*)^2
\eeq
Given a negligible change in $d^2S_E/dT^2$ during the PT, we can assume that $\beta'$ decreases monotonically with temperature during the PT. This yields:
\beq
\left|R_1(t_*;t)\right|<\frac{1}{8} \left| \frac{3\beta_*}{H_*} +\frac{\beta'^2_*}{H^2_*}\right| \left(\frac{T_*^2}{T^2}-1\right)^2.
\eeq

To ensure that the error in this approximation remains bounded by $\epsilon_1$ during the PT, specifically that $\left|R_1(t_*;t)\right|\leq\epsilon_1$, we set $T_*=T_n$ and $T=T_p$. This leads to the following condition:
\beq
\frac{1}{8} \left| \frac{3\beta_n}{H_n} +\frac{\beta'^2_n}{H^2_n}\right| \left(\frac{T_n^2}{T_p^2}-1\right)^2 \leq\epsilon_1 
\label{eq:exp}
\eeq

In addition, to distinguish this nucleation mode from constant nucleation, we impose the following condition:
\beq
\epsilon_0<\frac{1}{2} \left|\frac{\beta_p}{H_p}\right| \left(1-\frac{T_p^2}{T_n^2}\right) <\left|R_0(t_*;t)\right|
\label{eq:notcon}
\eeq

This criterion confirms the validity of the exponential nucleation approximation and differentiates it from constant nucleation. 
Exponential nucleation typically arises in the regime where the tunneling action is largely influenced by an exponentially suppressed rate, as depicted by the blue curve in Fig.~\ref{fig:nucl_mode}.

(iii) {\it Gaussian (Simultaneous) nucleation}. In the case of Gaussian (simultaneous) nucleation, when $t_*$ is selected at the extremum of $S_E(t)$, denoted as $t_m$, the first-order derivative of $S_E(t)$ becomes zero. Thus, $S_E(t)$ can be approximated to second order as follows:
\beq
S(t)=S(t_m)+\frac{1}{2}\frac{d^2S(t)}{dt^2}\Big|_{t_m}(t-t_m)^2+R_2(t_m;t) 
\eeq
where $R_2(t_m;t)$ is the second-order remainder term, expressed as:
\beq
R_2(t_m;t)= \frac{1}{3!}\frac{d^3S(t)}{dt^3}\Big|_{t_\xi} (t-t_m)^3
\eeq
where $t<t_\xi<t_m$.
In terms of $\beta$ and $\beta'$ parameters, it can be written as:
\beq
\left|R_2(t_m;t)\right|=\frac{1}{6}\left|15\beta_\xi H_\xi^2+9\beta_\xi^{'2}H_\xi+\beta^{''3}_\xi \right| (t_m-t)^3
\eeq
where $\beta''^3(T)=H(T)^3T^3\frac{d^3S_E}{dT^3}$. 
For the case where $\beta''>0$ during the PT within temperature range from $T_n$ to $T_p$, we have
\beq
\left|R_2(t_m;t)\right|<\frac{1}{48} \left|\left(\frac{15\beta}{H}+\frac{9\beta^{'2}}{H^2}\right)\right|\left(\frac{T^2}{T_m^2}-1\right)^3
\eeq

To ensure that the error in this approximation is bounded by $\epsilon_2$ during the PT from $T_n$ to $T_p$, specifically, that $\left|R_2(t_m;t)\right|\leq\epsilon_2$, we set $T=T_n$. Therefore, the following condition must be satisfied:
\beq
\frac{1}{16} \left|\left(\frac{5\beta_n}{H_n}+\frac{3\beta^{'2}_n}{H_n^2}\right)\right|\left(\frac{T_n^2}{T_m^2}-1\right)^3\leq\epsilon_2
\label{eq:sim}
\eeq

On the other hand, note that $R_0(t_m;t)=R_1(t_m;t)$ in this case. To distinguish this from constant nucleation, one should impose the condition $|R_0(t_m;t_n)|>\epsilon_0$, which is 
satisfied provided that
\beq
|R_0(t_m;t_n)|=\frac{1}{8}\frac{\beta_m^{'2}}{H_m^2}\left(1-\frac{T_m^2}{T_n^2}\right)^2>\epsilon_0.
\eeq

\bibliographystyle{JHEP}
\bibliography{biblio}

\end{document}